\documentclass{aa} 
\usepackage{graphicx,natbib,psfig} 
\newcommand{\ltsima} {$\; \buildrel < \over \sim \;$} 
\newcommand{\kms}{$\rm{\,km \,s}^{-1}$}
\defcitealias{paper1}{CB05}
\defcitealias{paper2}{BC06}
\defcitealias{paper3}{CB06}
\begin{document} 

\title{The host galaxy/AGN connection\thanks 
{Based on observations obtained at the Space
Telescope Science Institute, which is operated by the Association of
Universities for Research in Astronomy, Incorporated, under NASA
contract BAS 5-26555.}.} \subtitle{Brightness profiles of
early-type galaxies hosting Seyfert nuclei.}

\titlerunning{Brightness profiles of
Seyfert galaxies} \authorrunning{A. Capetti and B. Balmaverde}
 
\author{Alessandro Capetti \inst{1} \and Barbara Balmaverde \inst{1}}
 
\offprints{A. Capetti}
 
\institute{INAF - Osservatorio
Astronomico di Torino, Strada Osservatorio 20, I-10025 Pino Torinese,
Italy\\ \email{capetti@oato.inaf.it} 
\\ \email{balmaverde@oato.inaf.it} }

\date{}
 
\abstract{We recently presented evidence of a connection between the
brightness profiles of nearby early-type galaxies and the properties
of the AGN they host. The radio loudness of the AGN appears to be
univocally related to the host's brightness profile: radio-loud nuclei
are only hosted by ``core'' galaxies while radio-quiet AGN are only found
in ``power-law'' galaxies.
We extend our analysis here to a sample of 42 nearby (V$_{\rm
rec} <$ 7000 \kms) Seyfert galaxies hosted by early-type
galaxies. From the nuclear point of view, they show a large deficit of
radio emission (at a given X-ray or narrow line luminosity) with respect
to radio-loud AGN, conforming with their identification as
radio-quiet AGN. 

We used the available HST images to study their brightness
profiles. Having excluded complex and highly nucleated galaxies, in the
remaining 16 objects the brightness profiles can be successfully
modeled with a Nuker law with a steep nuclear cusp characteristic of
``power-law'' galaxies (with logarithmic slope $\gamma= 0.51 - 1.07$).
This result is what is expected for these radio-quiet AGN 
based on our previous findings, thus
extending the validity of the connection between brightness
profile and radio loudness to AGN of a far higher 
luminosity.

We explored the robustness of this result against a different choice
of the analytic form for the brightness profiles, using a S\'ersic law. In
no object  could we find evidence of a central light deficit with respect
to a pure S\'ersic model, the defining feature of ``core'' galaxies in
this modeling framework. 
We conclude that, regardless of the modeling strategy, 
the dichotomy of AGN radio loudness can be univocally related to
the host's brightness profile. 
Our general results can be re-phrased as 
``radio-loud nuclei are hosted by core galaxies, while radio-quiet AGN 
are found in non-core galaxies''.

} \maketitle

\section{Introduction.}
\label{intro}

The study of active galaxies attained a new role following recent
developments in our understanding of the nuclear structure of galaxies and of
its connection with the process of galactic formation and evolution. It is 
now clear that not only massive galaxies host a supermassive black hole
(SMBH) in their centers but also that SMBH and host galaxies follow a common
evolutionary path. This is suggested by the presence of tight relationships
between the SMBH mass and the stellar velocity dispersion
\citep{ferrarese00,gebhardt00}, as well as by the mass of the spheroidal
component of their hosts \citep[e.g.][]{marconi03,haring04}. There is also
increasing evidence that, in the co-evolution of the SMBH/galaxy system,
nuclear activity plays a major role in one of the 
many different forms known as active galactic
nuclei (AGN). In fact, the energy liberated in the
accretion process generates a feed-back process acting on the host galaxy
\citep{dimatteo05}, e.g. suppressing the star formation in massive galaxies
\citep{croton06}. Thus AGN obviously represent, on the one hand, our best
tool for investigating formation and growth of SMBH. But on the other, nuclear
activity also strongly influences galaxy evolution.

But, despite this breakthrough in our understanding of the SMBH/galaxy system,
we still lack a clear picture of the connection between the properties of AGN
and of their host galaxies. One of the crucial issues here is to explain the
so-called AGN radio loudness dichotomy \citep[e.g.][]{kellermann94} which
essentially corresponds to the ability of the central engine to produce highly
collimated relativistic jets. Understanding the origin of this dichotomy is
clearly important from the point of view of AGN physics.
But, since it corresponds to different modes of 
energy transfer from the AGN into the host galaxy, it is also related 
to different feedback processes that couple nuclear
activity and galaxy evolution.
In this context, previous studies have not provided 
clear-cut answers.
It is in fact well-established that
spiral galaxies preferentially harbor radio-quiet AGN,
but early-type galaxies can host both radio-loud and radio-quiet
AGN. Similarly, radio-loud AGN are generally associated with the most
massive SMBH as there is a median shift between the radio-quiet and
radio-loud distribution, but both distributions are broad and overlap
considerably \citep[e.g.][]{dunlop03}. 

In a series of three papers \citep[ hereafter CB05, CB06, BC06]
{paper1,paper3,paper2} we recently re-explored the classical
issue of the connection between the
multiwavelength properties of AGN in nearby early-type galaxies and the
characteristics of their hosts. Early-type galaxies appear to be the critical
class of objects, as they host AGN of both classes of radio loudness. 
Starting from an initial sample of 332 galaxies, we selected 116 AGN 
candidates
(requiring the detection of a radio source with a flux limit of $\sim$ 1 mJy,
as measured from 5 GHz VLA observations). In \citetalias{paper1} we 
analyzed the 65 objects with
available archival HST images where a classification 
into ``core'' and ``power-law'' galaxies was possible for 51 objects,
distinguishing them on the basis of the nuclear slope of their brightness
profiles following the modeling scheme proposed by \citet{lauer95}. 

We used HST and Chandra data to isolate the nuclear emission of these galaxies
in the optical and X-ray bands, thus enabling us (once combined with the radio
data) to study the multiwavelength behavior of their nuclei. The properties
of the nuclei hosted by the 29 ``core'' galaxies were presented in
\citetalias{paper2}. ``Core'' galaxies invariably host a radio-loud nucleus,
with a median radio loudness of Log R = 3.6 and an X-ray based radio loudness
parameter of Log R$_{\rm X}$ = -1.3. In \citetalias{paper3} we discussed the
properties of the nuclei of the 22 ``power-law'' galaxies. They show a
substantial excess of optical and X-ray emission with respect to ``core''
galaxies at the same level of radio luminosity. Conversely, their
radio loudness parameters, Log R $\sim$ 1.6 and Log R$_{\rm X} \sim$ -3.3, are
similar to those measured e.g. in Seyfert galaxies selected from the
Palomar survey \citep{ho97} for which Log R
$\sim$ 1.9 and Log R$_{\rm X} \sim$ -3.6 \citep{panessa07} . 

As already noted by \citet{ho01}, if we were to adopt the classical dividing
line between radio-loud and radio-quiet objects introduced by
\citet{kellermann94}, i.e. at Log R = 1, a substantial fraction of Seyfert and
``power-law'' galaxies should be considered as radio-loud.  However, the
recent work by \citeauthor{panessa07} shows that 
at low luminosities (with respect
to high-luminosity QSO for which the radio loudness definition was originally
proposed) AGN still separate into two different populations of 
radio loudness, but the two classes are
optimally distinguished by using Log R $\sim$ 2.4 and Log
R$_{\rm X} \sim$ -2.8 as thresholds. The difference between these values
and the traditional separation drawn at Log R = 1 
is most likely an indication of a luminosity evolution
of the level of radio loudness. 

Adopting this definition,
the radio loudness of AGN hosted by early-type galaxies 
can be univocally
related to the host's brightness profile: radio-loud AGN are only hosted by
``core'' galaxies, while radio-quiet AGN are found only in ``power-law''
galaxies.

Since the brightness profile is determined by the galaxy's evolution
through its merger history \citep[e.g.][]{faber97,ryden01,khochfar03}, 
our results suggest that the same process sets the AGN
flavor. In this scenario, the black holes hosted by the merging galaxies
rapidly sink toward the center of the newly formed object, setting its nuclear
configuration, described by e.g. the total mass, spin, mass ratio, or
separation of the SMBHs. These parameters are most likely at the origin of
the different levels of the AGN radio loudness. For example, it has been
proposed that a ``core'' galaxy is the result of (at least) one major merger
and that the core formation is related to the dynamical effects of the binary
black holes on the stellar component \citep[e.g.][]{milo02,ravi02}. From the
AGN point of view, \citet{wilson95} suggested that a radio-loud source can
form only after the coalescence of two SMBH of similar (high) mass, forming a
highly spinning nuclear object, from which the energy necessary to launch a
relativistic jet can be extracted. In this situation (the merging
of two large galaxies of similar mass), the expected outcome is a massive
``core'' galaxy in line with our results.
The connection of the radio loudness
with the host's brightness profile might open a
new path toward understanding the origin of the radio-loud/radio-quiet AGN
dichotomy, and it 
provide us with a further tool for exploring the co-evolution of
galaxies and supermassive black holes. A better understanding 
of this issue would put
us in the position of relating the manifestation of nuclear activity in a
given galaxy with its formation history.

We present here the analysis of the brightness profiles of a sample of 42
nearby (cz$\leq$ 7000 km s$^{-1}$) Seyfert galaxies hosted by early-type
galaxies. This study is motivated by the extremely low nuclear
luminosity of the objects we selected in \citetalias{paper1}, a 
characteristic of volume-limited surveys in which the rare high-luminosity
objects tend to be under-represented. Indeed the objects we considered
previously extend only, with just a few exceptions, to an X-ray luminosity
of up to $\sim$ 10$^{41}$ erg s$^{-1}$. This lead us to  mainly explore a
regime of a very low level of AGN activity. The nature of these 
low luminosity objects and
their relationship with brighter AGN is still a matter of much debate
\citep[e.g.][]{chiaberge05,maoz07}.
It is thus important to establish whether the
results we obtained for these weakly active galaxies,
relating the properties of the AGN with those of their host galaxy,
can be extended to
objects more representative of the overall AGN population. Considering
Seyfert galaxies we indeed extend the coverage up to an X-ray luminosity of
$\sim$ 10$^{43}$ erg s$^{-1}$. The analysis of a sample of Seyfert galaxies
thus represents a significant step forward for the study of the AGN/host galaxy
connection. In particular this study will enable us to test the prediction
based on our previous findings that these radio-quiet AGN should be hosted by
``power-law'' galaxies.

With respect to the initial series of 3 papers we include here a full
modeling of the brightness profiles also with a \citet{sersic68} model. There
are several reasons suggesting this approach. First of all, \citet{graham03}
argue that a S\'ersic model provides a better characterization of the
brightness profiles of early-type galaxies, in particular considering their
large-scale curvature (in a Log-Log space) and, furthermore, that a S\'ersic
fit can reproduce the brightness profiles of dwarf ellipticals
\citep{graham03b} with very low values for the nuclear slope. They also
suggest a new definition of ``core'' galaxy as the class of objects showing a
light deficit toward the center with respect to the S\'ersic law
\citep{trujillo04}. Recently, an analysis of the brightness profiles of 100
early-type galaxies in the Virgo cluster (the Virgo Cluster Survey, VCS) 
has been presented by
\citet{ferrarese06}. This work provides us with a useful benchmark for 
interpreting our results within this modeling scheme not previously
available in the literature. On the other hand, \citet{lauer06} question
these results, arguing that S\'ersic models are not a good representation of
the central regions of the surface brightness profiles of early-type galaxies.
Since the situation is far from settled and there is significant controversy
on this issue, we preferred to use both analytic forms. Quite reassuringly, we
show that our results are independent of the fitting scheme and we
recover a unique correspondence between the host's brightness profile and the
AGN properties.

The structure of the paper is as follows. In Sect. 
\ref{sample} we describe the sample studied and in Sect. \ref{fit}
we present the HST images used to derive the surface brightness
profiles that are modeled using both a Nuker and a S\'ersic model.
The multiwavelength properties of our sample are discussed
in Sect. \ref{nuc} where we show that they all conform
to the definition of radio-quiet AGNs. In Sect. \ref{summary}
we summarize and discuss our results.

We adopt a Hubble constant H$_0$= 75 km s$^{-1}$ Mpc$^{-1}$.

\section{Sample selection and basic data}
\label{sample}

A detailed description of the selection criteria of the sample
analyzed in this paper can be found in 
\citet{mulchaey96} and \citet{nagar99}. Briefly, \citeauthor{mulchaey96} 
selected their
sample of Seyfert galaxies with early-type hosts from the catalogue of
\citet{hewitt91}, J. P. Huchra (1989, private communication) and
\citet{veron91}, restricting it to a range of magnitude and recessional
velocity for which morphological classification is complete and
reliable, i.e. total magnitude m$_V\leq$ 14.5 and recessional velocity
cz$\leq$ 7000 km s$^{-1}$.

High-quality VLA radio observation of the galaxies of the sample were
presented by \citet{nagar99}, with the further requirement of a
declination greater than $\delta$ =-41$^{\circ}$. Three objects later
found to satisfy the selection criterion (namely NGC 7743, Mrk 335, and
Mrk 612) were added to the original list by \citeauthor{nagar99} Since
the radio data are a crucial ingredient for our study, we restrict the
analysis to this group of 42 galaxies\footnote{We also excluded
MCG-2-27-9 since it was not observed by \citeauthor{nagar99}} 
that form our final sample.

In Table \ref{tabsample1} we report the basic data of these Seyfert
galaxies: name, Seyfert type, recession velocity (corrected from Local
Group infall into Virgo), total K-band magnitude from 2MASS, the radio
flux at 3.6 and 20 cm from \citet{nagar99} (when the radio-source is
extended, we also give the radio-core flux) and the stellar velocity
dispersion from the HYPERLEDA database (available for 26 sources).

\begin{table*}
\caption{Basic properties of the Seyfert sample}
\label{tabsample1}
\begin{tabular}{l l c c c c c c c c}
\hline\hline
Name        & Host Type & Sy Type & V  & m$_K$ & F$_{\rm 3.6 cm}$ & F$_{\rm 20cm}$ & F$_{\rm core 3.6 cm}$ & F$_{\rm core 20 cm}$  & $\sigma$ \\
\hline		      		
MRK 335      & S0/a    & 1     &   7698      & 10.059 $\pm$ 0.030  &  2      & 6.8     &       &         &         \\
MRK 348      & SA0/a   & 2   &   4624 	& 10.097 $\pm$ 0.047  &	238.0  & 302.2   &   	 &         & 118  \\
NGC 424      & SB0/a   & 2   &   2154	& 9.129 $\pm$  0.021  &	13.1   & 23.9    &  12.2 &  23.9   &      \\	
NGC 526A     & S0 pec? & 2   &   5466      & 10.436$\pm$  0.050  &  5.0    & 10.5    &  5.0  &  5.9    &      \\
NGC 513      & S0?     & 2   &   5510	& 9.914 $\pm$  0.020  &	...    & 41.2    &  ...  &  4.2	   & 152  \\
MRK 359      & SB0 pec & 1.5 &   5125	& 10.461 $\pm$ 0.026  &  0.5    & 2.4	 &	 &	   &         \\
MRK 1157     & SB0/a   & 2   &   4674	& 10.063$\pm$  0.028  &	4.9    & 25.7    &  	 &         & 95   \\
MRK 573      & SAB0    & 2   &   5139	& 10.385$\pm$  0.030  &	1.8    & 14.3    &  	 &         & 123  \\ 
NGC 788      & SA0/a   & 2   &   3806	& 9.071 $\pm$  0.025  &	0.7    & 2.9     &       &     	   & 140  \\
ESO 417-G6   & SA0/a?  & 2   &   4699	& 10.237$\pm$  0.043  &	1.0    & 3.1     &  	 & 	   &      \\
MRK 1066     & SB0     & 2   &   3733	& 9.793 $\pm$  0.023  &	16.4   & 96.3    &  4.8  &  95.3   & 105  \\	
MRK 607      & S0/a    & 2   &   2612	& 9.359 $\pm$  0.029  &	1.3    & 3.7     &  	 & 	   & 116  \\
MRK 612      & SBa     & 2   &   6053	& 10.650$\pm$  0.036  &	2.2    & 8.2     &  	 & 	   &      \\
NGC 1358     & SAB0/a  & 2   &   3924	& 8.948 $\pm$  0.032  &	0.9    & 3.4     &   	 &         & 173  \\
NGC 1386     & Sa/S0   & 2   &   610	& 8.066 $\pm$  0.014  &	9.1    & 28.8    &  	 & 	   & 120  \\
ESO 362-G8   & S0?     & 2   &   4520	& 9.568 $\pm$  0.024  &	0.8    & 2.7     &  0.4  &  2.7    &      \\
ESO 362-G18  & S0/a    & 1.5 &   3550	& 10.025 $\pm$ 0.033  &  2.8    & 6	 &	 &	   &         \\
NGC 2110     & SA0/SBa?& 2   &   2064	& 8.144 $\pm$  0.019  &	130.1  & 289.0   &  81.2 &  289.0  & 220  \\
MRK 3        & E2 pec  & 2   &   4248	& 8.970 $\pm$  0.019  &	79.0   & 1060    &  50.1 &  1060   & 269  \\
MRK 620      & SAB0/a,SBab & 2   &   2049	& 8.480 $\pm$  0.020  &	10.2   & 52.0    &  7.6	 &  52.0   & 124  \\
MRK 6        & SAB0+,Sa& 1.5 &   6101	& 9.560  $\pm$ 0.017  &  30     & 253	 &	 &	   &          \\
MRK 10       & SAb,SBbc& 1.2 &   8968	& 10.345 $\pm$ 0.038  &  ...    & 0.5	 &	 &	   & 143$^{a}$\\
MRK 622      & S0 pec  & 2   &   7094	& 11.078$\pm$  0.064  &	1.7    & 6.0     &   	 & 	   & 100  \\
MCG -5-23-16 & S0      & 2   &   2294	& 9.349 $\pm$  0.021  &	...    & 11.0    &   	 & 	   & 210$^{a}$  \\
MRK 1239     & compact & 1.5 &   5684	& 9.603  $\pm$ 0.015  &  7.9    & 56.5	 &	 &	   & 263$^{a}$\\
NGC 3081     & SAB0/a  & 2   &   2243	& 8.910 $\pm$  0.028  &	1.0    & 3.5     &   	 & 	   &      \\
NGC 3516     & SB0,SB0/a& 1.2 &   2902	& 8.512  $\pm$ 0.027  &  4.1    & 9.4	 &	 &	   &  235      \\
NGC 4074     & S0 pec  & 2   &   6875	& 10.566$\pm$  0.028  &	0.8    & 2.0     &   	 &         & 192  \\
NGC 4117     & S0      & 2   &   1143	& 10.047$\pm$  0.031  &$<$0.1   & 2.2     &   	 & 	   & 95   \\
NGC 4253     & SB0/a,SBa & 1.5 &   4038	& 9.839  $\pm$ 0.022  &  8.6    & 39.3	 &	 &	   &           \\
ESO 323-G77  & SAB0    & 1.2 &   4352	& 8.802  $\pm$ 0.017  &  1.3    & 31.9	 &	 &	   &           \\
NGC 4968     & SAB0    & 2   &   2858	& 9.481 $\pm$  0.037  &	6.5    & 32.3    &   	 & 	   &      \\
MCG -6-30-15 & E-S0    & 1.2 &   2168	& 9.582  $\pm$ 0.017  &  0.9    & 4	 &	 &	   & 162$^{a}$  \\
NGC 5252     & S0      & 2   &   6773      & 9.768 $\pm$  0.037  &  9.3    & 17.2    &  7.9  &  13.5   & 190  \\
MRK 270      & SAB0    & 2   &   3152      & 9.974 $\pm$  0.028  &	3.1    & 13.9    &  	 & 	   & 148  \\
NGC 5273     & SA0     & 1.5 &   1316	& 8.665  $\pm$ 0.024  &  0.6    & 2.4	 &	 &	   &  79        \\
IC 4329A     & S0      & 1.2 &   4660	& 8.805  $\pm$ 0.013  &  10.7   & 60	 &	 &	   &  235$^{a}$ \\
NGC 5548     & SA0/a,Sa & 1.2 &   5310	& 9.387  $\pm$ 0.019  &  3.1    & 23	 & 3.1   &  6	   &  48$^{a}$  \\
ESO 512-G20  & SB0     & 1   &   3309	& 10.447 $\pm$ 0.049  &  1.2    & 3.6	 &	 &	   &   \\
IC 5169      & SAB0 pec& 2   &   2943	& 9.776 $\pm$  0.015  &	3.7    & 17.6    &   	 & 	   &      \\
NGC 7465     & SB0     & 2   &   2046	& 9.542 $\pm$  0.021  &	1.2    & 6.0     &   	 &         &      \\
NGC 7743     & SB0     & 2   &   1725	& 8.418 $\pm$  0.028  &	0.9    & 5.3     &   	 &         & 83   \\
\hline
\end{tabular}

Column description: (1) Name, 
(2) Host type
(3) Seyfert type
(4) recession velocity in km s$^{-1}$ corrected for LG infall onto Virgo from HYPERLEDA, 
(5) total K band galaxy's magnitude from 2MASS, 
(6) total radio-flux [mJy] at 3.6 cm, 
(7) total radio-flux [mJy] at 20 cm,
(8) nuclear radio-flux[mJy] at 3.6 cm, 
(9) nuclear radio-flux[mJy] at 20 cm, 
(10) stellar velocity dispersion in km s$^{-1}$ 
from \citet{nelson95} or from $^{a}$ HYPERLEDA.
\end{table*}
	    
\begin{figure*}
\centerline{ 
\psfig{figure=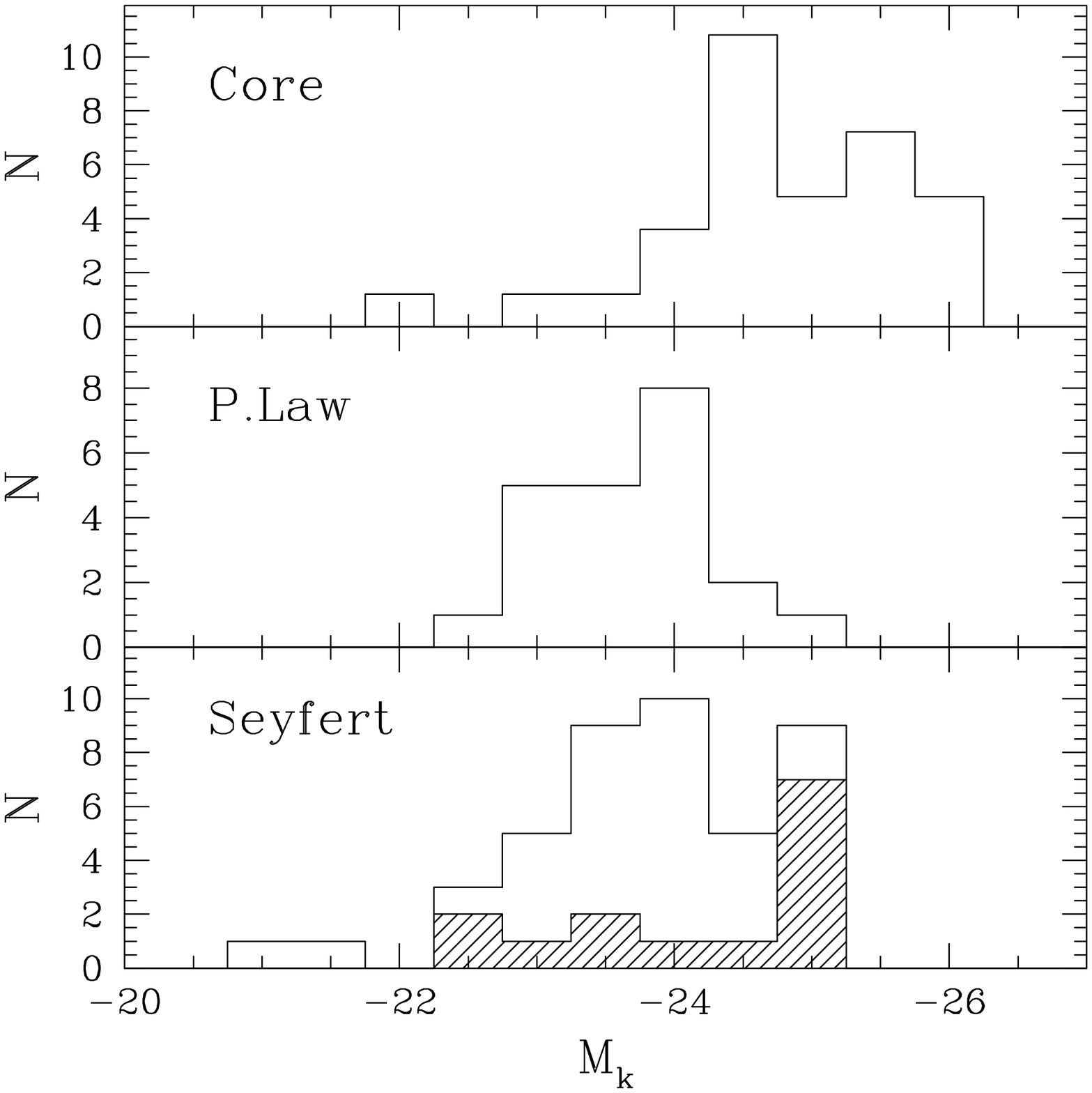,width=0.5\linewidth}
\psfig{figure=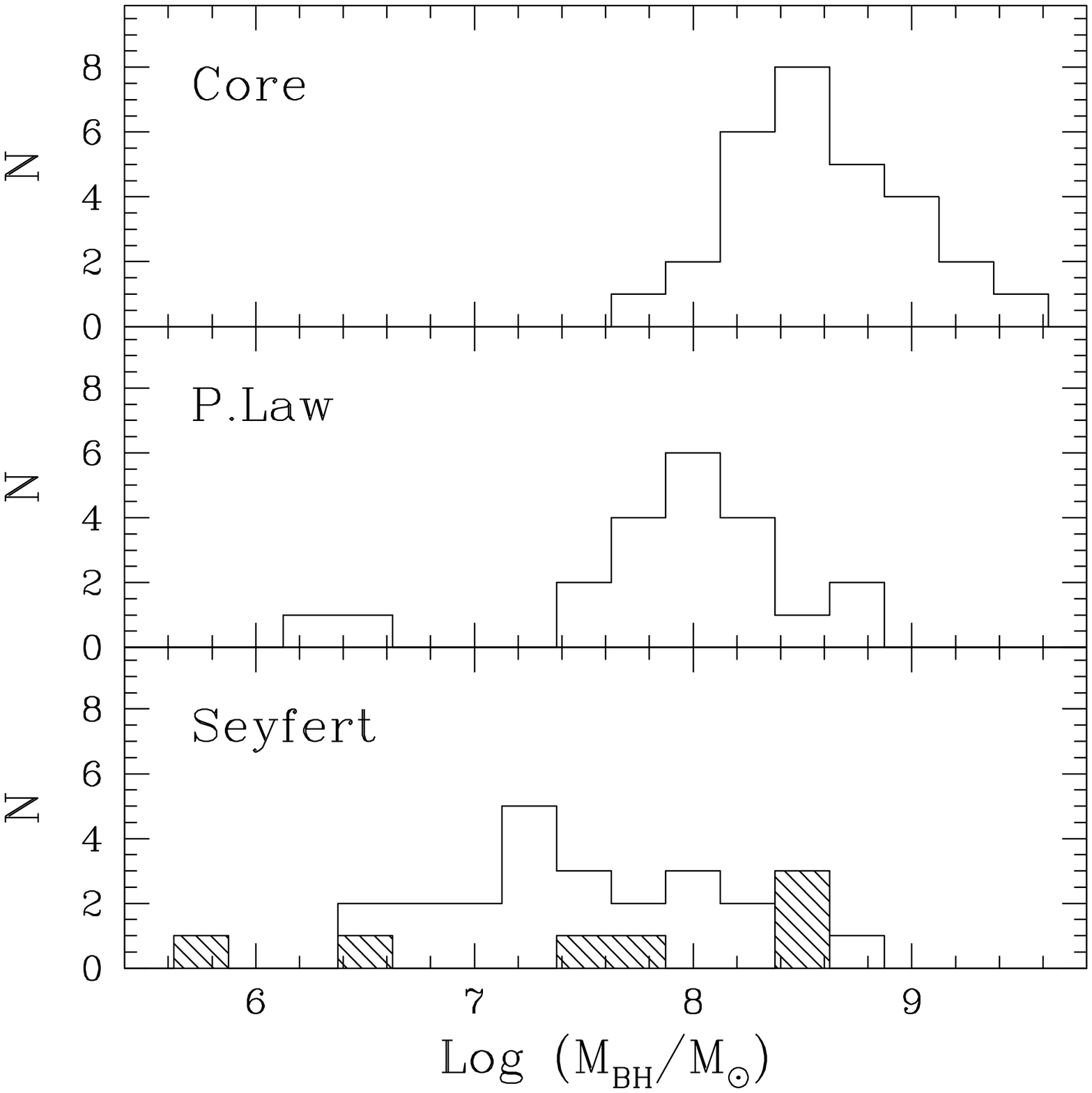,width=0.5\linewidth} }
\caption{Distribution of absolute magnitudes M$_K$ (left panel) and
of black hole masses M$_{BH}$ (right panel) compared for the
two sub-samples of the early-type galaxies 
we analyzed in \citetalias{paper1} (separated
into ``core'' and ``power-law'' galaxies) and for the present sample
of Seyfert galaxies. The shaded region in the bottom panels marks 
the contribution of type I Seyfert.}
\label{hist}
\end{figure*}

In Fig. \ref{hist} we compare the K band absolute magnitudes 
and the estimated black-hole masses\footnote{Estimated using
the relationship with the stellar velocity dispersion 
in the form given by \citet{tremaine02}.} for the sample of Seyfert galaxies
against the two sub-samples of early-type galaxies we analyzed 
in \citetalias{paper1}, separated
into ``power-law'' and ``core'' galaxies.
The luminosity distributions of Seyfert hosts 
and of ``power-law'' galaxies match closely, with the same median value 
of $M_K=-23.8$, while they are both fainter (by about 1 mag) 
than ``core'' galaxies for which the
median magnitude is $M_K=-24.8$.
Nonetheless a large overlap between the three classes is found
in the region $-23.5 < M_K < -25$. A slightly different result
is found when comparing the distribution of black-hole masses:
Seyfert, ``power-law'', and ``core'' galaxies 
have median values of Log $M_{BH} = 7.6$, 8.0,
and 8.5 respectively. But again, there is a broad region of overlap.
We note that stellar velocity dispersion measurements
are available for only 26 Seyfert galaxies of our sample 
and, in general, they are affected by substantial uncertainties due to the
dilution of the absorption lines caused by the active nucleus.
The comparison between the $M_{BH}$ distributions should be then treated
with some caution.

\section{Surface brightness profile analysis}
\label{fit}

\begin{figure*}
\psfig{figure=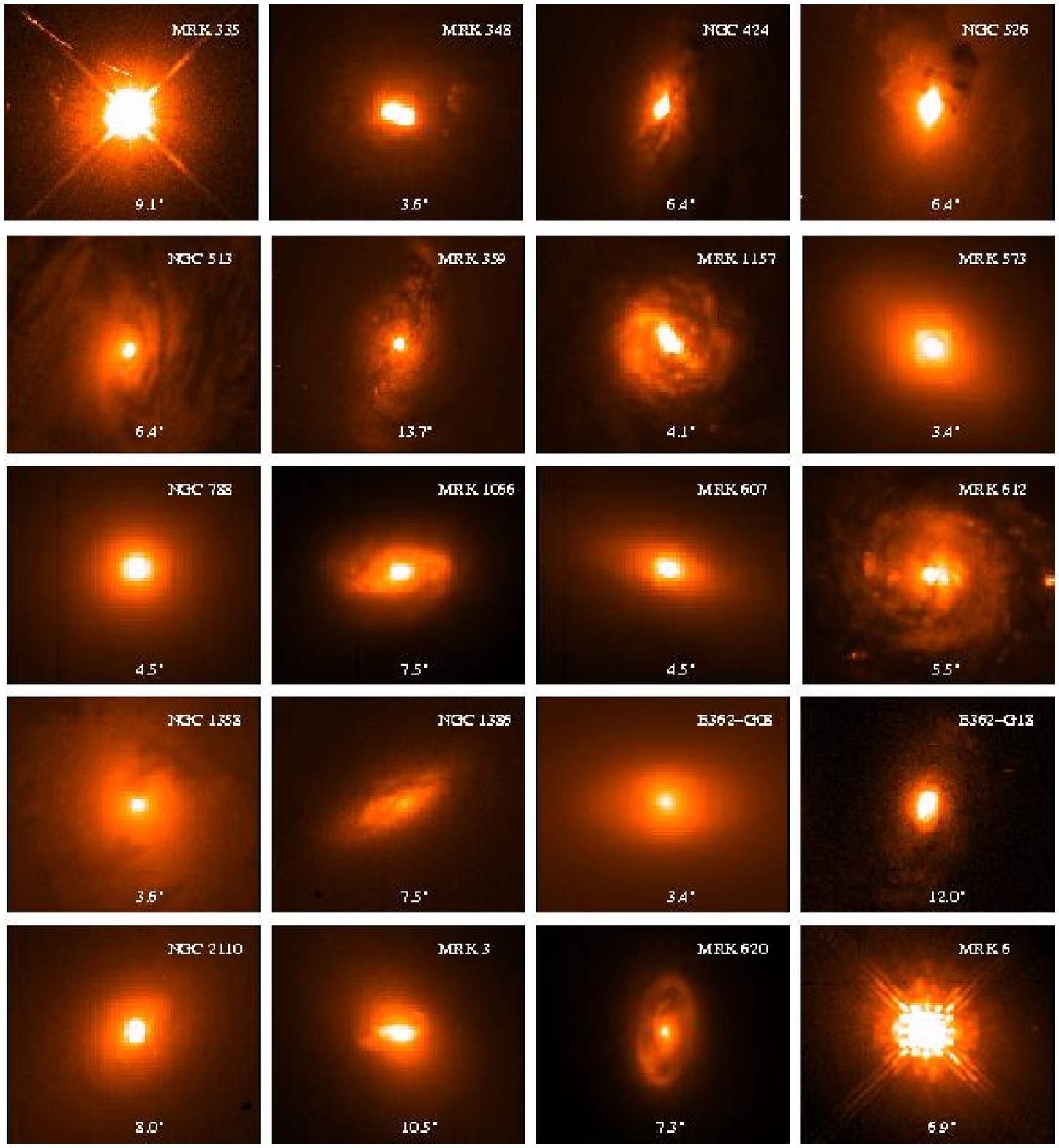,width=1.00\linewidth}
\vskip 0.5cm
\caption{\label{f1} HST images of the Seyfert galaxies of the sample.
The image size is given in the bottom. 
The instrument/filter combination is reported 
in Table \ref{nuker}.} 
\end{figure*}

\begin{figure*}
\psfig{figure=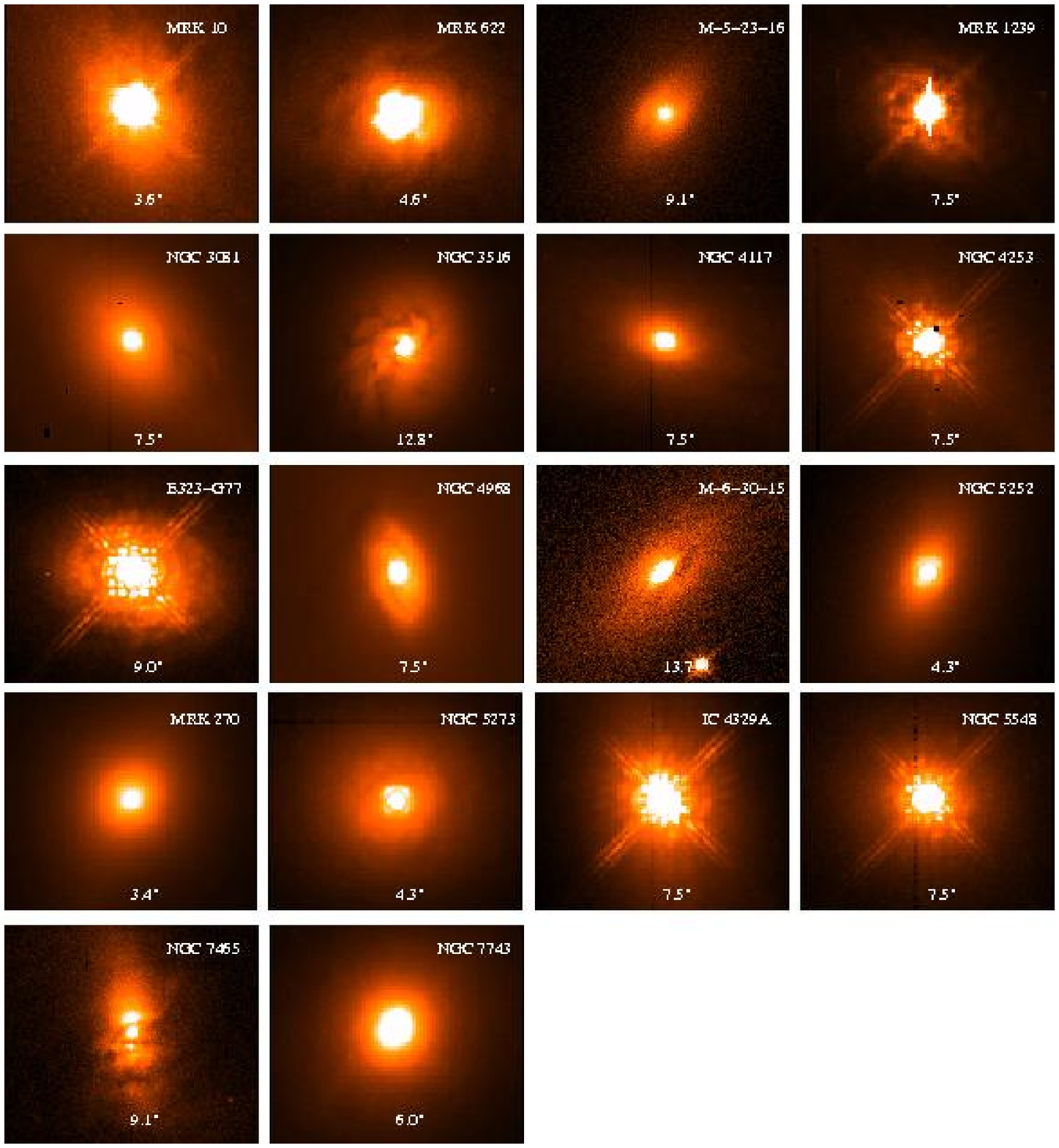,width=1.00\linewidth}
\vskip 0.5cm
\caption{\label{f2} HST images of the Seyfert galaxies of the sample.
The image size is given in the bottom. 
The instrument/filter combination is reported 
in Table \ref{nuker}.} 
\end{figure*}

We explored the properties of the surface brightness profiles
of the galaxies of the sample retrieving images from 
the HST archive, available for all but 4 objects. 
All data were calibrated by the standard on the fly
re-processing (OTFR) system. The resulting images are presented
in Figs. \ref{f1} through \ref{f2}. 
The HST images of the sample of Seyfert galaxies were obtained in
different instruments and filters as reported in Table \ref{nuker},
but most images were obtained using the broad-band filter F606W (R band)
on the WFPC2 or the F160W filter (H band) on the NICMOS. 

We derived a one-dimensional surface-brightness 
profile by fitting elliptical isophotes to the images using the
IRAF task `ellipse' \citep{jedrzejewski87}. This step of the analysis
is often compromised by the presence of complex structures that cannot
be reproduced with an ellipse fitting (this is the case for most
optical images). The 16 galaxies discarded at this stage are marked
in Table \ref{nuker} as `complex'.

As explained in the Introduction,
we performed a fit on these profiles by using both a 
Nuker law \citep{lauer95} (see Sect. \ref{nukfit}) and 
a \citet{sersic68} model (see Sect. \ref{serfit}).
We followed the same approach as in \citetalias{paper1},
e.g. minimizing the residuals between the data and the models convolved
with the appropriate point spread function, produced with the
TINYTIM software. Given the
widespread presence of bright nuclear point sources, we preferred to
include an unresolved nuclear source in the fit, while in
\citetalias{paper1} we excluded the central regions from the fit
in the case of a nucleated galaxy.

\subsection{Nuker fit to the brightness profiles.}
\label{nukfit}

\begin{figure*}
\centerline{ 
\psfig{figure=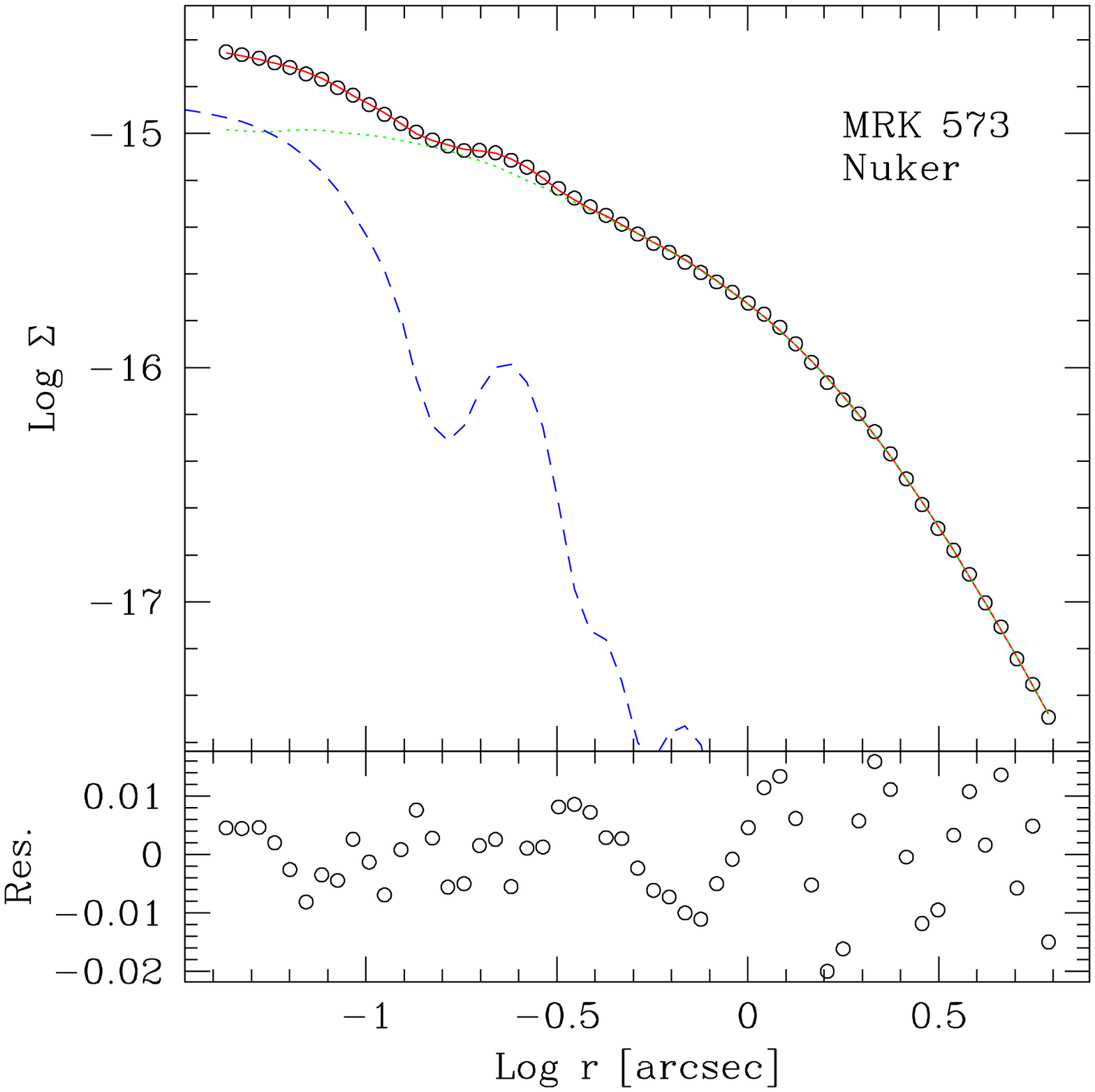,width=0.25\linewidth,height=0.22\linewidth} 
\psfig{figure=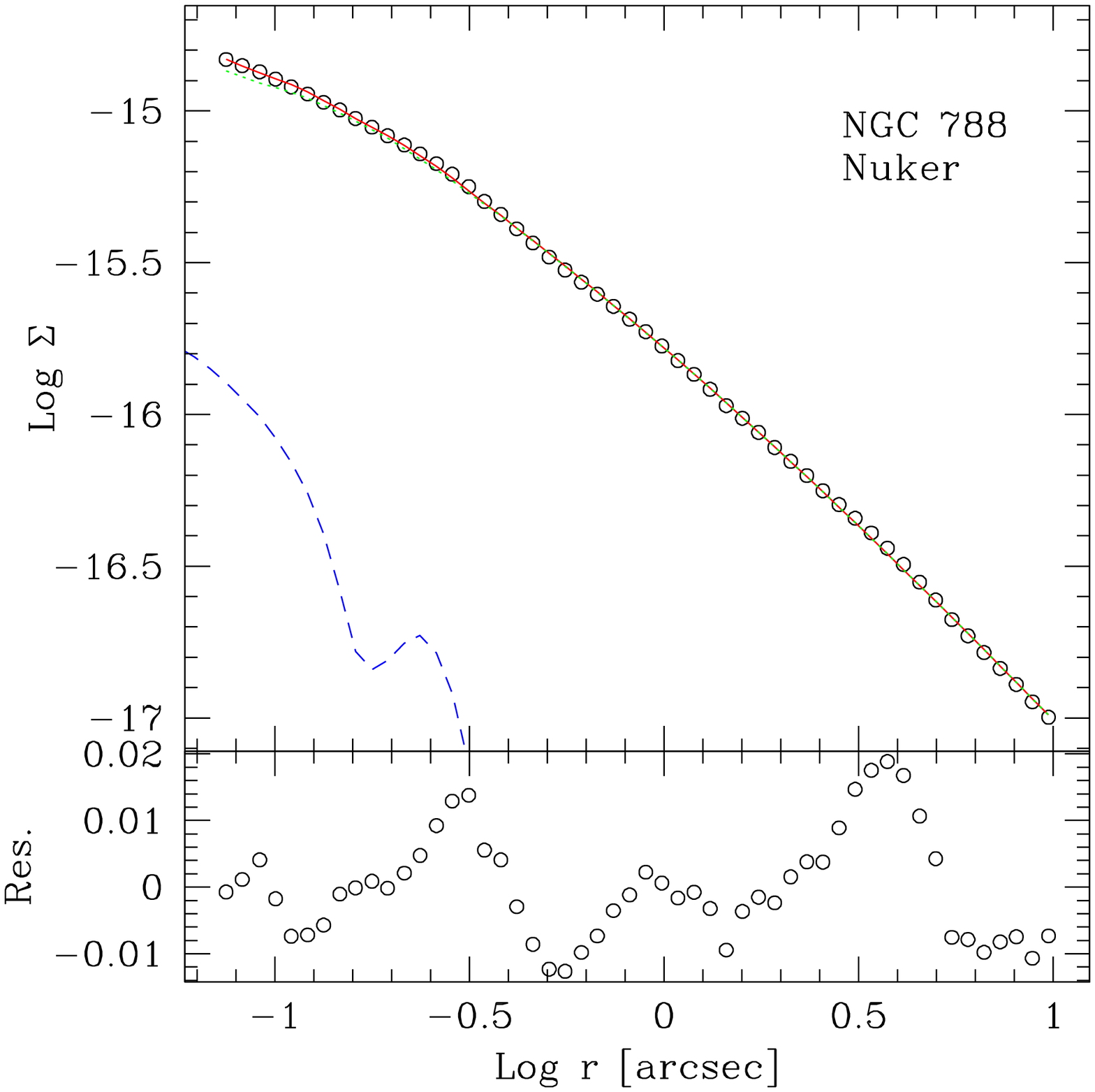,width=0.25\linewidth,height=0.22\linewidth}
\psfig{figure=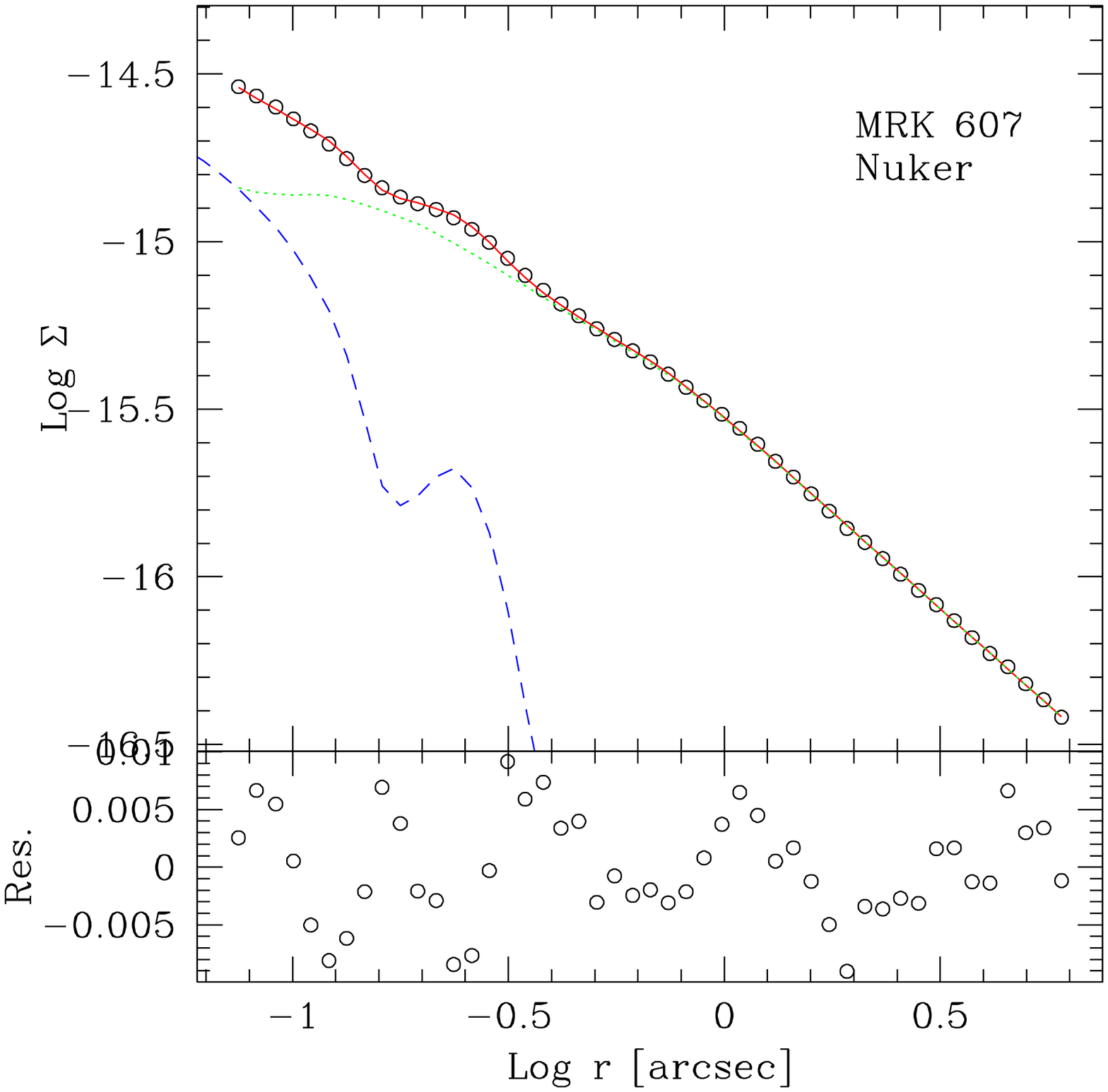,width=0.25\linewidth,height=0.22\linewidth}
\psfig{figure=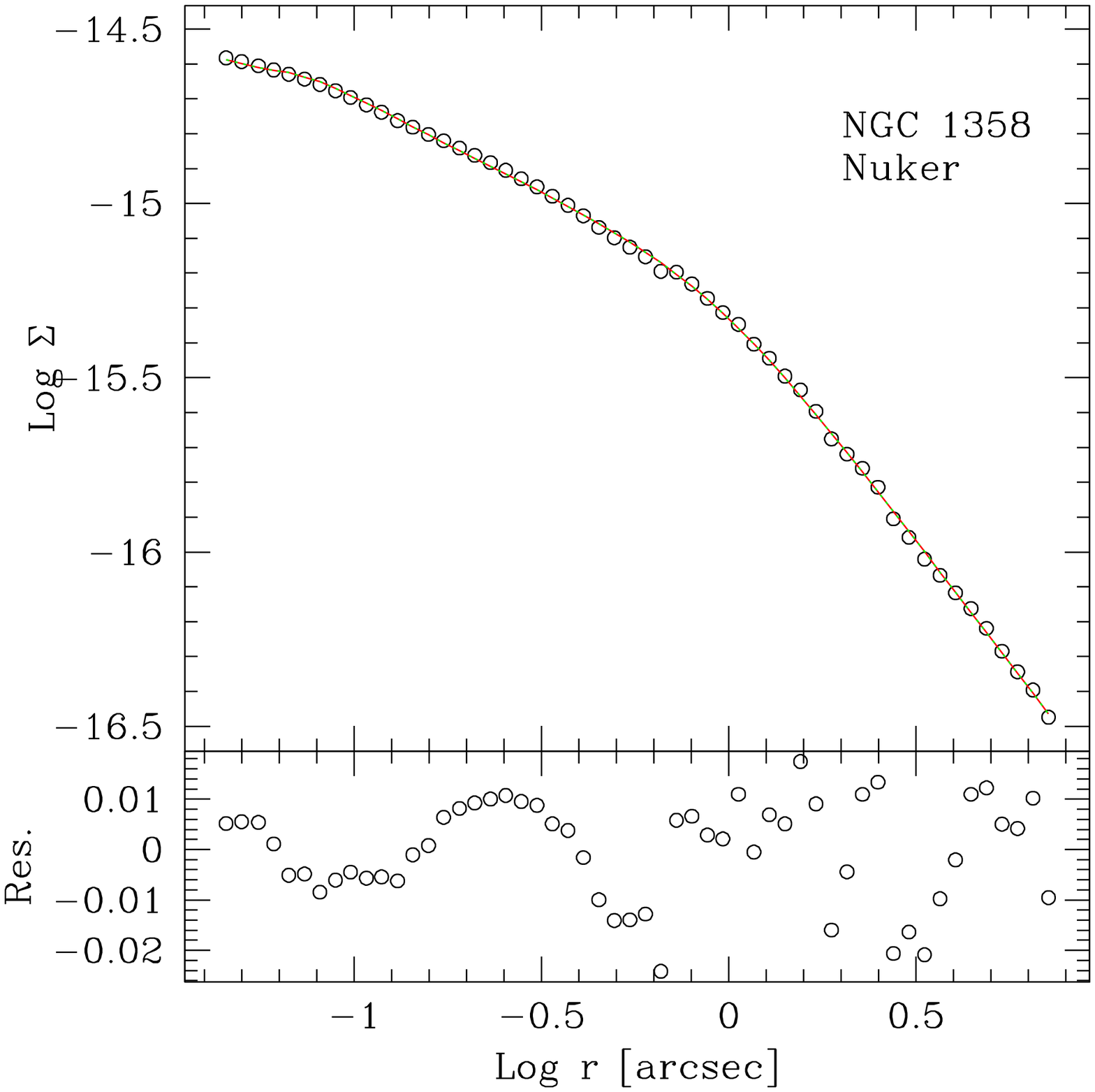,width=0.25\linewidth,height=0.22\linewidth} 
}             	    
\centerline{  	    
\psfig{figure=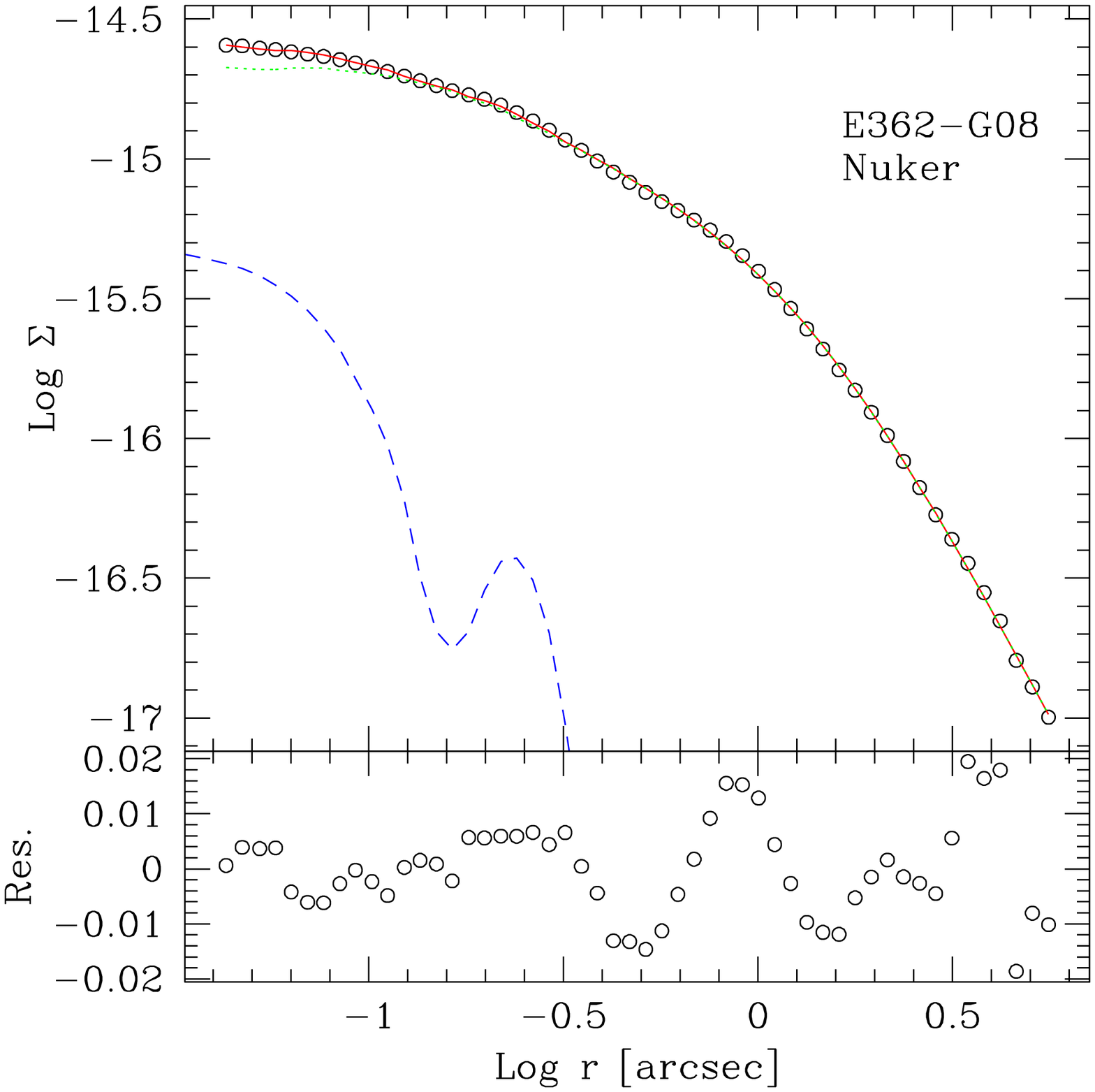,width=0.25\linewidth,height=0.22\linewidth}
\psfig{figure=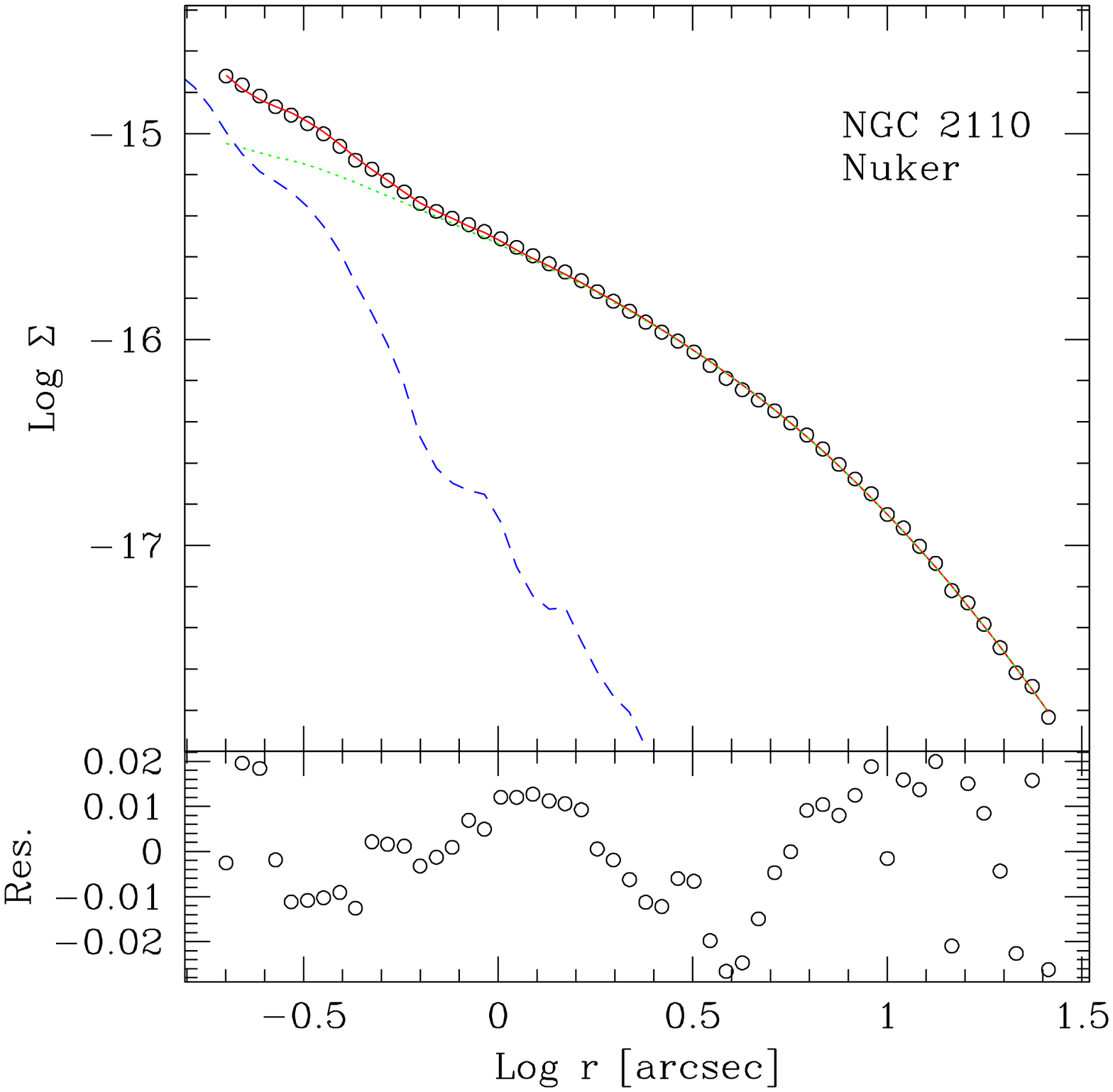,width=0.25\linewidth,height=0.22\linewidth}
\psfig{figure=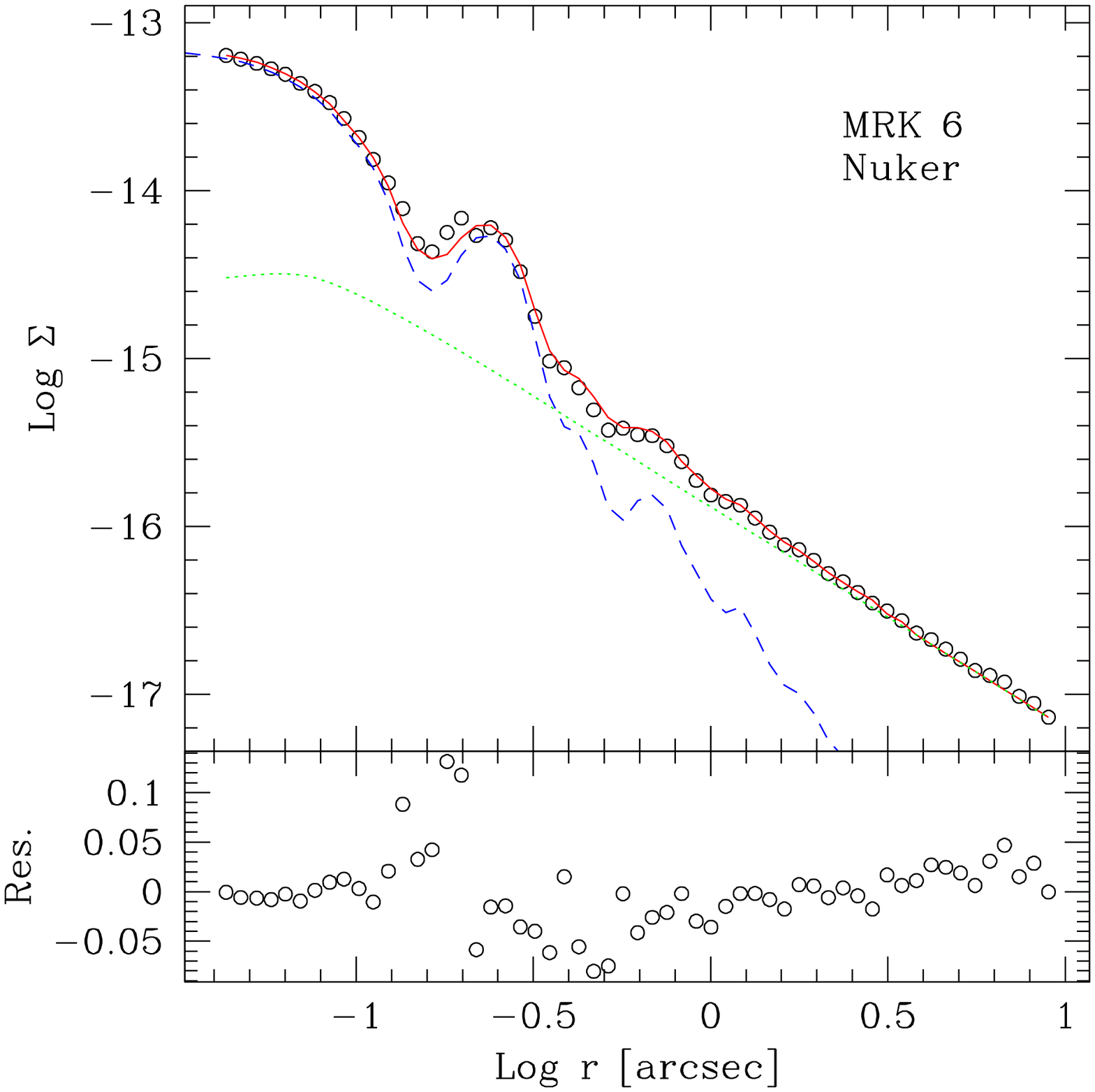,width=0.25\linewidth,height=0.22\linewidth} 
\psfig{figure=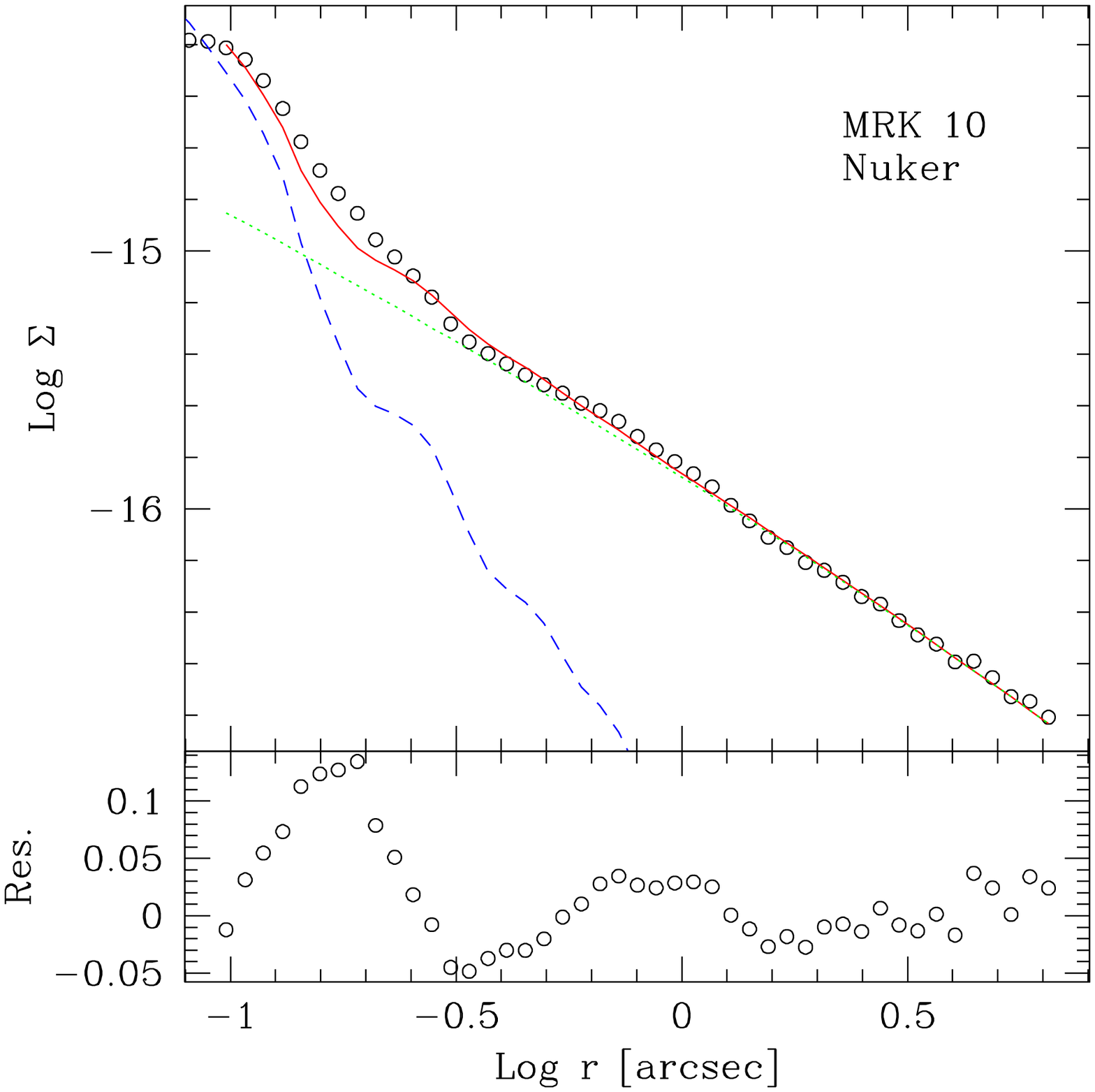,width=0.25\linewidth,height=0.22\linewidth}
}             	    
\centerline{  	    
\psfig{figure=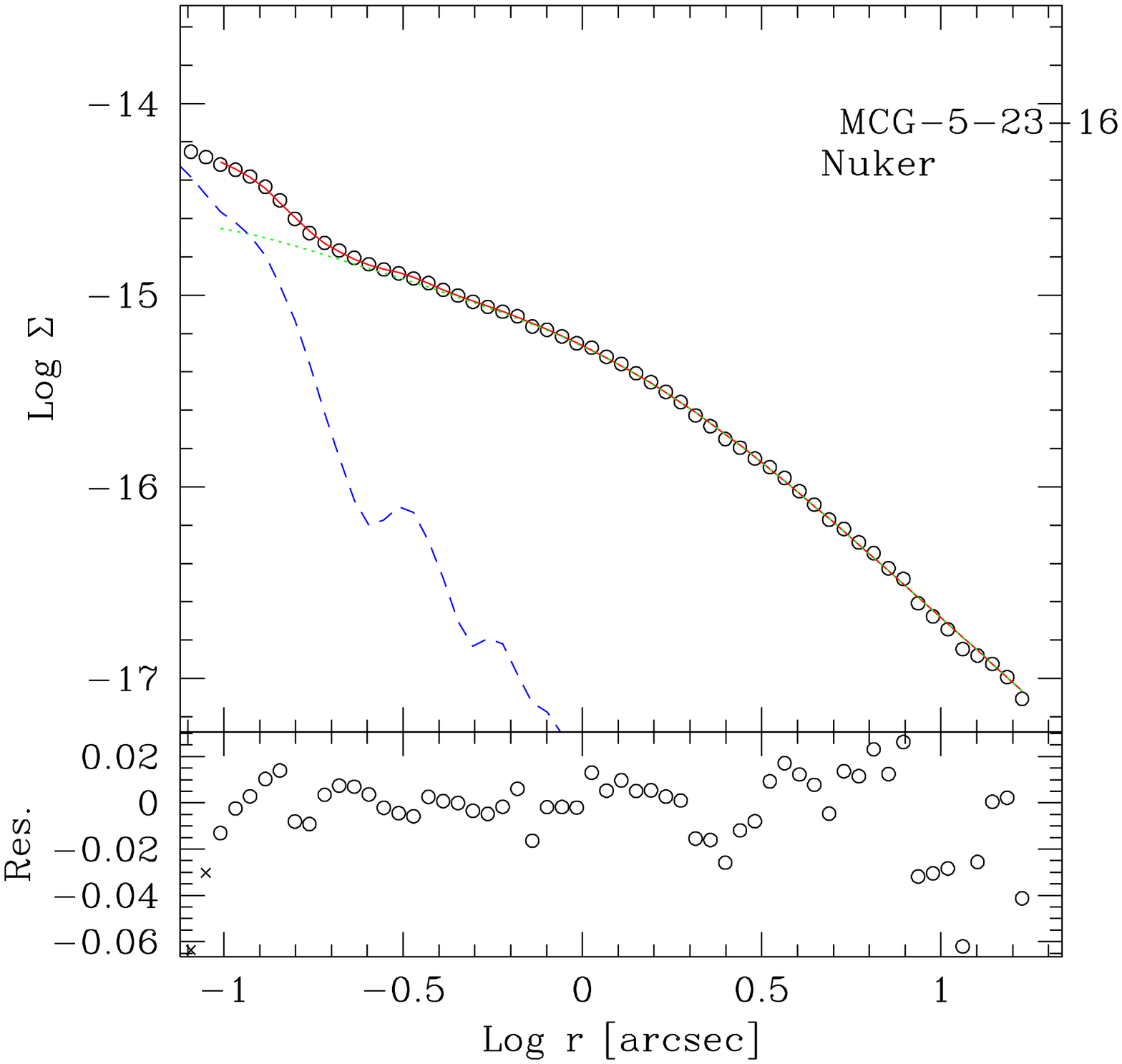,width=0.25\linewidth,height=0.22\linewidth}
\psfig{figure=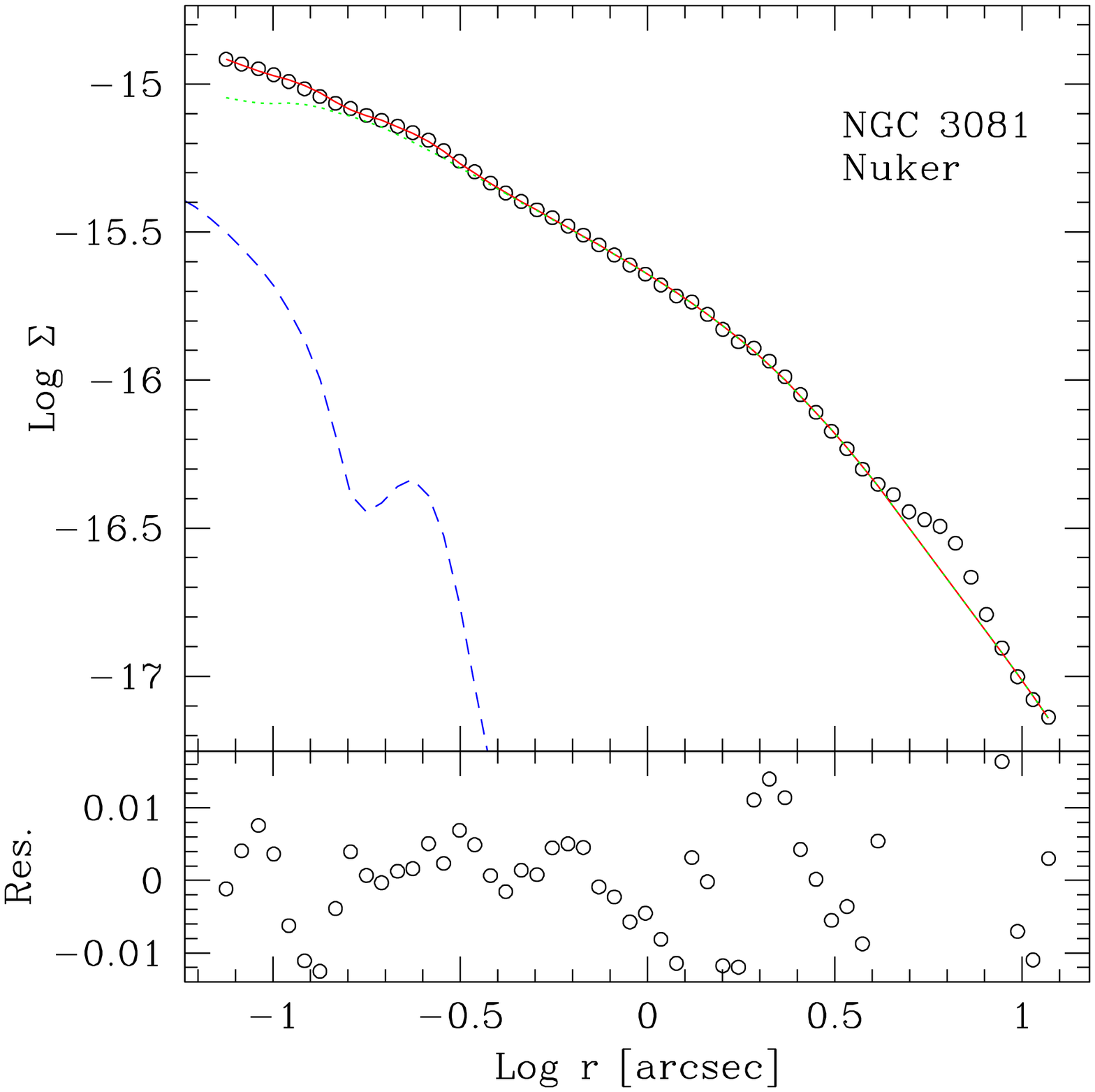,width=0.25\linewidth,height=0.22\linewidth} 
\psfig{figure=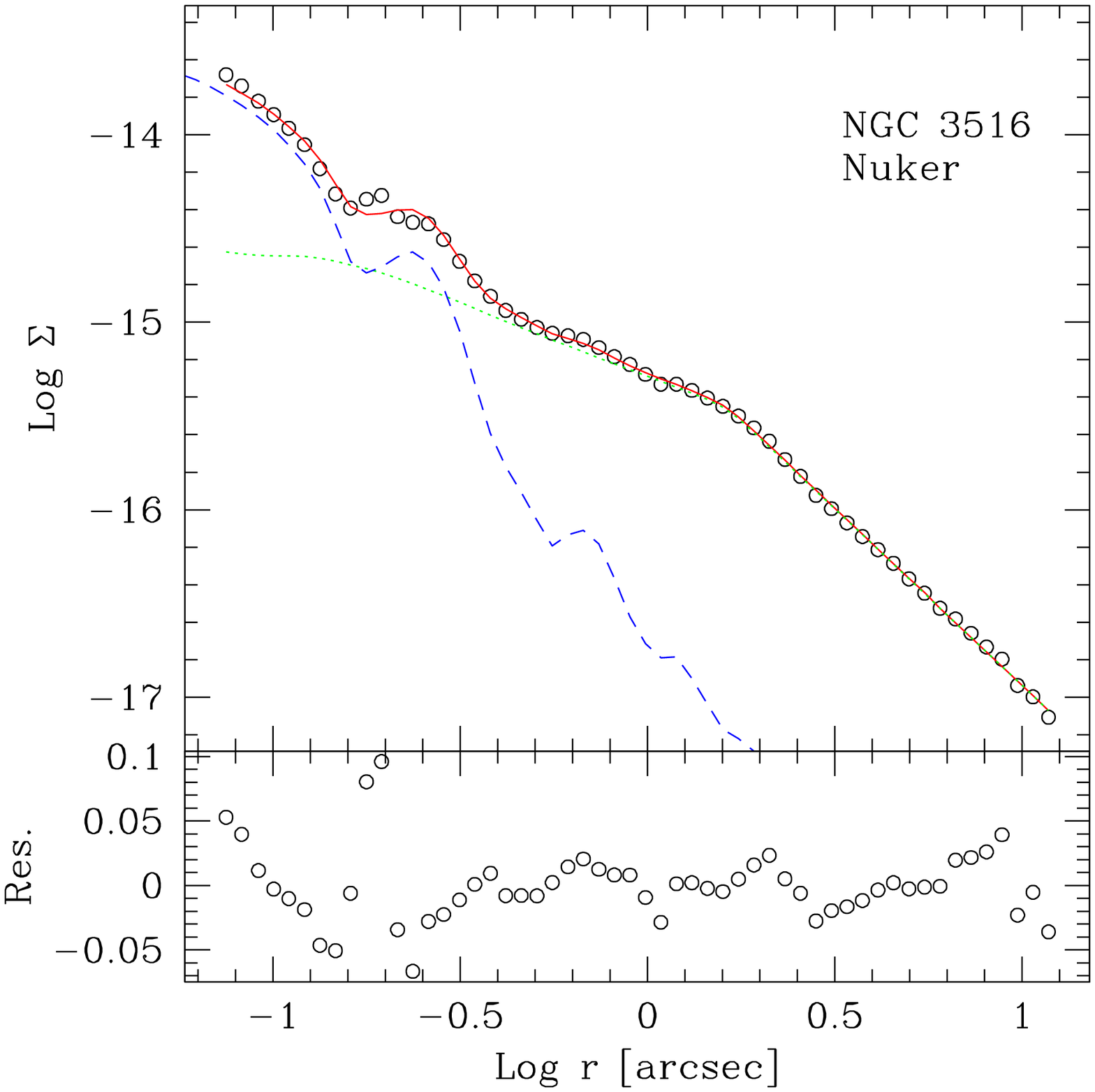,width=0.25\linewidth,height=0.22\linewidth} 
\psfig{figure=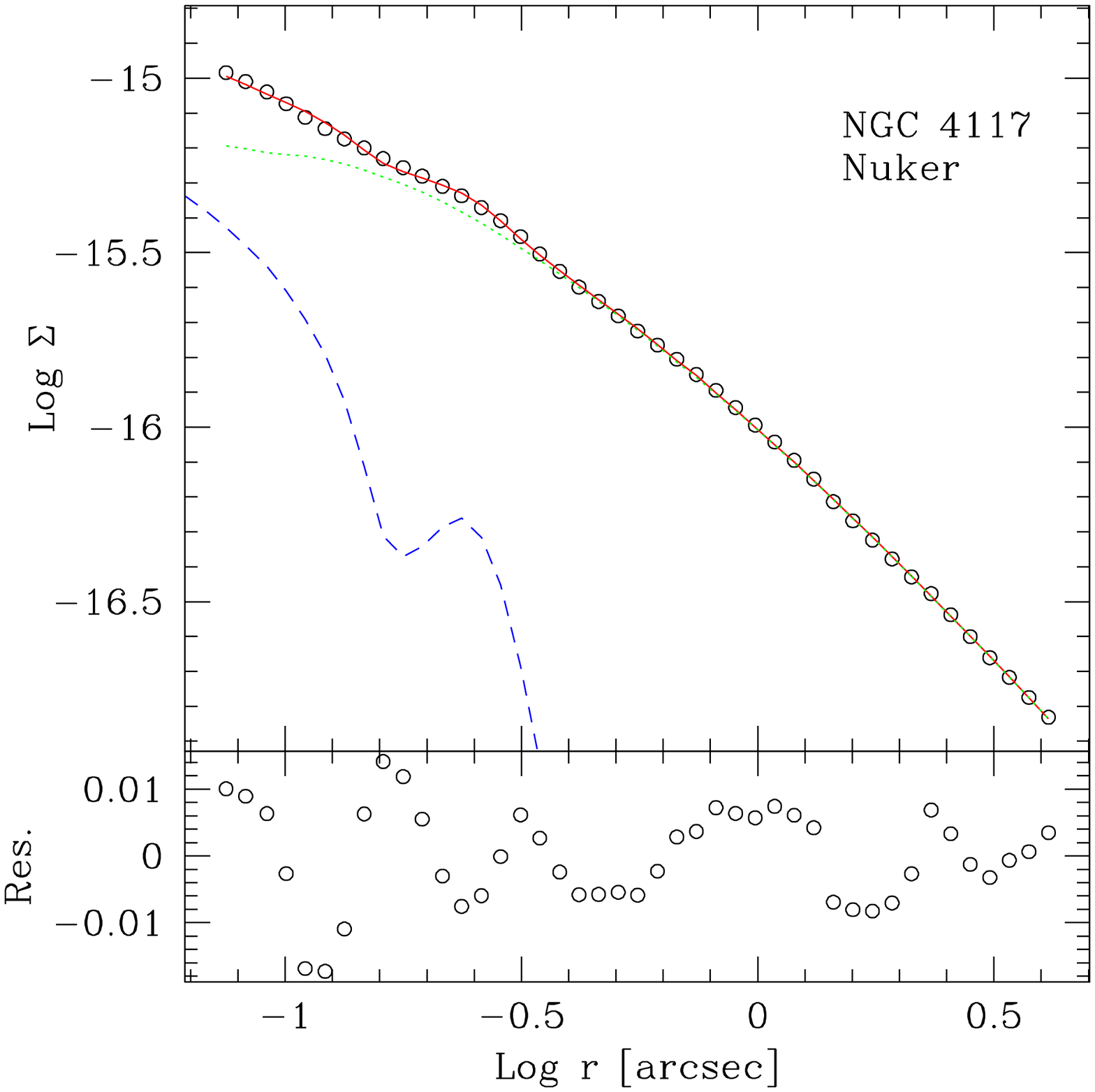,width=0.25\linewidth,height=0.22\linewidth}
}             	    
\centerline{  	    
\psfig{figure=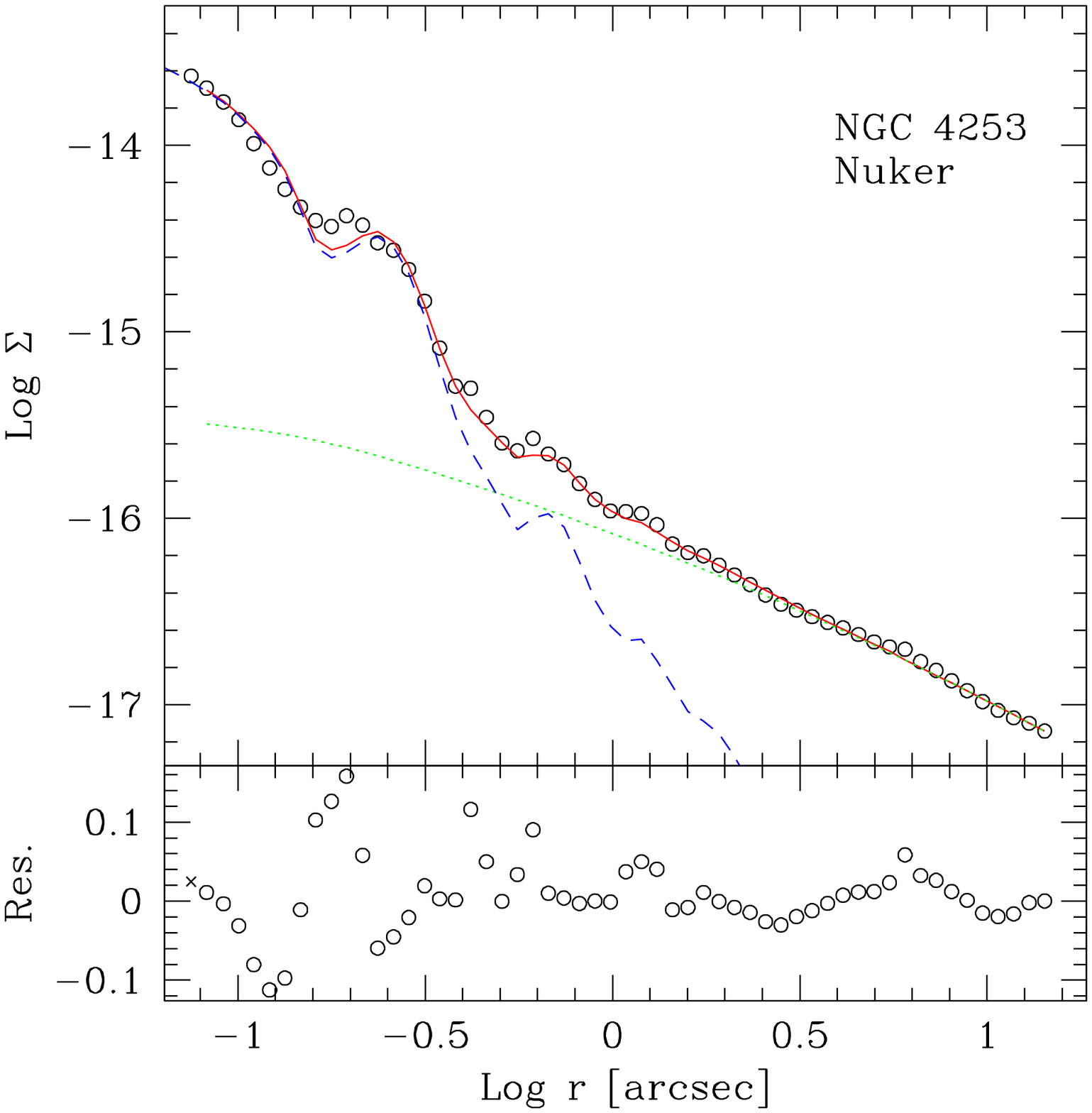,width=0.25\linewidth,height=0.22\linewidth}
\psfig{figure=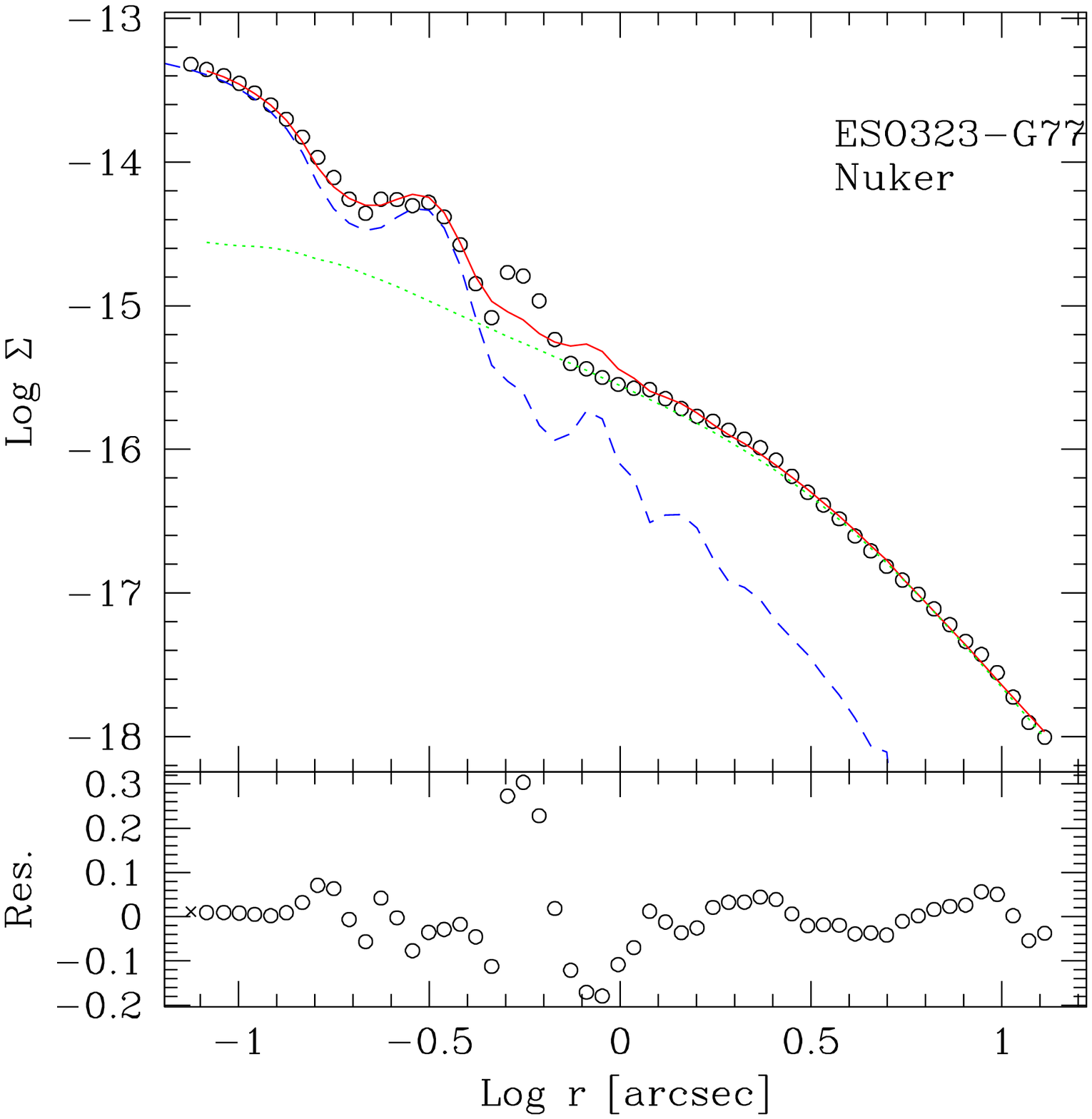,width=0.25\linewidth,height=0.22\linewidth}
\psfig{figure=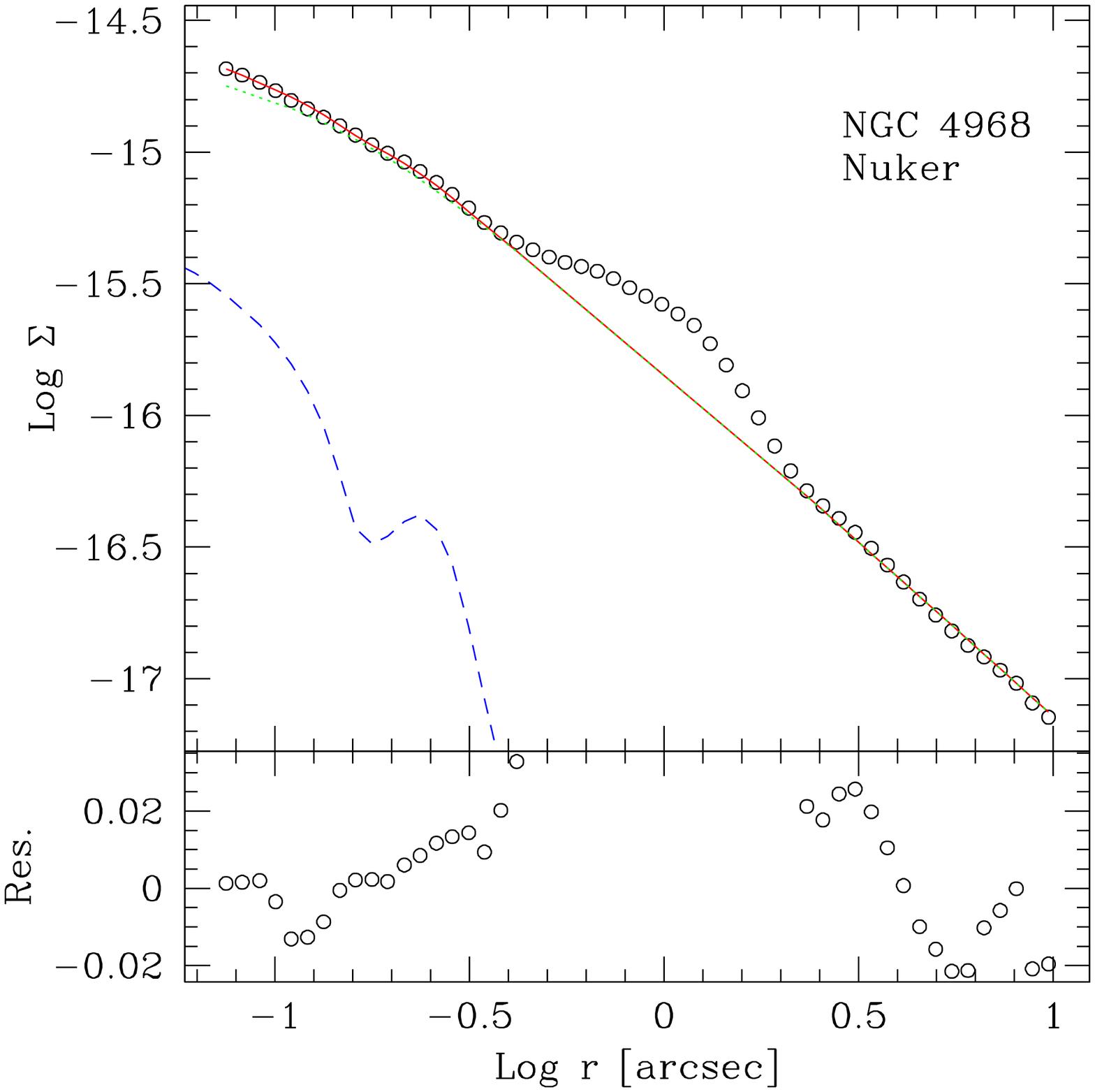,width=0.25\linewidth,height=0.22\linewidth} 
}             	    
\centerline{  	    
\psfig{figure=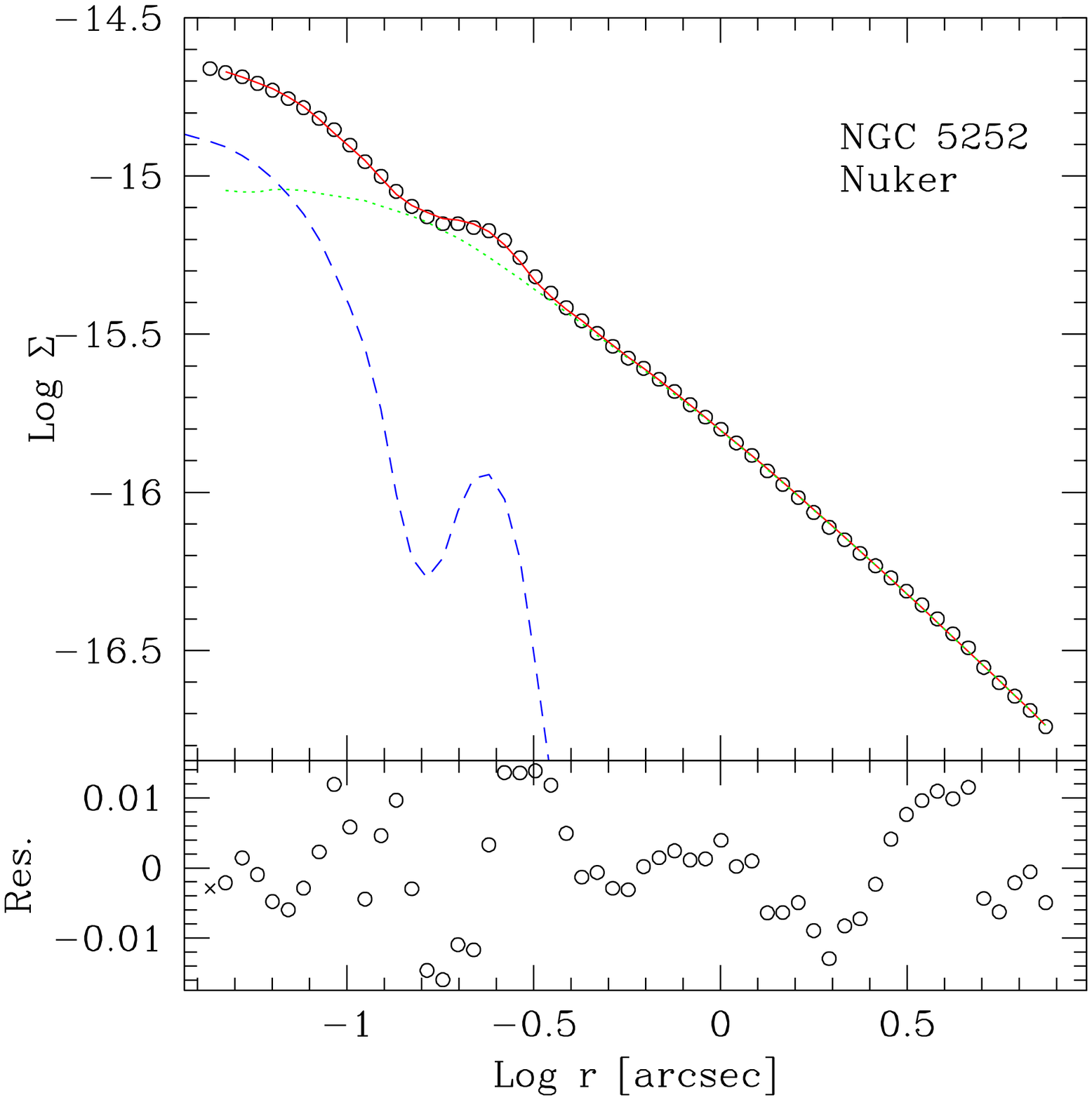,width=0.25\linewidth,height=0.22\linewidth}
\psfig{figure=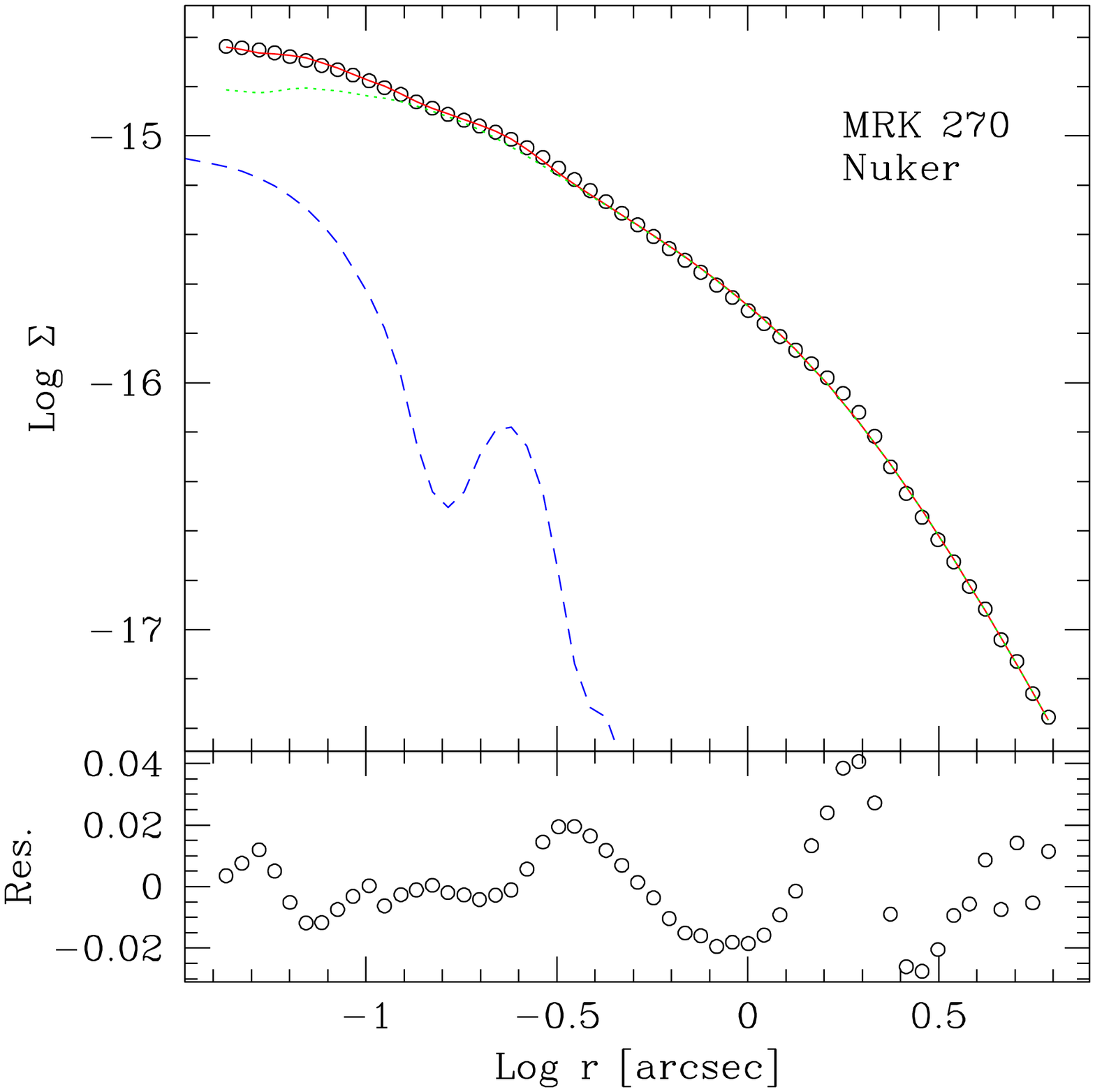,width=0.25\linewidth,height=0.22\linewidth}
\psfig{figure=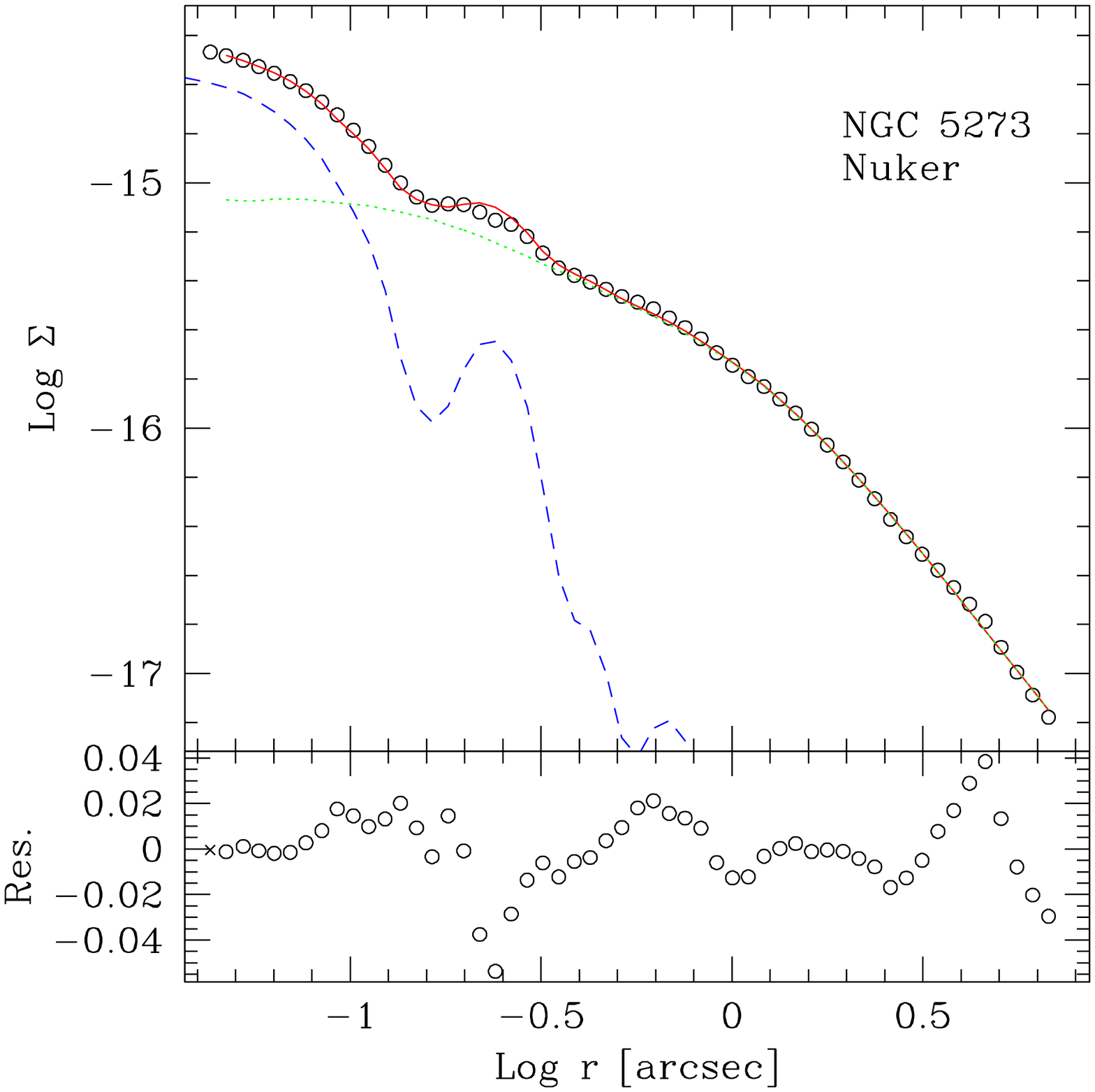,width=0.25\linewidth,height=0.22\linewidth}
}             	    
\centerline{  	    
\psfig{figure=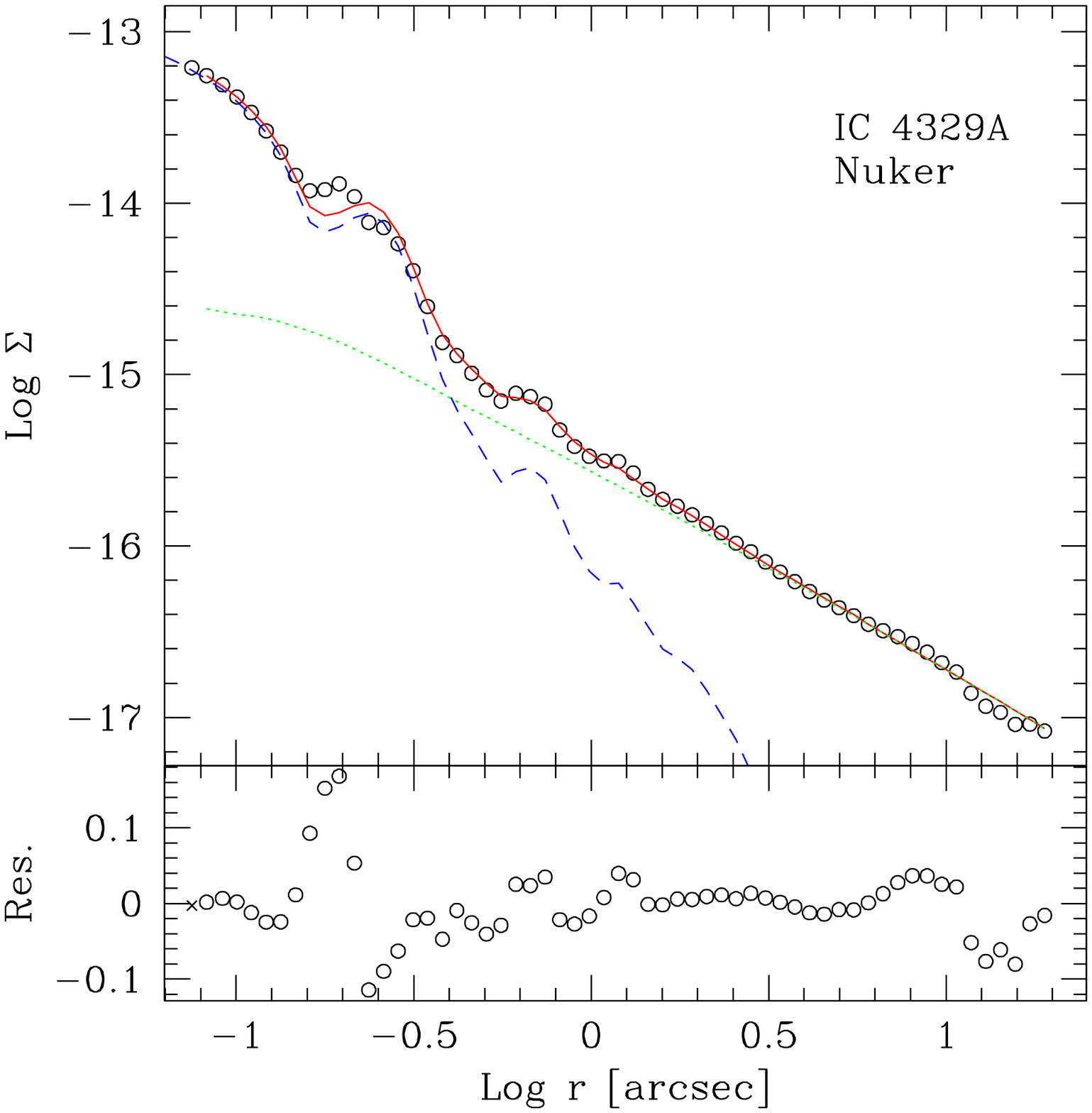,width=0.25\linewidth,height=0.22\linewidth} 
\psfig{figure=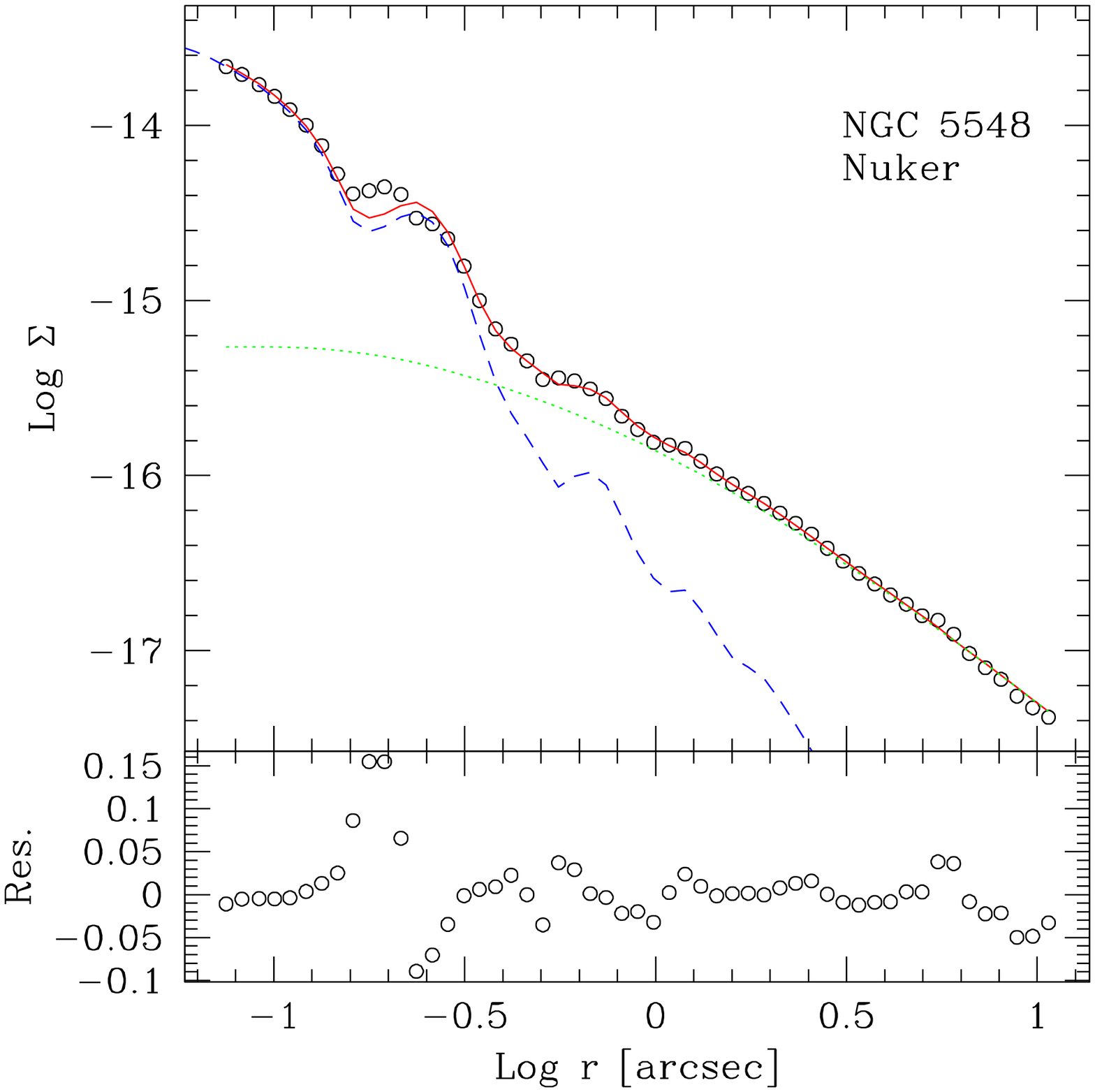,width=0.25\linewidth,height=0.22\linewidth} 
\psfig{figure=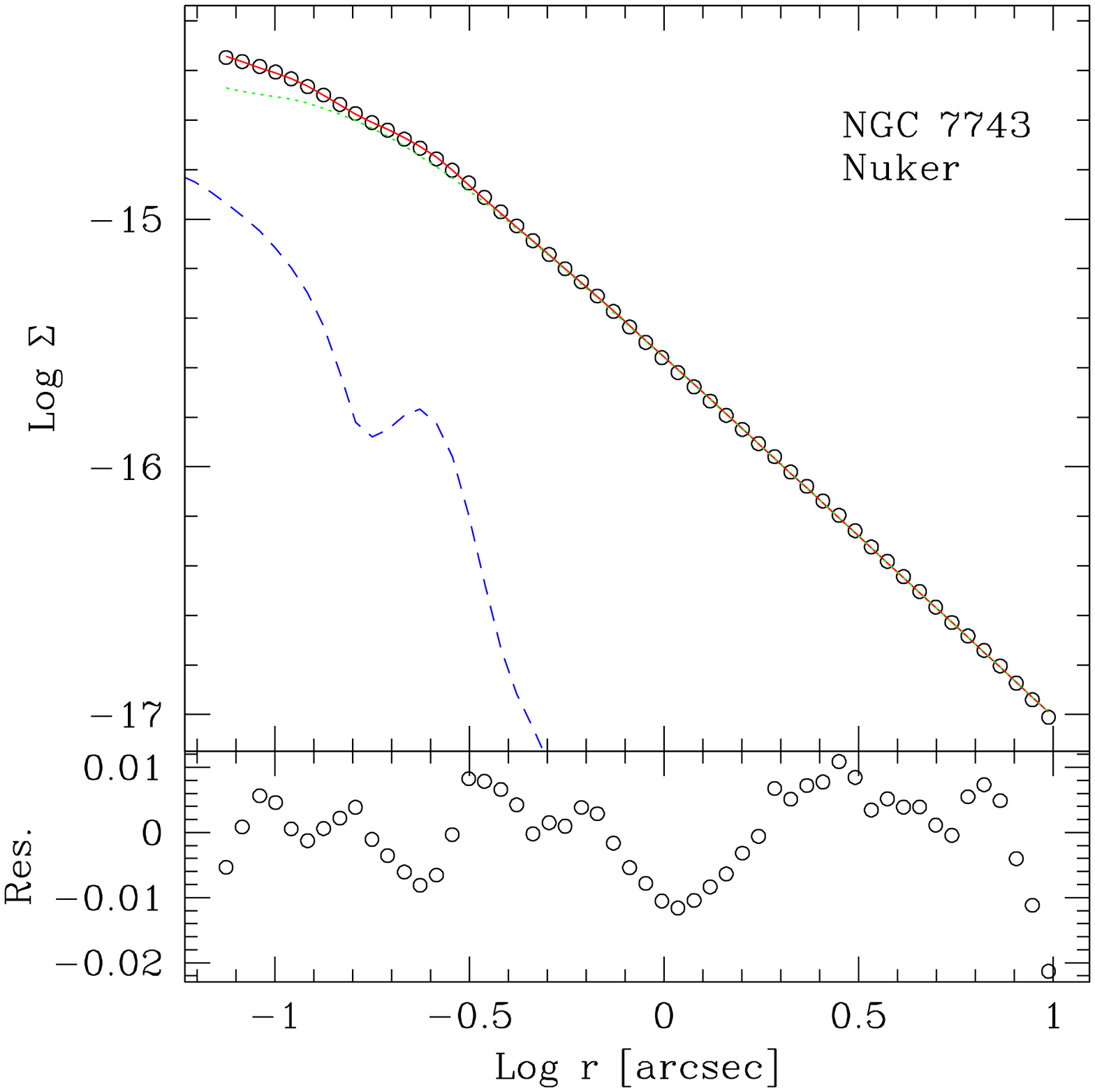,width=0.25\linewidth,height=0.22\linewidth}
}
\caption{Fit (solid line) to the observed brightness profiles 
(in erg s$^{-1}$ cm$^{-2}$ \AA$^{-1}$ units)
obtained with a Nuker law. The dotted and dashed lines 
represent the contribution
of the galaxies light and nuclear sources respectively. 
Residuals of the fit are given in the bottom panel.
No residuals are given in the regions of 
the brightness profiles of NGC 4968 and NGC 3081 
excluded from the fit.}

\label{sb1}
\end{figure*}

\begin{table*}
\caption{Nuker parameters for the Seyfert sample.}
\label{nuker}
\centering
\begin{tabular}{l c c c c c c c c l}
\hline\hline
Name   & Sy Type  & HST Image   &$\alpha$&$\beta$&$\gamma$&$r_b$&$\mu_b$&   & \\
\hline	 
MRK 335     & 1   & WFPC2/F606W & \multicolumn{5}{l}{Saturated} & & \\
MRK 348     & 2   & WFPC2/F606W & \multicolumn{5}{l}{Complex} & & \\
NGC 424     & 2   & WFPC2/F606W & \multicolumn{5}{l}{Complex} & & \\
NGC 526A    & 1.9 & WFPC2/F606W & \multicolumn{5}{l}{Complex} & & \\  	  
NGC 513     & 2   & WFPC2/F606W & \multicolumn{5}{l}{Complex} & & \\
MRK 359     & 1.5 & WFPC2/F606W & \multicolumn{5}{l}{Complex} & & \\
MRK 1157    & 2   & WFPC2/F606W & \multicolumn{5}{l}{Complex} & & \\
MRK 573     & 2   & NIC1/F160W  & 1.93 & 3.14 & 0.62 & 1.74 & -16.10 & \\ 
NGC 788     & 2   & NIC2/F160W  & 0.67 & 1.39 & 0.51 & 0.26 & -15.19 & \\ 
ESO 417-G06 & 2   & Unobserved  & & & & & & \\  
MRK 1066    & 2   & NIC2/F160W  & \multicolumn{5}{l}{Complex} & & \\
MRK 607     & 2   & NIC2/F160W  & 6.27 & 1.15 & 0.73 & 0.81 & -15.44 & \\ 
MRK 612     & 2   & WFPC2/F606W & \multicolumn{5}{l}{Complex} & & \\
NGC 1358    & 2   & WFPC2/F606W & 3.11 & 1.41 & 0.52 & 0.92 & -15.29 & \\
NGC 1386    & 2   & NIC2/F160W  & \multicolumn{5}{l}{Complex} & & \\ 
ESO 362-G08 & 2   & NIC1/F160W  & 2.14 & 2.67 & 0.60 & 1.33 & -15.60 & \\ 
ESO 362-G18 & 1.5 & WFPC2/F547M & \multicolumn{5}{l}{Complex} & & \\
NGC 2110    & 2   & NIC3/F160W  & 1.04 & 3.43 & 0.64 & 10.80 & -16.91 & \\ 
MRK 3       & 2   & WFPC2/F814W & \multicolumn{5}{l}{Complex} & & \\
MRK 620     & 2   & NIC2/F160W  & \multicolumn{5}{l}{Complex} & & \\
MRK 6       & 1.5 & NIC1/F160W  & \multicolumn{5}{l}{Nucleated} & & \\
MRK 10      & 1.2 & WFPC2/F606W & \multicolumn{5}{l}{Nucleated} & & \\
MRK 622     & 2   & WFPC2/F606W & \multicolumn{5}{l}{Complex} & & \\
MCG -5-23-16& 2   & WFPC2/F791W &  2.06 & 1.70 & 0.53 & 1.48 & -15.44 & \\
MRK 1239    & 1.5 & WFPC2/F606W & \multicolumn{5}{l}{Complex} & & \\
NGC 3081    & 2   & NIC2/F160W  & 2.64 & 1.79 & 0.66 & 2.27 & -15.99 & \\ 
NGC 3516    & 1.2 & NIC2/F160W  & 23.5 & 1.67 & 0.92 & 1.62 & -15.46 & \\ 
NGC 4074    & 2   & Unobserved  & &  & & & & \\  
NGC 4117    & 2   & NIC2/F160W  & 1.31 & 1.50 & 0.53 & 0.55 & -15.72 & \\ 
NGC 4253    & 1.5 & NIC2/F160W  & \multicolumn{5}{l}{Nucleated} & & \\
ESO 323-G77 & 1.2 & NIC2/POL0L  & 2.24 & 3.24 & 1.07 & 3.39 & -16.39 & \\
NGC 4968    & 2   & NIC2/F160W  & --   & 1.34 & -- & $<$ 0.2 & -- & \\
MCG -6-30-15& 1.2 & WFPC2/F791W & \multicolumn{5}{l}{Complex} & & \\
NGC 5252    & 1.9 & NIC1/F160W  & 1.26 & 1.16 & 0.66 & 0.74 & -15.68 &\\ 
MRK 270     & 2   & NIC1/F160W  & 2.05 & 2.87 & 0.80 & 1.74 & -16.06 & \\
NGC 5273    & 1.5 & NIC1/F160W  & 2.90 & 1.97 & 0.63 & 1.27 & -15.86 & \\ 
IC 4329A    & 1.2 & NIC2/F160W  & \multicolumn{5}{l}{Nucleated} & & \\
NGC 5548    & 1.2 & NIC2/F160W  & \multicolumn{5}{l}{Nucleated} & & \\
ESO 512-G20 & 1   & Unobserved  & &  & & & & \\  
IC 5169     & 2   & Unobserved  & &  & & & & \\  
NGC 7465    & 2   & WFPC2/F791W & \multicolumn{5}{l}{Complex} & & \\
NGC 7743    & 2   & NIC2/F160W  & --   & 1.46 & -- & $<$ 0.2 & -- & \\
\hline	     
\end{tabular}

Notes: the break radius $r_b$ is given in arcsec; the brightness
at the break radius is in a Log scale in 
erg s$^{-1}$ cm$^{-2}$ \AA$^{-1}$ units.
\end{table*}

On the surface brightness 
profiles we performed a fit with a Nuker law in the form 

$$ I(r) = I_b 2^{(\beta-\gamma)/\alpha} 
\left({{r_b}\over r}\right)^\gamma 
\left[1 + \left( {r \over {r_b}} \right)^\alpha \right]
^{(\gamma-\beta)/ \alpha}. $$

\noindent
The parameter $\beta$ measures the slope of the outer region of the
brightness profile, $r_b$ is the break radius (corresponding to a
brightness $I_b$), where the profile flattens to a smaller 
slope measured by the
parameter $\gamma$ and $\alpha$ sets the sharpness of the
transition between the inner and outer profile. 

In Fig. \ref{sb1} we superposed the best fit with a
Nuker law to the actual data in the top panel, while the lower panel
presents the residuals for the 21 galaxies for which an ellipse fitting
was possible. In Table \ref{nuker} we report the best-fit
parameters. In 6 objects, all type I Seyfert, the presence
of bright nuclei prevents us to explore the properties of their host
galaxies in the innermost regions: only a large scale emission tail,
well described by a single power-law, can be seen in these
objects (marked as `nucleated' in Tab. \ref{nuker}\footnote{For one
nucleated galaxy, Mrk 335, the optical image is saturated at the center and no
brightness profile is given.}). 

Nonetheless there are 3 exceptions to this general behavior among
Seyfert 1, namely NGC 3516, ESO 323-G77, NGC 5273, all observed with
NICMOS in the infrared band. Here a clear change in the slope (with a
difference between the slope of the outer and inner power-law is
$\beta - \gamma > \sim 0.8$) of the brightness profile occurs at
sufficiently large radius (with $r_b \sim 1\farcs3 - 2\farcs4 $), in a
region where the nuclear emission provides a negligible contribution.

Conversely, the nuclear points sources in Sy 2, although often
present, are not as prominent as in Sy 1 and this allows us to study
in more detail this sub-sample. Eleven Seyfert 2 galaxies can be
reproduced with a Nuker law with a well resolved break radius (we
conservatively adopt a minimum value for the break radius of $r_b \geq
0\farcs2$ to consider it sufficiently resolved to yield an estimate of
the cusp slope $\gamma$). It two cases (namely NGC 4968 and NGC 3081) 
we were only able to set an
upper limit to the break radius of 0\farcs2. 

Summarizing, the Nuker law provides an accurate description
for 16 galaxies of the sample, 3 Sy 1 and
13 Sy 2. The typical amplitude of the largest residuals of the modeling are
in the range of 0.01-0.03 dex. Only in the two 
objects (namely NGC 4968 and NGC 3081) the brightness profile shows
a 'bump' at relatively large radii and a Nuker fit is possible only
excluding this region. 
The nuclear cusps are reproduced with a slope in the range
$\gamma = 0.51 \,-\, 1.07$, typical of ``power-law'' galaxies.

\subsection{S\'ersic fit to the brightness profiles.}
\label{serfit}

\begin{figure*}
\centerline{ 
\psfig{figure=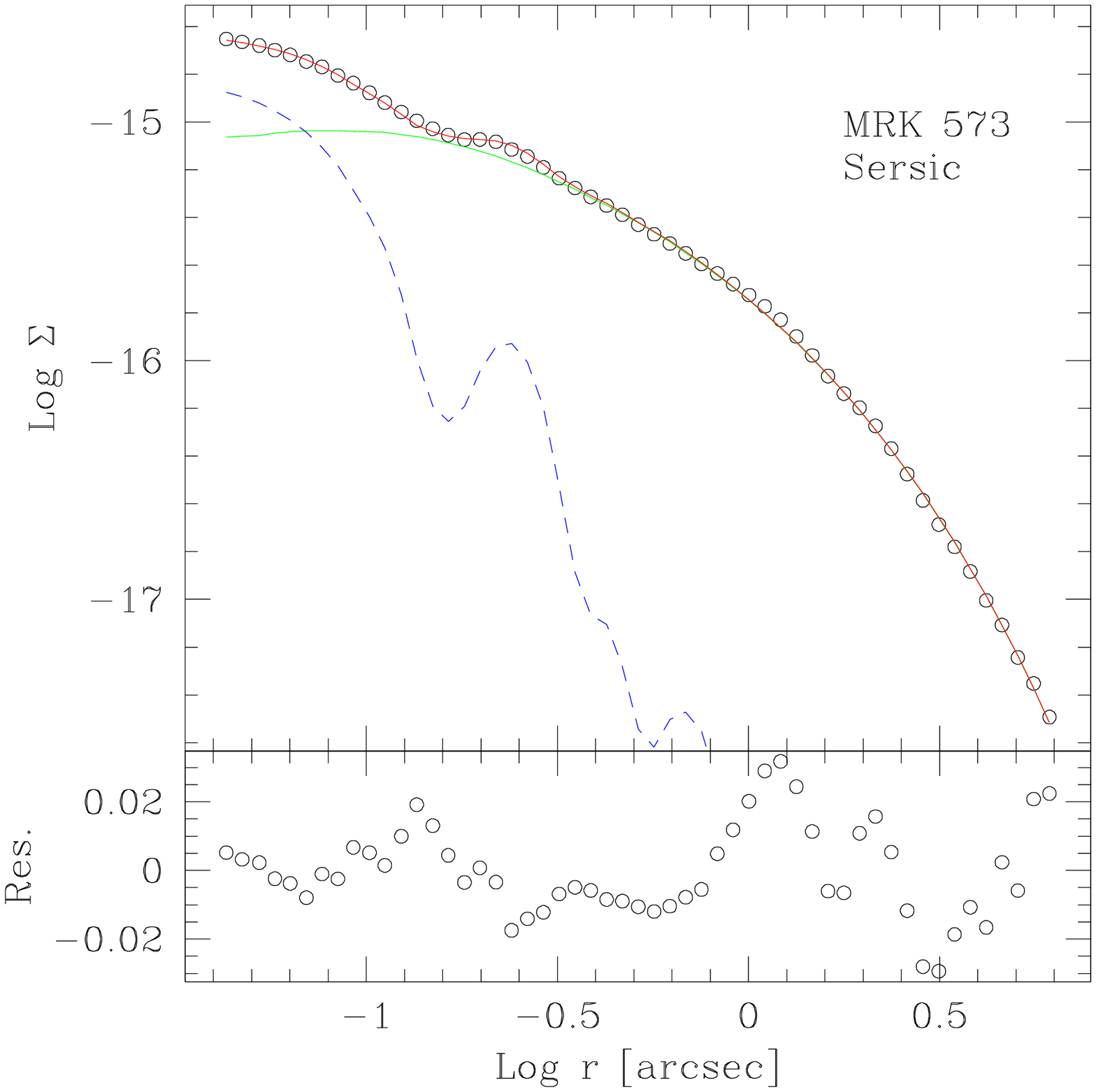,width=0.25\linewidth,height=0.22\linewidth} 
\psfig{figure=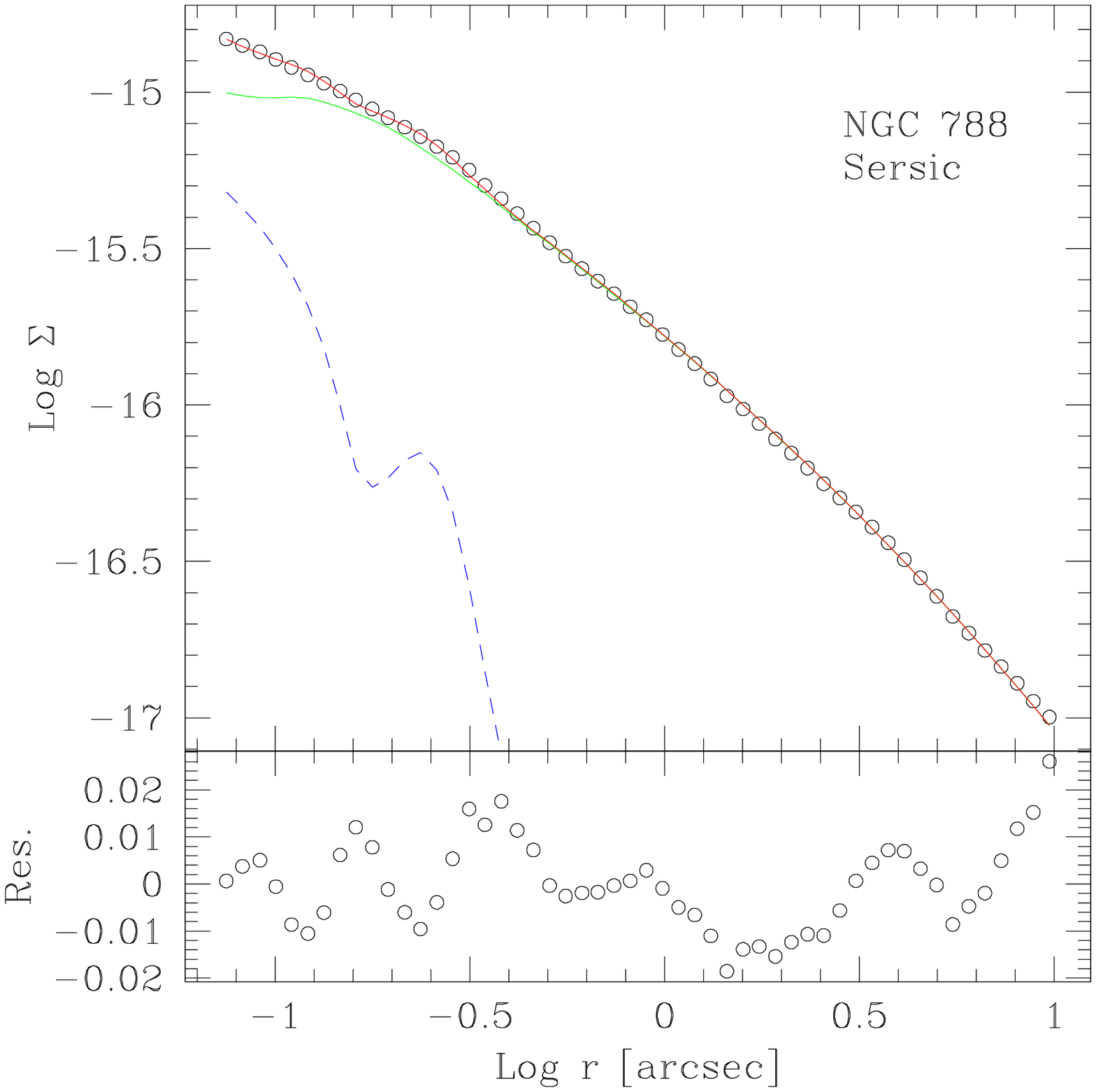,width=0.25\linewidth,height=0.22\linewidth}
\psfig{figure=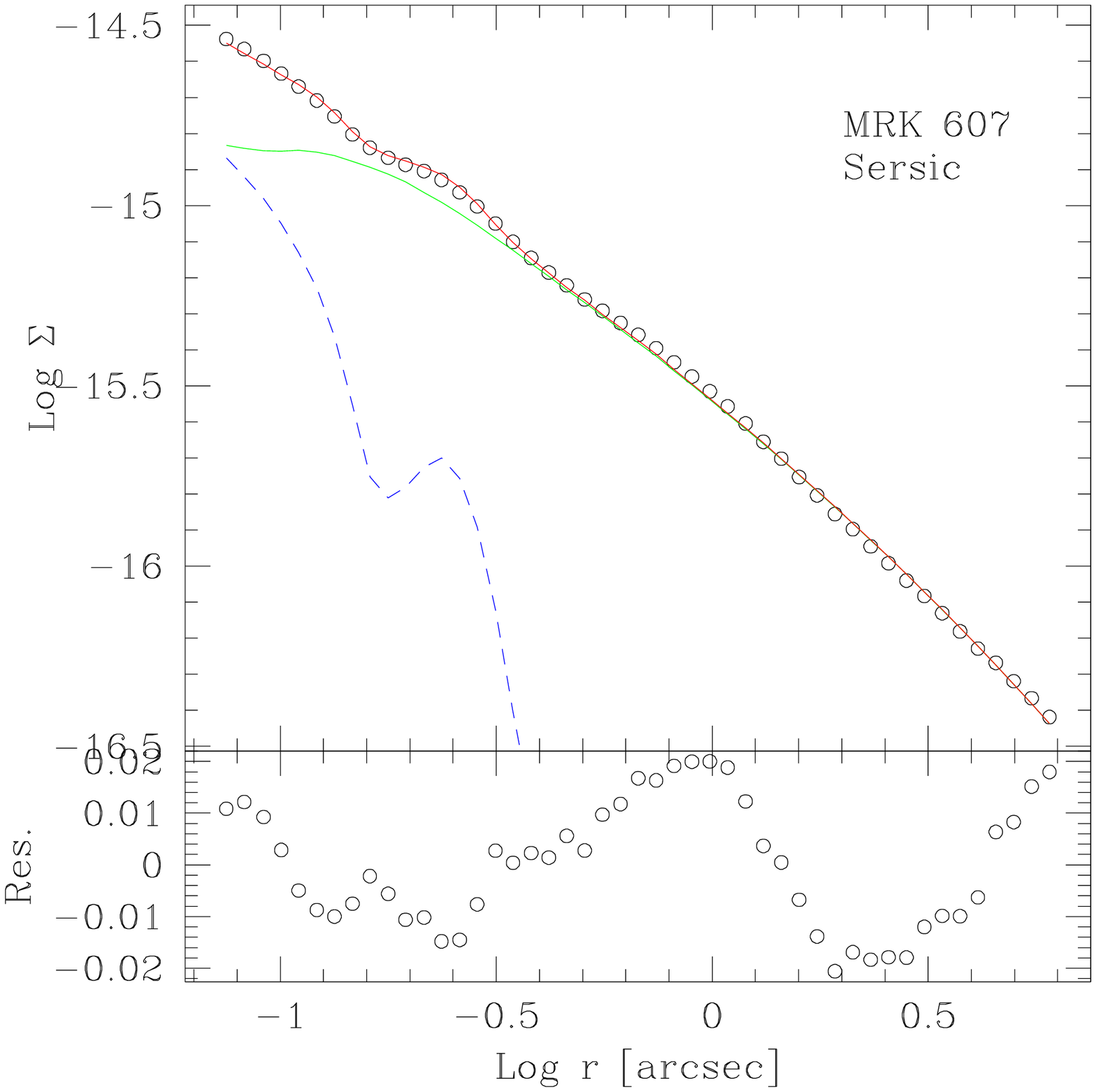,width=0.25\linewidth,height=0.22\linewidth}
\psfig{figure=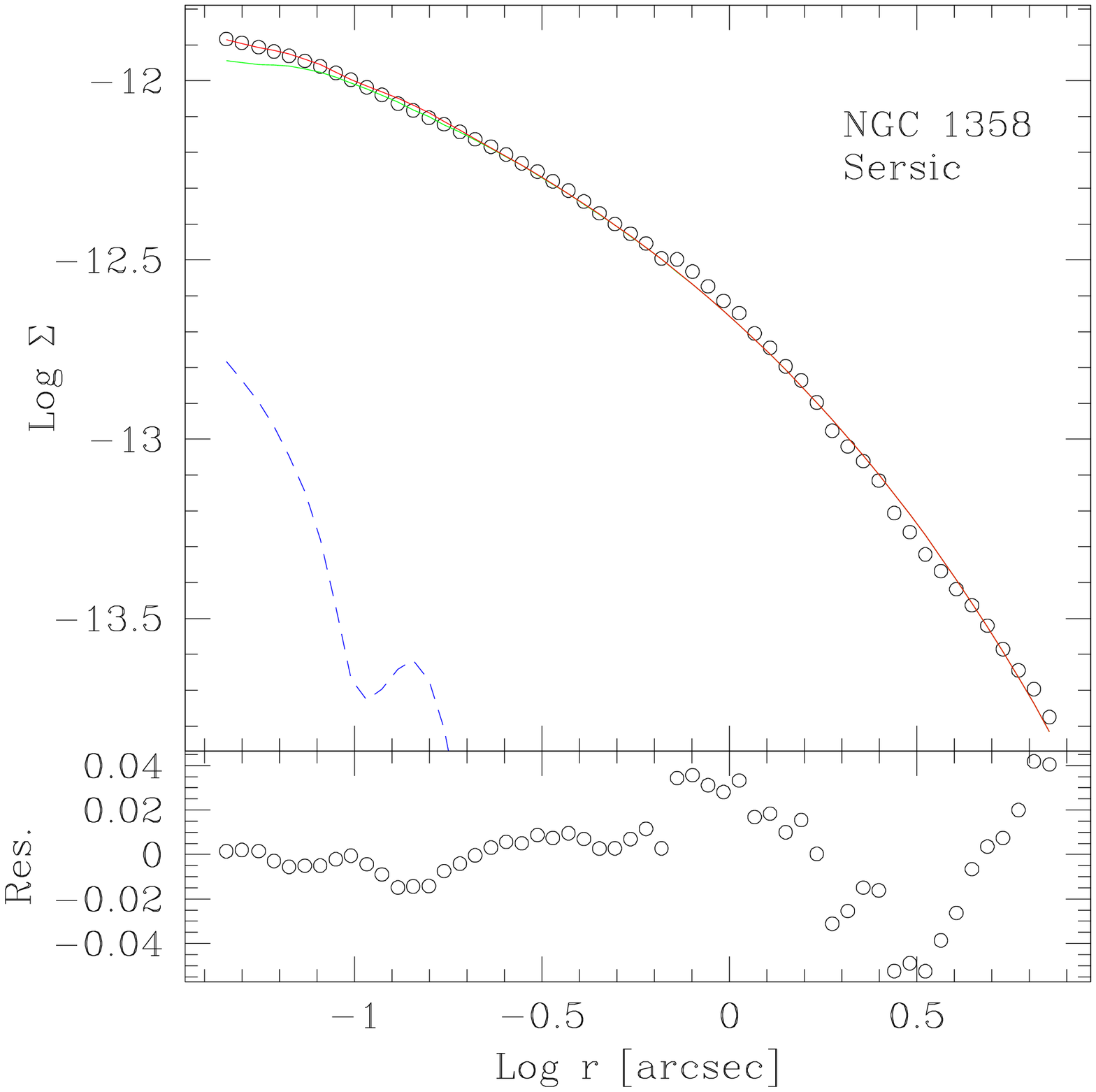,width=0.25\linewidth,height=0.22\linewidth} 
}                    
\centerline{         
\psfig{figure=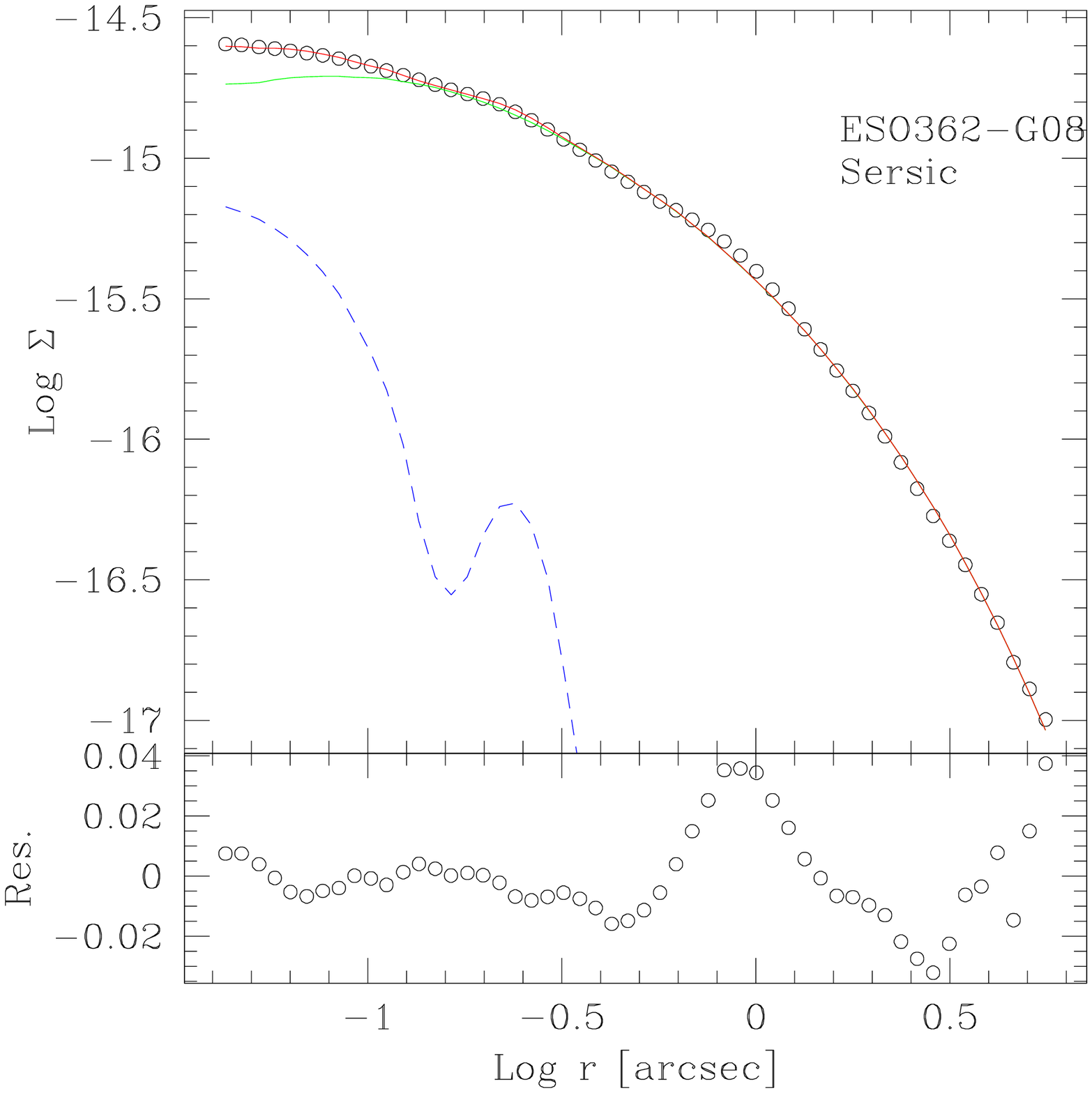,width=0.25\linewidth,height=0.22\linewidth}
\psfig{figure=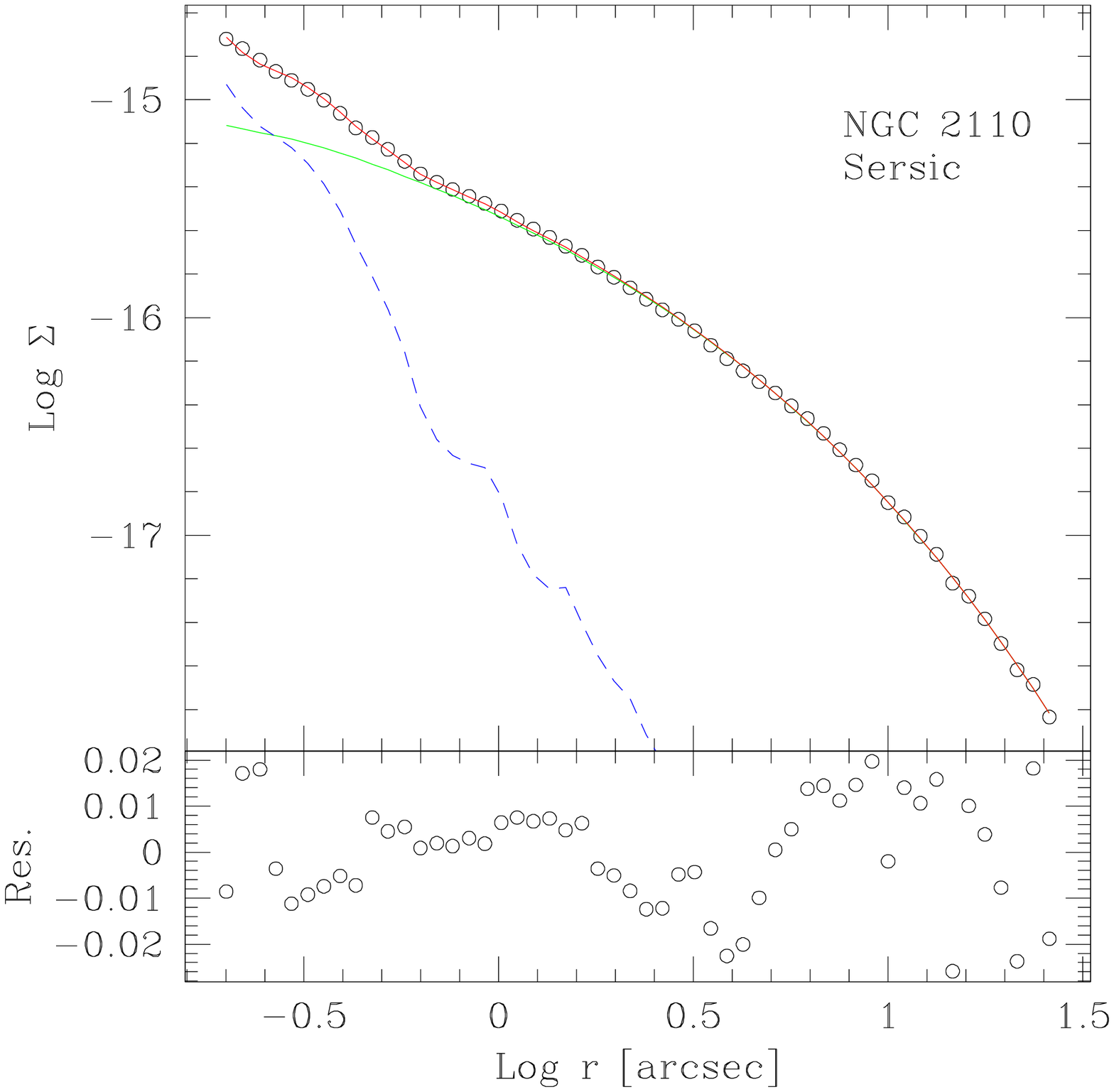,width=0.25\linewidth,height=0.22\linewidth} 
\psfig{figure=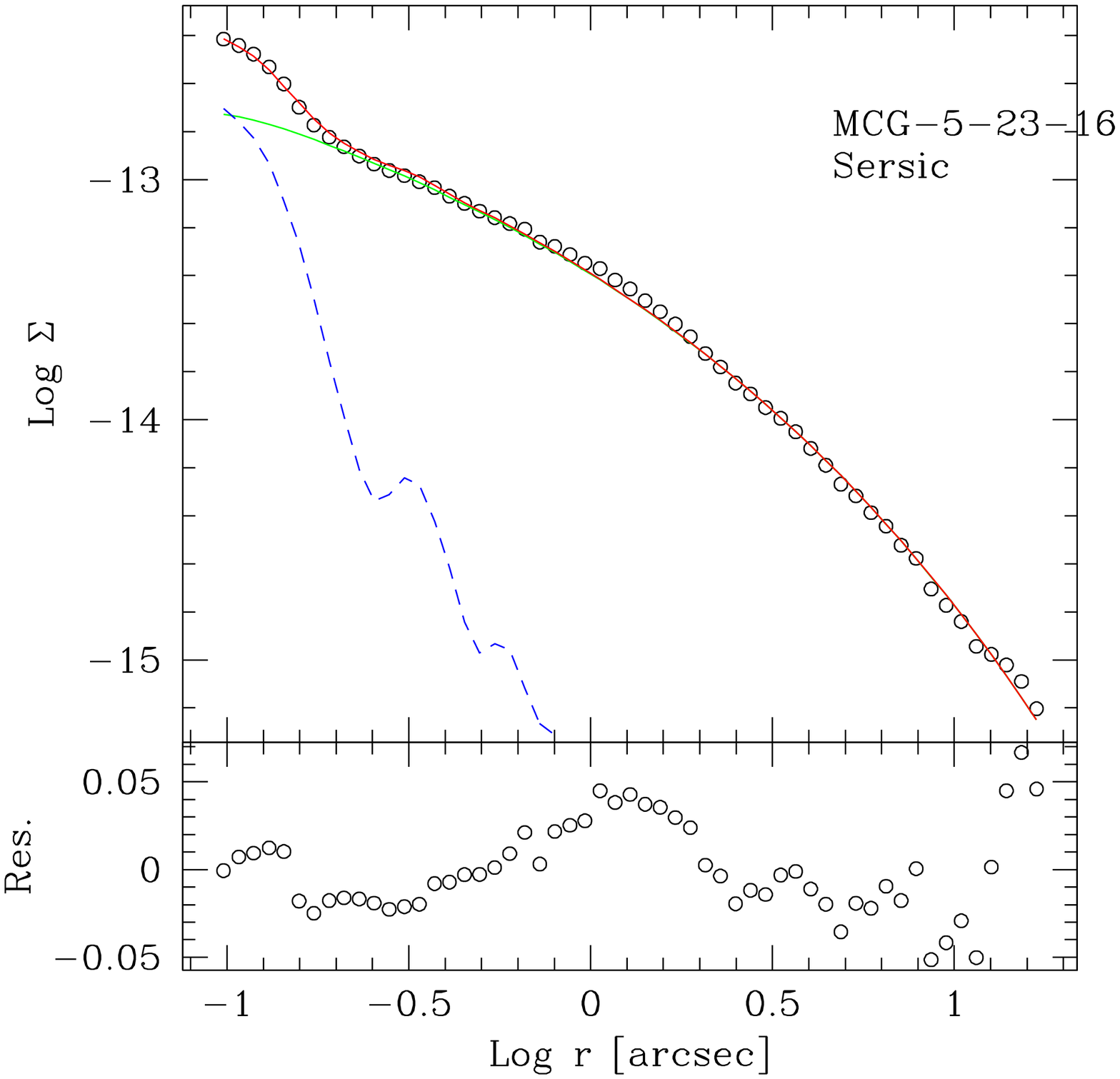,width=0.25\linewidth,height=0.22\linewidth}
\psfig{figure=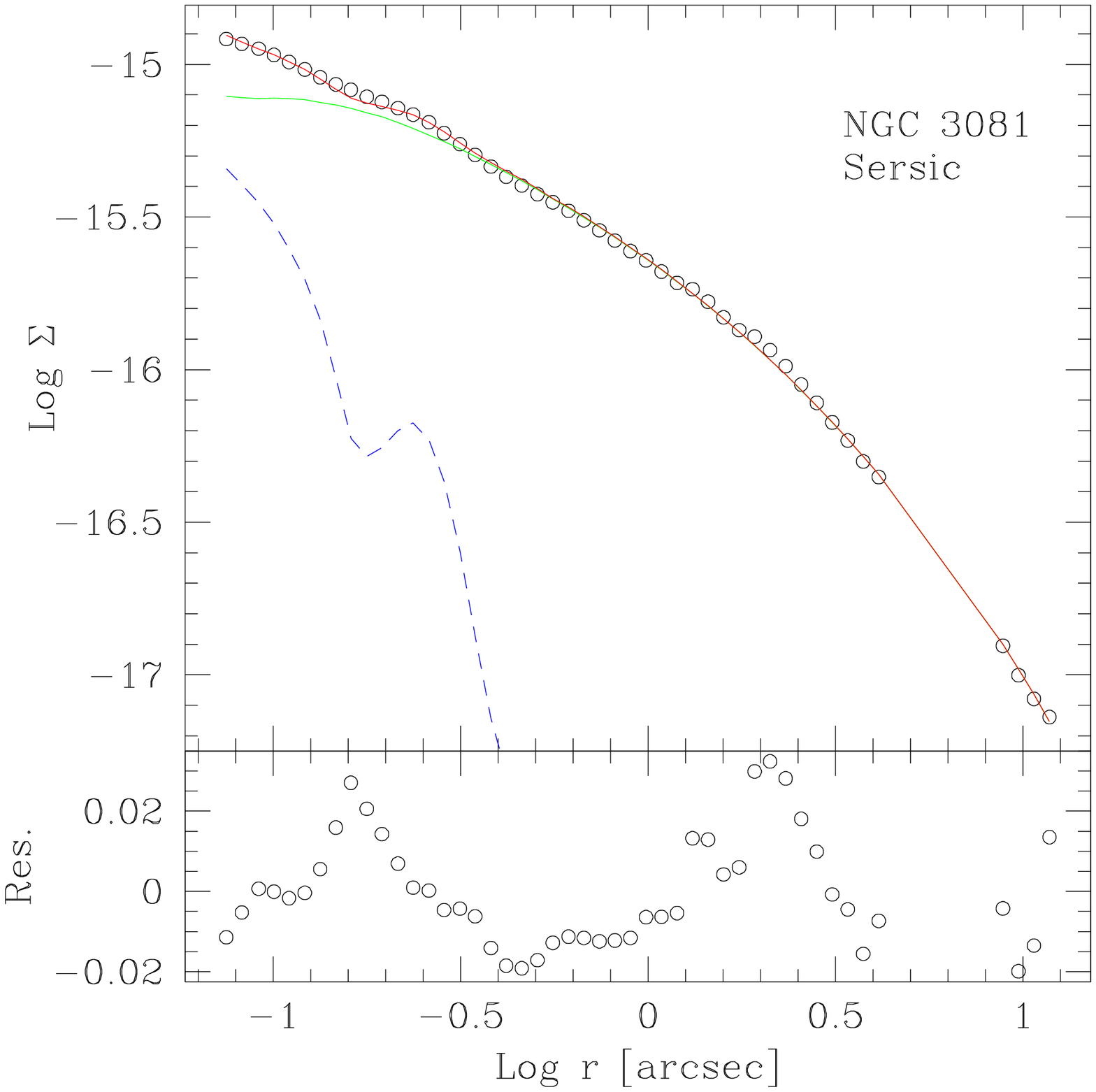,width=0.25\linewidth,height=0.22\linewidth}
}                    
\centerline{         
\psfig{figure=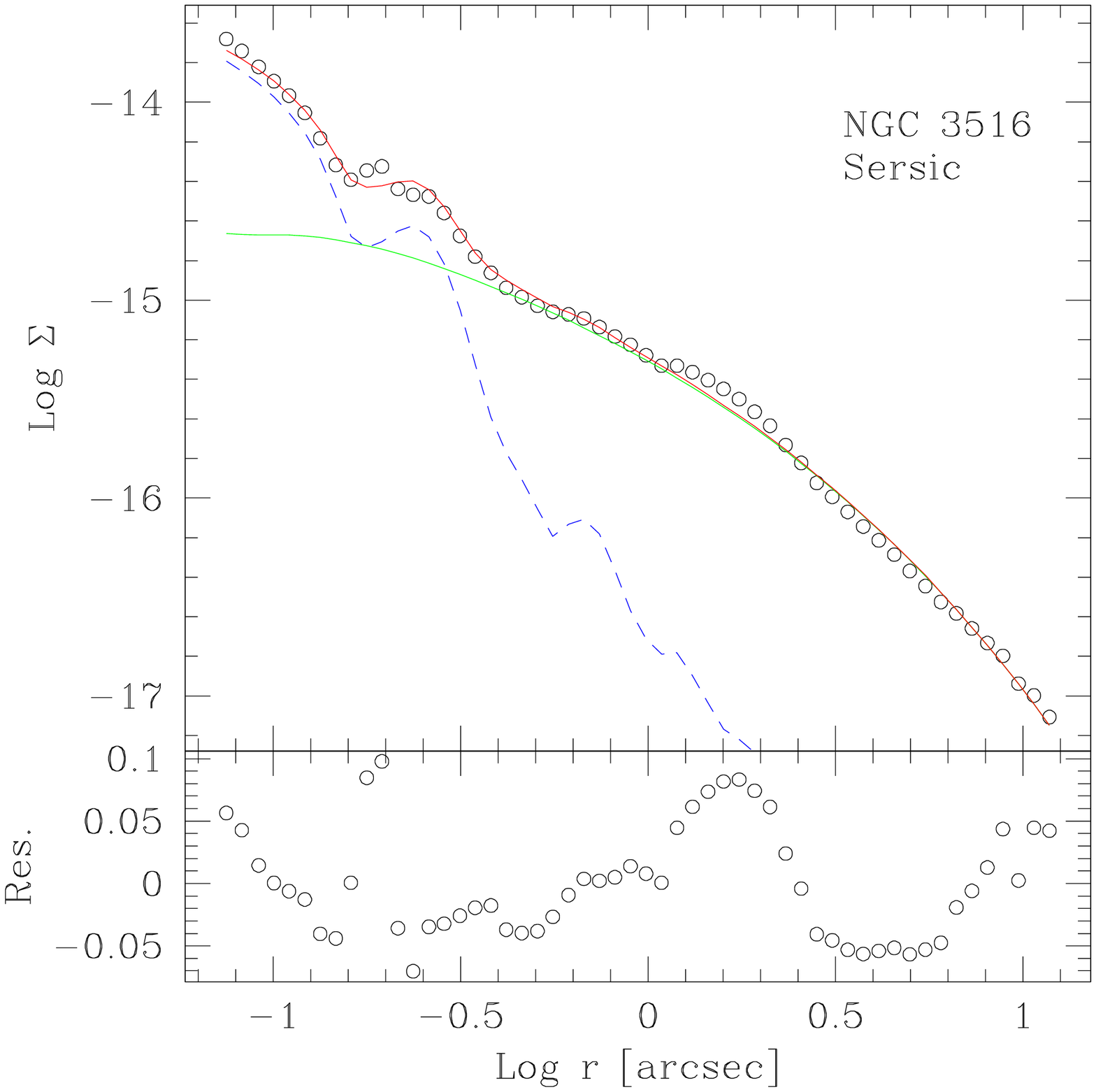,width=0.25\linewidth,height=0.22\linewidth} 
\psfig{figure=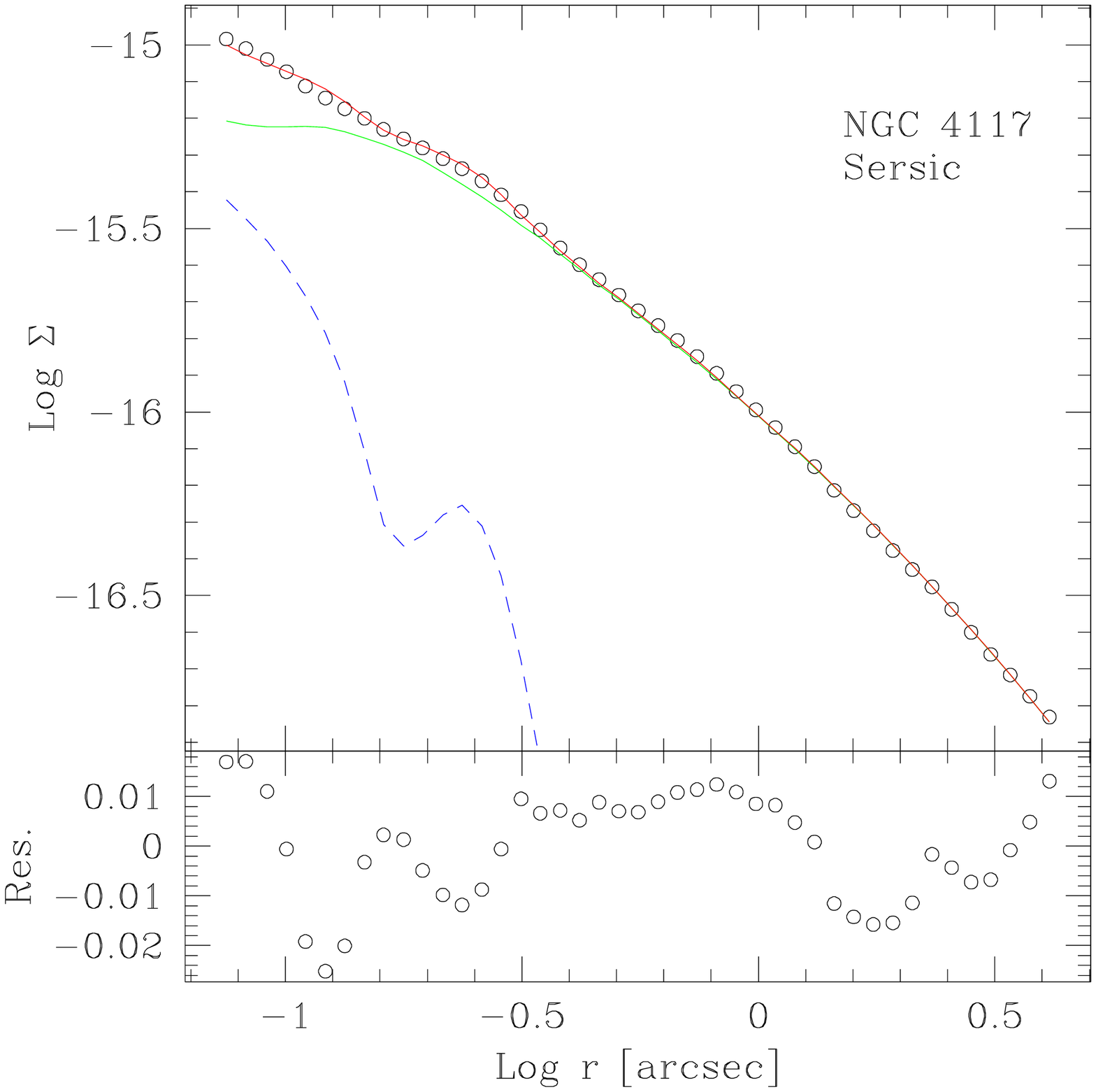,width=0.25\linewidth,height=0.22\linewidth}
\psfig{figure=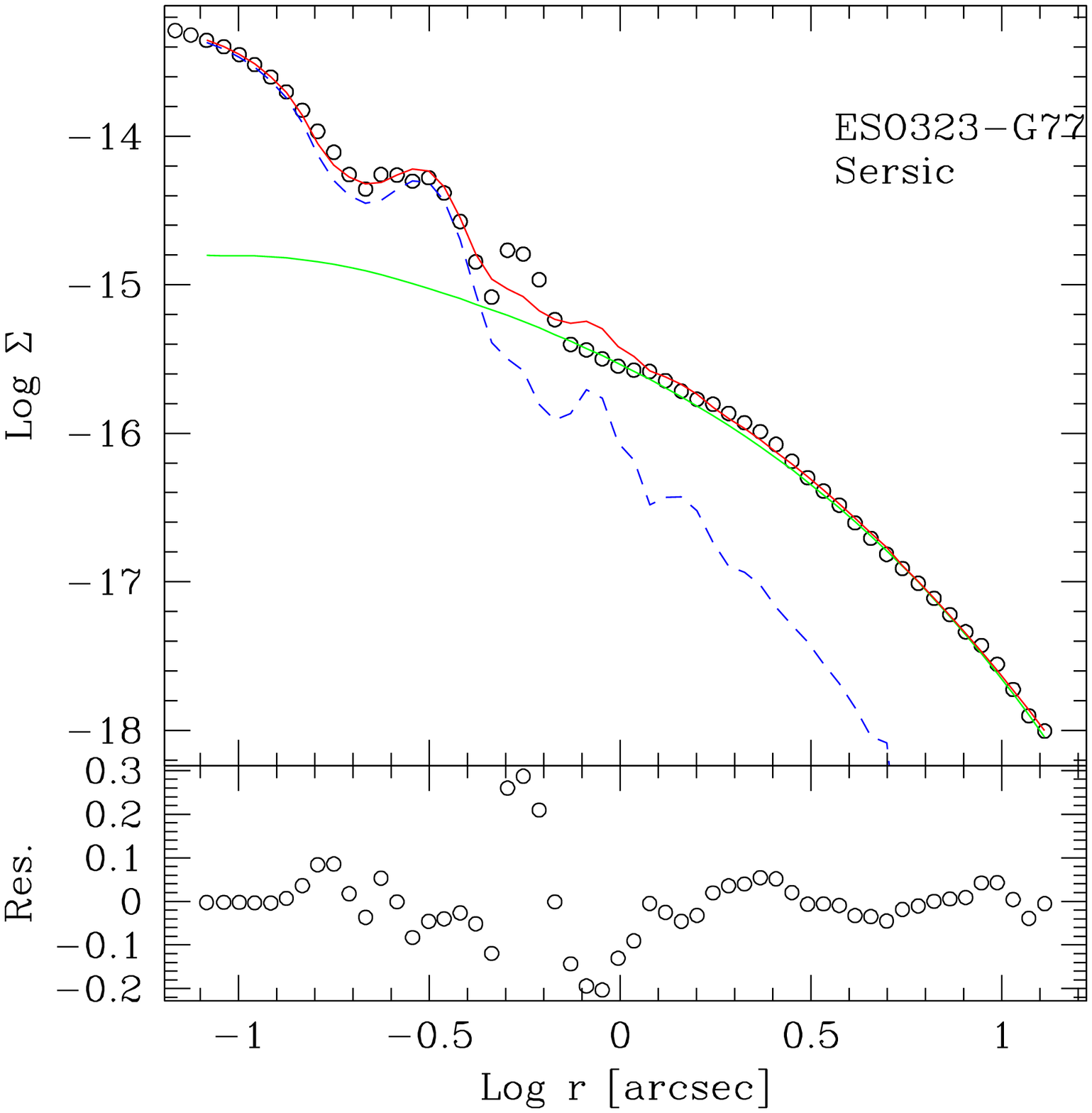,width=0.25\linewidth,height=0.22\linewidth}
\psfig{figure=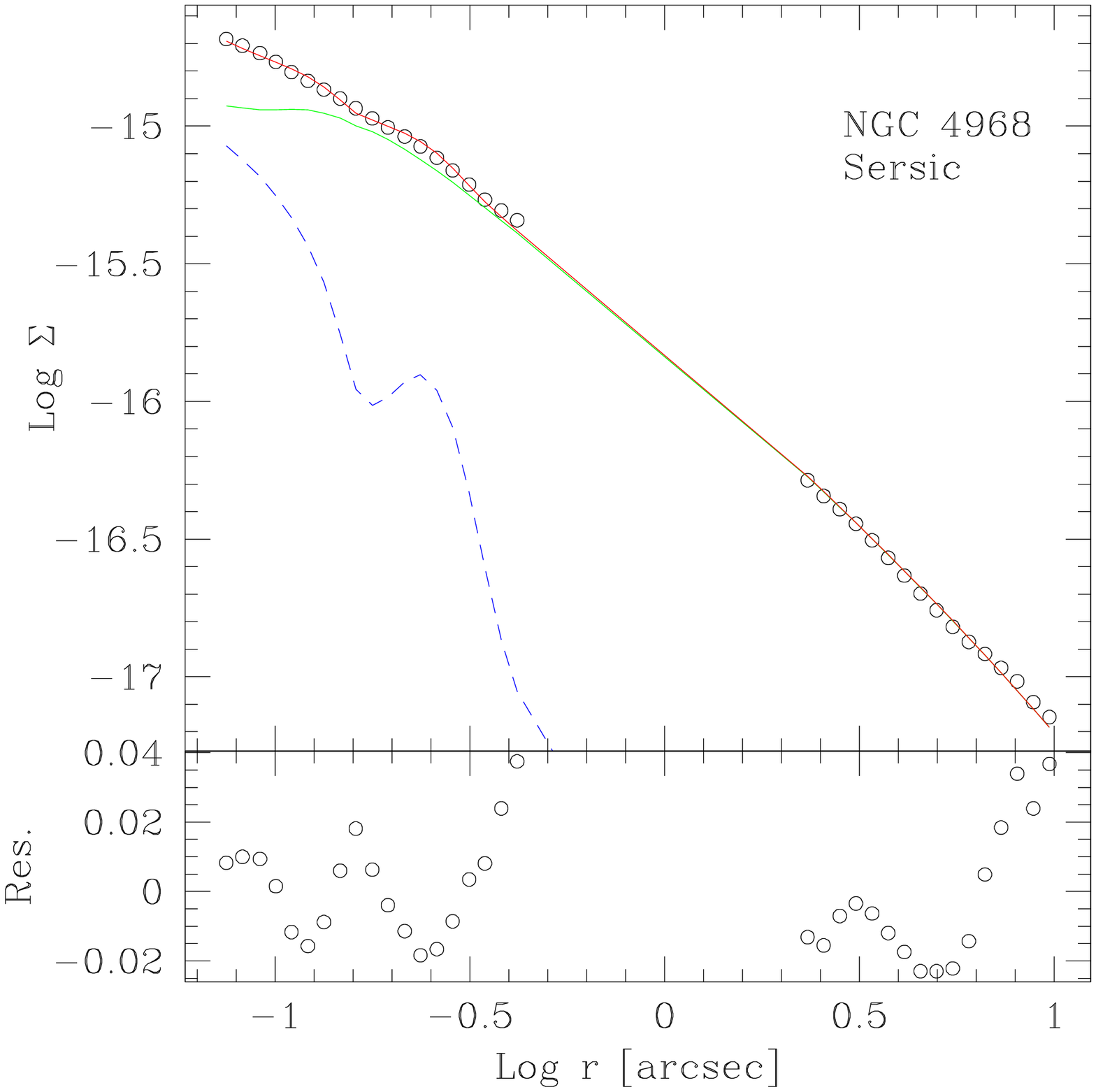,width=0.25\linewidth,height=0.22\linewidth} 
}                    
\centerline{         
\psfig{figure=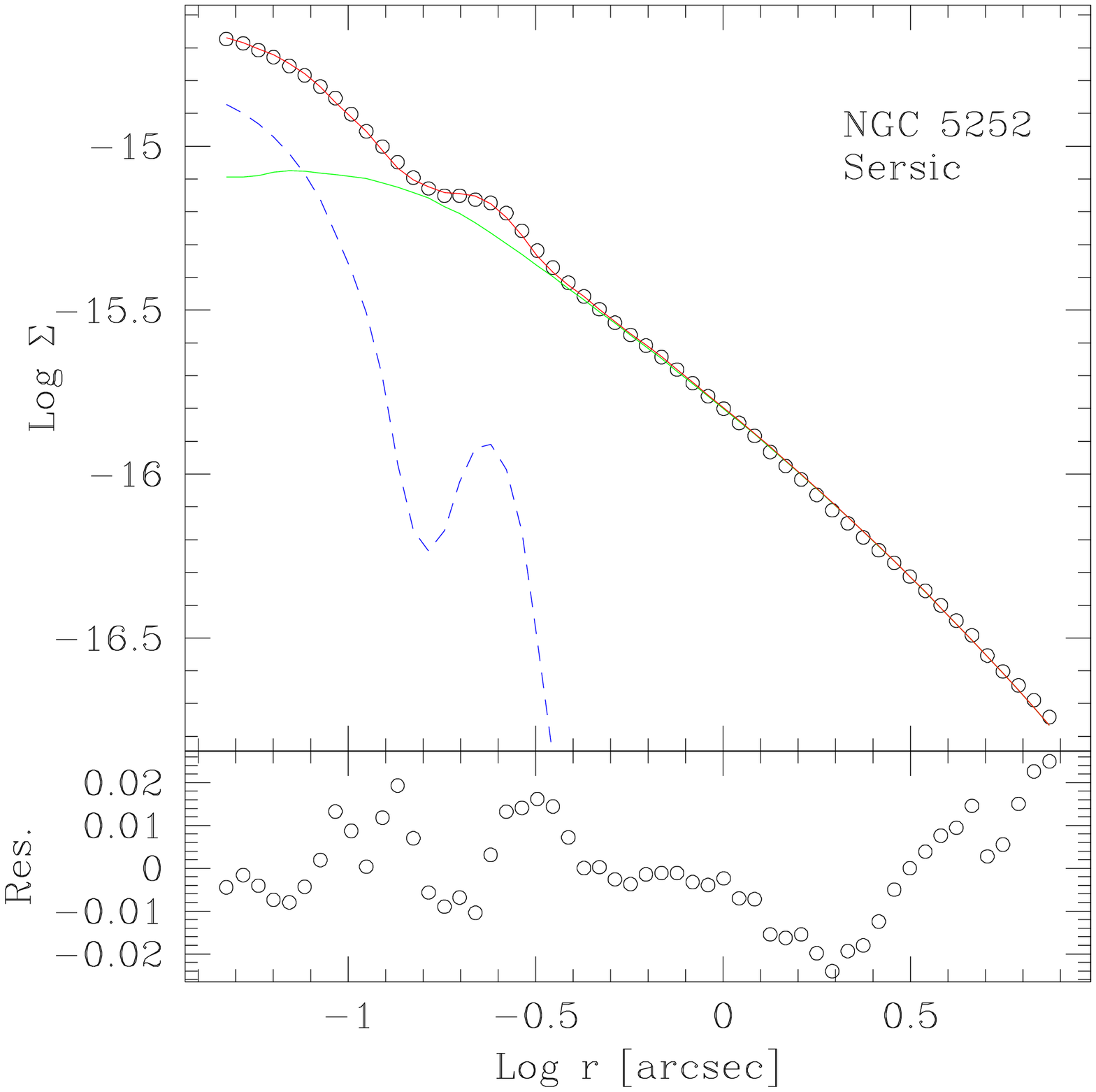,width=0.25\linewidth,height=0.22\linewidth}
\psfig{figure=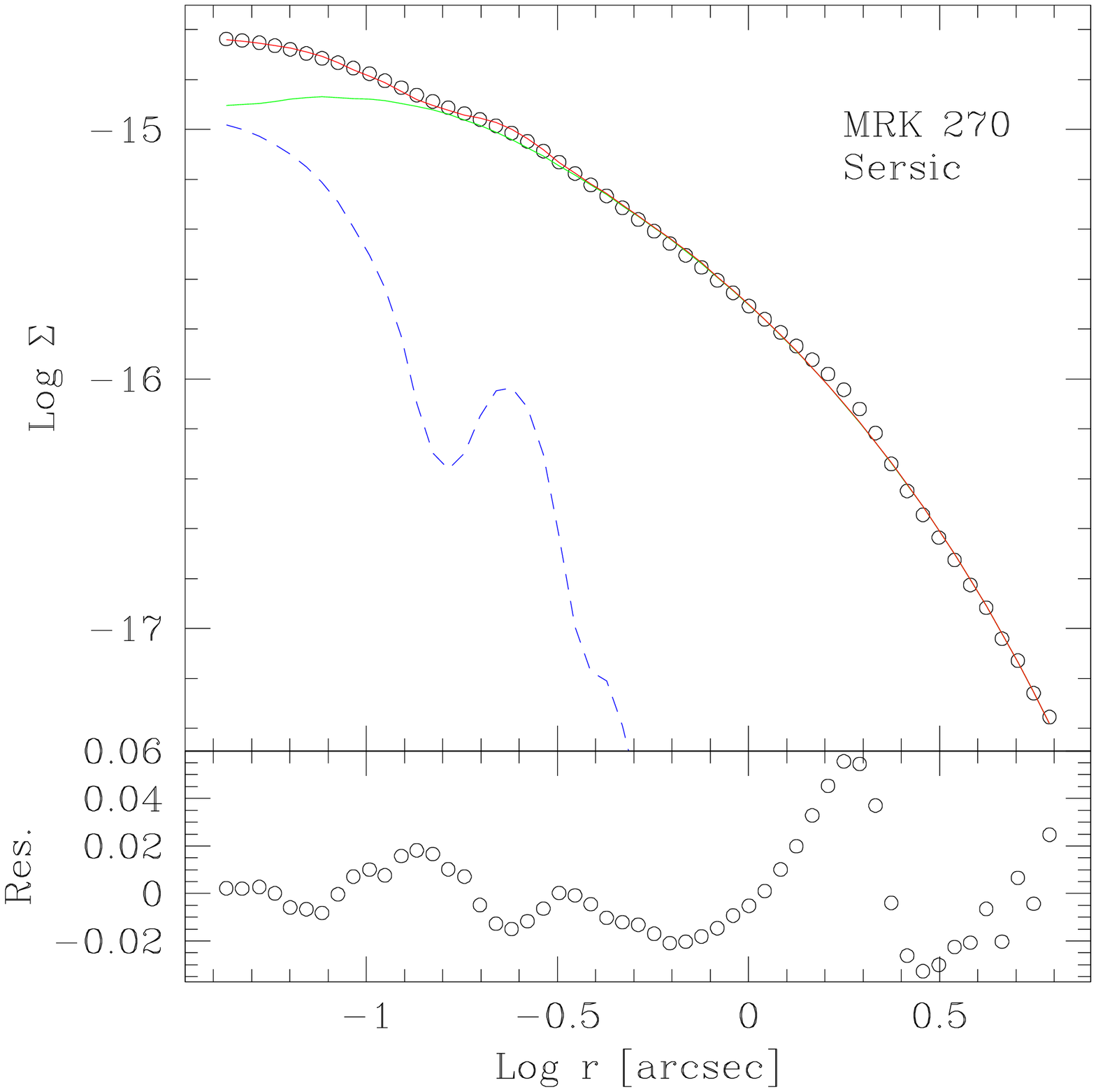,width=0.25\linewidth,height=0.22\linewidth}
\psfig{figure=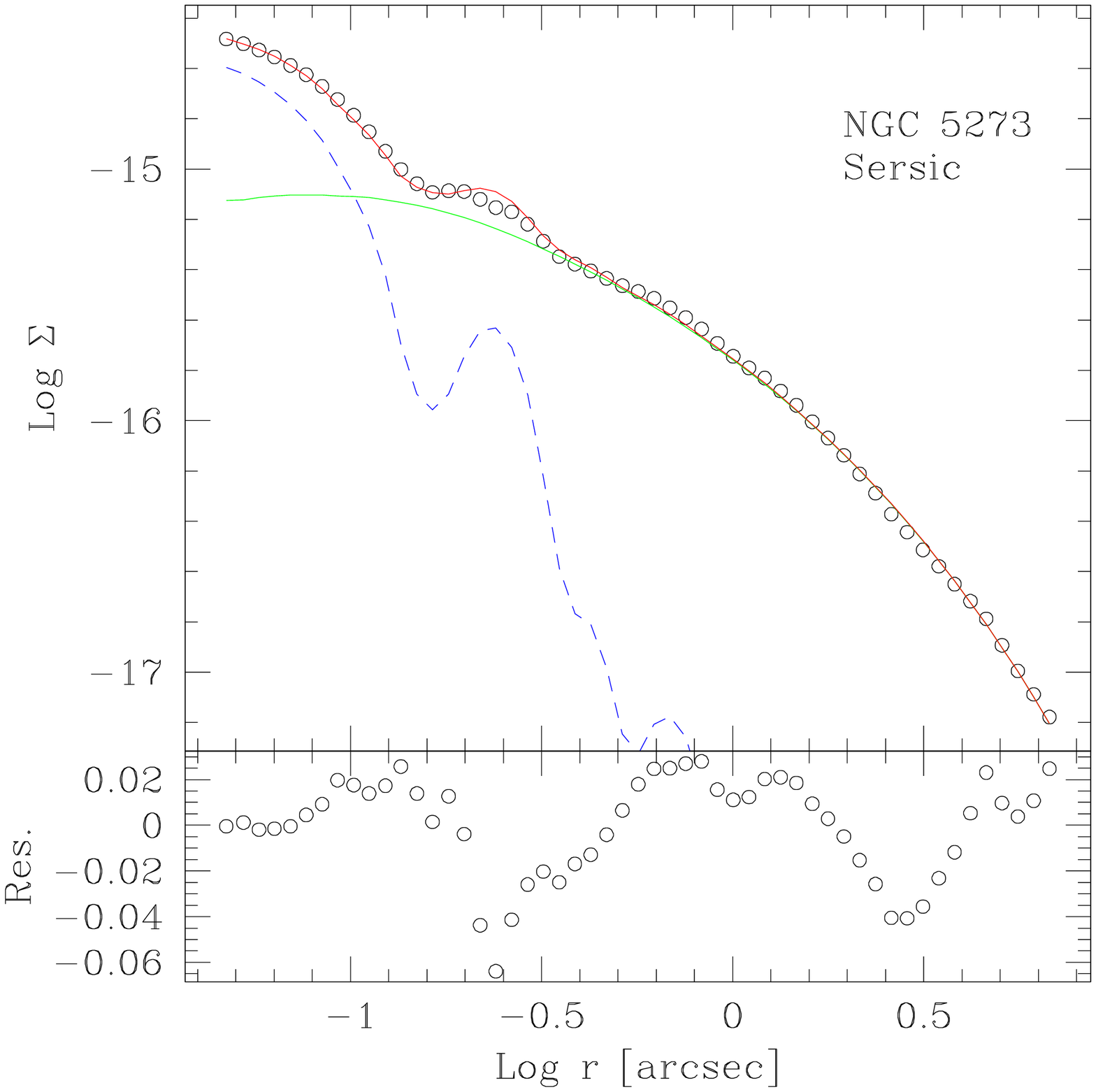,width=0.25\linewidth,height=0.22\linewidth} 
\psfig{figure=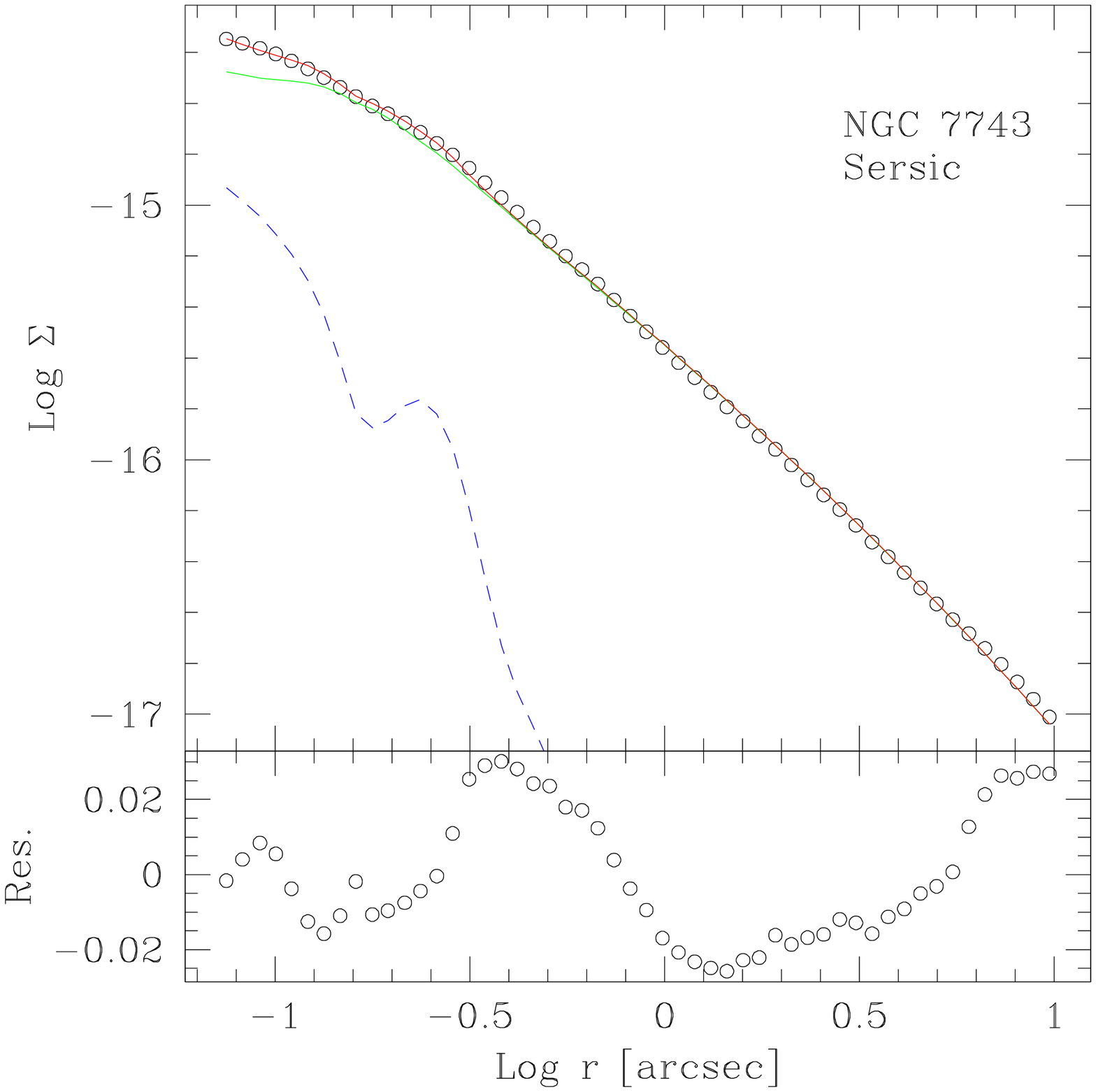,width=0.25\linewidth,height=0.22\linewidth}
}
\caption{Fit (solid line) to the observed brightness profiles 
(in erg s$^{-1}$ cm$^{-2}$ \AA$^{-1}$ units)
obtained with a S\'ersic law, having excluded complex
and nucleated sources. The dotted and dashed lines 
represent the contribution
of the galaxies light and nuclear sources respectively. 
Residuals of the fit are given in the bottom panel.}
\label{sbser1}
\end{figure*}

Leaving aside the 6 nucleated galaxies, we performed a fit
on the remaining 16 galaxies of the sample using a \citet{sersic68} 
law for the brightness distribution given by the expression 
$$
I(r)=I_e \exp \left\{ -b_n \left[ \left( \frac{r}{r_e}\right)^{1/n}-1
 \right] \right\},
$$
where $n$ is the shape parameter, 
$I_{e}$ is the intensity at the half-light radius $r_{e}$.
The quantity $b_n$ is a function of $n$, and is 
defined so that $r_{e}$ is the radius enclosing half 
the light of the galaxy model; it can be approximated by 
$b_n\approx 1.9992n-0.3271$, for 1 \ltsima n \ltsima 10 
\citep[see e.g.][]{graham03}.

The fitting results are presented
graphically in Fig. \ref{sbser1}, 
and tabulated in Table \ref{tabser}.
All profiles are well fitted with a S\'ersic law,
with an accuracy similar to that obtained 
with a Nuker law. In several cases, however, the
behavior at large radii, while still substantially well reproduced
by the model, shows residuals in the form of
large scale fluctuations with a typical amplitude of 0.05 dex.

\begin{table}
\caption{ S\'ersic parameters for the Seyfert sample.}
\label{tabser}
\centering
\begin{tabular}{l c c c c}
\hline\hline
Name          & Log $I_e$ & $r_e$ & $n$ &  $\gamma_{0.1}$ \\
\hline	 
MRK 573      & -16.15 &  1.81 & 1.85  & 0.38\\
NGC 788      & -18.23 & 49.26 & 6.22  & 0.72\\
MRK 607      & -17.90 & 48.16 & 5.58  & 0.64\\
NGC 1358     & -13.82 &  7.14 & 2.83  & 0.42\\
ESO 362-G08  & -15.85 &  1.85 & 1.93  & 0.40\\
NGC 2110     & -16.76 &  9.04 & 2.71  & 0.36\\
MCG -5-23-16 & -14.72 &  9.37 & 3.20  & 0.46\\
NGC 3081     & -16.85 &  8.35 & 2.77  & 0.38\\
NGC 3516     & -16.28 &  4.80 & 2.72  & 0.45\\
NGC 4117     & -17.50 &  9.75 & 4.39  & 0.68\\
ESO 323-G77  & -16.16 &  2.54 & 2.40  & 0.48\\
NGC 4968     & -18.19 & 35.63 & 6.95  & 0.84\\
NGC 5252     & -18.43 & 80.61 & 5.99  & 0.63\\
MRK 270      & -16.16 &  1.92 & 2.31  & 0.51\\
NGC 5273     & -16.49 &  3.20 & 2.28  & 0.40\\
NGC 7743     & -18.09 & 36.69 & 9.10  & 1.02\\
\hline	     
\end{tabular}

Notes: the effective radius $r_e$ is given in arcsec, while $I_e$ 
is in a Log scale in 
erg s$^{-1}$ cm$^{-2}$ \AA$^{-1}$ units.
$\gamma_{0.1}$ is the corresponding logarithmic slope
of the brightness profile evaluated at 0\farcs1.
\end{table}

The defining element of
a core-S\'ersic galaxy is the presence of a light deficit
with respect to the S\'ersic law \citep{trujillo04}. 
This should manifest itself in a characteristic S-shaped pattern
in the residuals when using a pure S\'ersic model, with a central deficit
and a larger scale excess. In no case we see such a signature.

Nonetheless,
since our sample is formed by galaxies which are at larger
distances than those considered by us in CB05 (the
limit on the recession velocity for this sample is 3000 \kms, while the median
here is 2900 \kms) and by \citet{ferrarese06} 
there is the possibility that genuine ``core'' galaxies
might be misclassified as pure S\'ersic
due to an insufficient physical resolution of the images.
However, the median core size measured for the sample of nearby early-type 
galaxies studied in CB05 is $\sim$ 200 pc, similar
to the values found by \citet{faber97} et al. (250 pc) and
by \citet{ferrarese06} in the Virgo survey (150 pc), 
considering only ``core'' galaxies. Given the distances
to each of the 16 galaxies considered here, a scale of 200 pc
corresponds to 0\farcs46 - 2\farcs8, with a median of 1\farcs1,
far larger than the HST resolution.

The presence of point sources also limit our ability to see
shallow cores. We then estimated the size of the region
significantly compromised by the presence of the nucleus as the radius
$r_{\rm nuc}$ at which it produces a contribution 
of 10 \% of the galaxy's starlight. 
The value of $r_{\rm nuc}$ is always smaller than 70 pc, with a median of only 
25 pc, indicating that in general the nuclear emission
does not hamper the detection of a stellar core with a typical size
of 200 pc.

We conclude that the presence of a light deficit 
extended over the typical core size seen in other samples 
studied in the literature would have been easily seen in the light profiles
presented here, despite the larger distances of the galaxies of
the sample and the presence of prominent nuclear point sources.
The lack of such signature indicates that the Seyferts hosts 
can be considered as pure-S\'ersic galaxies,
without evidence for the presence of core-S\'ersic 
galaxies in this sample.

\section{Multiwavelength properties 
of the Seyfert nuclei}
\label{nuc}

To study the nuclear properties of our Seyfert sample we searched the
literature for published emission-line and X-ray nuclear luminosities.
With respect to our previous works, we will not use the information on
the optical nuclei. In fact, in the case of Seyfert 2 galaxies
that form the majority of this sample, it is well established that
their optical emission is substantially affected by nuclear
obscuration. Conversely, observational evidence suggests that the X-ray
(once corrected for absorption) and narrow emission line luminosities
provide sound orientation-independent measures of the intrinsic
luminosity of the nuclei of AGNs \citep[e.g.][]{mulchaey94}.

\begin{table*}
\caption{X-ray and optical spectral information}
\label{tab2}

\begin{tabular}{l | c l c c|c c}

\hline		      	    
\hline		      	    
Name &\multicolumn{4}{|c|} {X-ray data summary} & \multicolumn{2}{c}{Line information} \\
     &  Observation date & Instrument & F(2-10 keV)  & Ref. &  F$_{[OIII]}$ & Ref.\\
\hline		      	    
MRK 335     & 1993Dec          & ASCA          & 9.2E-12    & (23) & 2.0E-13     & (30)   \\
MRK 348     & 1995Aug          & ASCA          & 4.80E-12   & (2)  &  2.8E-13    & (12)   \\
NGC 424     & 2002Feb/2001Dec  & Chandra/XMM   & 1.60E-12   & (3)  &  2.4E-13    & (18)   \\
NGC 526A    & 1995Nov          & ASCA          & 3.44E-11   & (20) &  3.1E-13    & (12)   \\
NGC 513     &                  &               &  	    &      &  1.5E-13    & (12)   \\
MRK 359     & 2000Jul	       & XMM           & 1.26E-11   & (24) & 1.1E-13     & (12)   \\
MRK 1157    &                  &               & 	    &      &  2.1E-13    & (12)   \\  
MRK 573     & 1979Jul          & Einstein      & 5.6E-13    & (4)  &  3.5E-12    & (4)    \\  
NGC 788     &                  &               & 4.6e-12    &      &  2.9E-13    & (12)   \\
ESO 417-G6  &                  &               &	    &      &  1.1E-13    & (12)   \\
MRK 1066    & 1997Aug          & ASCA          & 3.60E-13   & (5)  &  1.7E-13    & (12)   \\
MRK 607     &                  &               & 	    &      &  2.4E-13    & (12)   \\
MRK 612     &                  &               & 	    &      &  1.8E-13    & (16)   \\
NGC 1358    & 1980Aug          & Einstein      & 1.40E-13   &(6)   &  1.1E-13    & (12)   \\
NGC 1386    & 1995Jan          & ASCA          & 3.88E-13   & (7)  &  7.9E-13    & (13)   \\
ESO 362-G8  &                  &               &	    &      &  3.1E-13    & (12)   \\
ESO 362-G18 &	               &               &            &	   & 3.4E-13     & (12)   \\
NGC 2110    & 1997Oct          & BeppoSAX      & 3.00E-11   & (8)  &  1.8E-13    & (14)   \\
MRK 3       & 1997Apr          & BeppoSAX      & 6.50E-12   & (9)  &  2.8E-12    & (17)   \\
MRK 620     & 1996Oct          & ASCA          & 1.20E-13   & (7)  &  1.1E-13    & (12)   \\
MRK 6       & 1997Apr          & ASCA          & 1.0E-11    & (25) & 1.48E-12	 & (29)   \\
MRK 10      & 1990/1991	       & ROSAT         & 4.82E-12   & (28) & 1.4E-13     & (32)    \\
MRK 622     &                  &               &  	    &      &  3.0E-14    & (12)   \\
MCG -5-23-16& 1998Apr          & BeppoSAX      & 10.50E-11  & (10) &   1.7E-13   & (19)   \\
MRK 1239    & 1990/1991        & ROSAT         & 4.03E-14   & (28) & 2.1E-13     & (12)   \\
NGC 3081    & 1996Dec          & BeppoSAX      & 13.30E-13  & (11) &   4.9E-13   & (15)   \\
NGC 3516    & 1994Apr          & ASCA          & 7.80E-11   & (23) & 4.8E-13     & (32)   \\
NGC 4074    &                  &               &	    &      &   7.0E-14   & (12)   \\
NGC 4117    & 1997Dec          & ASCA          & 3.71E-12   & (21) &   7.0E-14   & (12)   \\
NGC 4253    & 1990Dec          & ROSAT         & 1.33E-11   & (27) & 4.3E-13     & (12)   \\
ESO 323-G77 &	               &               &            &	   & 1.8E-13     & (12)   \\
NGC 4968    & 1994Feb          & ASCA          & 3.56E-12   & (20) &   2.1E-13   & (19)   \\
MCG -6-30-15& 1994Jul          & ASCA          &  4.6E-11   & (23) & 3.6E-15     & (16)   \\
NGC 5252    & 1994Jan          & ASCA          & 5.72E-12   & (20) &  2.7E-14    & (1)    \\
MRK 270     & 1979Apr          & Einstein      &$<$2.85E-11 & (22) &   2.7E-13   & (12)   \\
NGC 5273    & 1990/1991        & ROSAT         &  1.0E-13   & (26) & 1.2E-13     & (19)   \\
IC 4329A    & 1993Aug          & ASCA          &  7.8E-11   & (23) & 2.7E-13     & (12)   \\
NGC 5548    & 1993Jul          & ASCA          &  4.3E-11   & (23) & 1.1E-13	 & (31)	  \\
ESO 512-G20 &	               &               &            &	   & 6.0E-15     & (12)   \\
IC 5169     &                  &	       &            &      &   1.0E-14   & (12)   \\
NGC 7465    &                  &	       &            &      &   2.9E-13   & (12)   \\
NGC 7743    & 1998Dec          & ASCA          & 7.30E-14   & (7)  &   5.7E-14   & (13)   \\
\hline						 
\end{tabular}

Column description:
(1) optical name,
(2) observation date, 
(3) Instrument,
(4) X-ray flux in the 2-10 keV band,
(5) reference for the X-ray analysis (see below for the list),
(6) [O III] emission line flux [erg cm$^{-2}$ s$^{-1}$],
(7) reference for the line flux.

References:
(1)\citet{gu06},
(2) \citet{awaki00}, (3) \citet{matt03},
(4) \citet{ulvestad83}, (5) \citet{levenson01},
(6) \citet{fabbiano92}, (7) \citet{terashima02}, (8) \citet{malaguti99},
(9) \citet{cappi99}, (10) \citet{risaliti02}, (11) \citet{maiolino98}, (12)\citet{mulchaey96},
(13)\citet{bergmann89}, (14)\citet{shuder80}, (15)\citet{durret86}, (16)\citet{shuder81},
(17)\citet{koski78}, (18)\citet{murayama98}, (19)\citet{ferruit00}, (20)\citet{turner97},
(21) \citet{terashima00}, (22)\citet{kriss80}, (23)\citet{reynolds97}, (24)\citet{obrien01},
(25)\citet{feldmeier99}, (26)\citet{roberts00}, (27)\citet{molendi93}, (28)\citet{pfefferkorn01},
(29)\citet{whittle88},(30)\citet{kuraszkiewicz00}, (31)\citet{wilson89}, (32)\citet{whittle92}
\end{table*}

The X-ray data available in the literature are 
tabulated in Table \ref{tab2}. 
The X-ray measurements are
very heterogeneous, based on observations of different satellites
(ASCA, ROSAT BeppoSax, Einstein, XMM, and Chandra) obtained over more
than 20 years. Although these observations have a large range of
resolution and sensitivity, Seyfert galaxies are associated to
sufficiently bright X-ray sources that they can be measured at a
sufficient level of accuracy for our purposes even with X-ray
telescopes of past generations. We only discarded ROSAT data for the
Seyfert 2 galaxies since in its energy range (0.1-2.5 keV) the thermal
host emission, and not the highly absorbed non-thermal nucleus, is
likely to be the dominant component at these low energies. When more
than one measurement was available, we referred the more recent
measurement or the one at higher resolution. We rescaled the X-ray
luminosities (see Table \ref{lumfr1}) to our adopted distance and
converted to the 2-10 keV band, using the published power-law index.

From the literature we also collected the [OIII] emission line fluxes.
They are given in Table \ref{tab2}, which includes the relative
references, while the derived emission line luminosities can be found
in Table \ref{lumfr1}.

As for the X-ray data, line luminosities are obtained with different 
methods (imaging and spectrophotometry) and apertures. Nonetheless,
since the [O III] emission tends to be strongly nuclearly concentrated
\citep[see e.g. ][]{mulchaey96} it does not strongly depend on
aperture. Furthermore, 
our analysis is based on orders of magnitude effects
and as such it is quite stable against changes 
within a factor of a few in the line (or X-ray) luminosities.

\begin{table}
\caption{Multiwavelength luminosity of the Seyfert galaxies}
\label{lumfr1}
\begin{tabular}{l c c c c c}
\hline \hline
Name     & Log $\nu$ L$_{r}$  & Log L$_{x}$  & Log L$_{[OIII]}$  & M$_K$  & \\
\hline  
MRK 335     &   38.10  &  43.06    &  41.40    &  -25.00 & \\     
MRK 348     &   39.73  &  42.34    &  41.10    &  -23.85 &\\
NGC 424     & 	37.78  &  41.20	   &  40.37    &  -23.16 &\\
NGC 526A    &   38.20  &  43.34    &  41.29    &  -23.88 &\\
NGC 513     & 	35.51  &  --	   &  40.98    &  -24.42 &\\
MRK 359     &	37.14  &  42.85	   &  40.79    &  -23.71 & \\
MRK 1157    & 	38.06  &  --	   &  40.99    &  -23.91 &\\
MRK 573     & 	37.70  &  41.50	   &  42.29    &  -23.79 &\\
NGC 788     & 	37.03  &  42.15    &  40.95    &  -24.46 &\\
ESO 417-G6  & 	37.37  &  --	   &  40.71    &  -23.75 &\\
MRK 1066    & 	37.85  &  41.03	   &  40.70    &  -23.69 &\\
MRK 607     & 	36.97  &  --	   &  40.54    &  -23.35 &\\
MRK 612     & 	37.93  &  --	   &  41.14    &  -23.88 &\\
NGC 1358    & 	37.17  &  40.66	   &  40.55    &  -24.65 &\\
NGC 1386    & 	36.56  &  39.49	   &  39.79    &  -21.49 &\\
ESO 362-G8  & 	36.94  &  --	   &  41.13    &  -24.33 &\\
ESO 362-G18 &	37.57  &  --	   &  40.96    &  -23.35 & \\
NGC 2110    & 	38.56  &  42.43	   &  40.21    &  -24.05 &\\
MRK 3       & 	38.98  &  42.40	   &  42.03    &  -24.80 &\\
MRK 620     & 	37.53  &  40.03	   &  39.99    &  -23.70 &\\
MRK 6       &	39.07  &  42.90	   &  42.07    &  -24.99 & \\
MRK 10      &	37.63  &  42.91    &  41.38    &  -25.04 & \\
MRK 622     & 	37.96  &  --	   &  40.50    &  -23.80 &\\
MCG -5-23-16& 	34.75  &  43.07	   &  40.28    &  -23.08 &\\
MRK 1239    &	38.43  &  40.44    &  41.16    &  -24.79 & \\
NGC 3081    & 	36.73  &  41.15	   &  40.72    &  -23.47 &\\
NGC 3516    &	37.56  &  43.14	   &  40.93    &  -24.43 & \\
NGC 4074    & 	37.60  &  --	   &  40.85    &  -24.25 &\\
NGC 4117    & 	35.14  &  41.01	   &  39.29    &  -20.87 &\\
NGC 4253    &	38.17  &  42.66    &  41.17    &  -23.82 & \\
ESO 323-G77 &	37.42  &  --	   &  40.86    &  -25.02 & \\
NGC 4968    & 	37.75  &  41.79	   &  40.56    &  -23.42 &\\
MCG -6-30-15&	36.65  &  42.66	   &  38.55    &  -22.72 & \\
NGC 5252    &   38.58  &  42.74    &  40.42    &  -25.01 &\\
MRK 270     & 	37.51  &  42.78	   &  40.75    &  -23.14 &\\
NGC 5273    &	36.04  &  39.56    &  39.64    &  -22.56 & \\
IC 4329A    &	38.39  &  43.55	   &  41.09    &  -25.16 & \\
NGC 5548    &	37.97  &  43.41	   &  40.82    &  -24.86 & \\
ESO 512-G20 &	37.14  &  --	   &  39.14    &  -22.78 & \\
IC 5169     & 	37.53  &  --	   &  39.26    &  -23.19 &\\		      
NGC 7465    & 	36.73  &  --	   &  40.41    &  -22.64 &\\		      
NGC 7743    & 	36.45  &  39.66	   &  39.56    &  -23.39 &\\
\hline 
\end{tabular}

Column description:
(1) name, (2) nuclear radio luminosity (5GHz), Log [erg s$^{-1}$],
(3) intrinsic nuclear X-ray luminosity (2-10 keV), Log  [erg s$^{-1}$],
(4) [O III] emission line luminosity, Log  [erg s$^{-1}$],
(5) total K band absolute magnitude.
\end{table}

\begin{figure*}
\centerline{ 
\psfig{figure=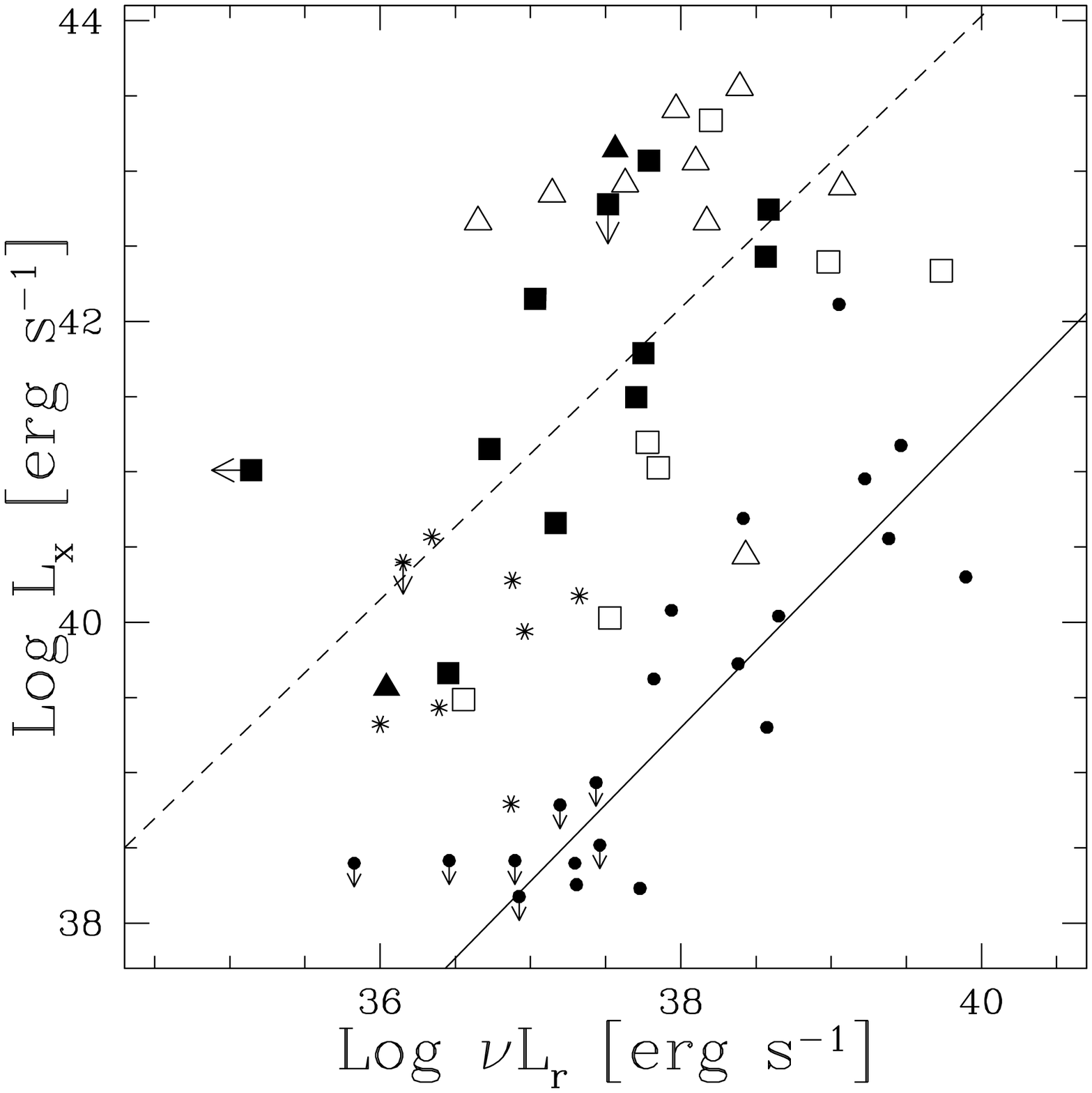,width=0.5\linewidth}
\psfig{figure=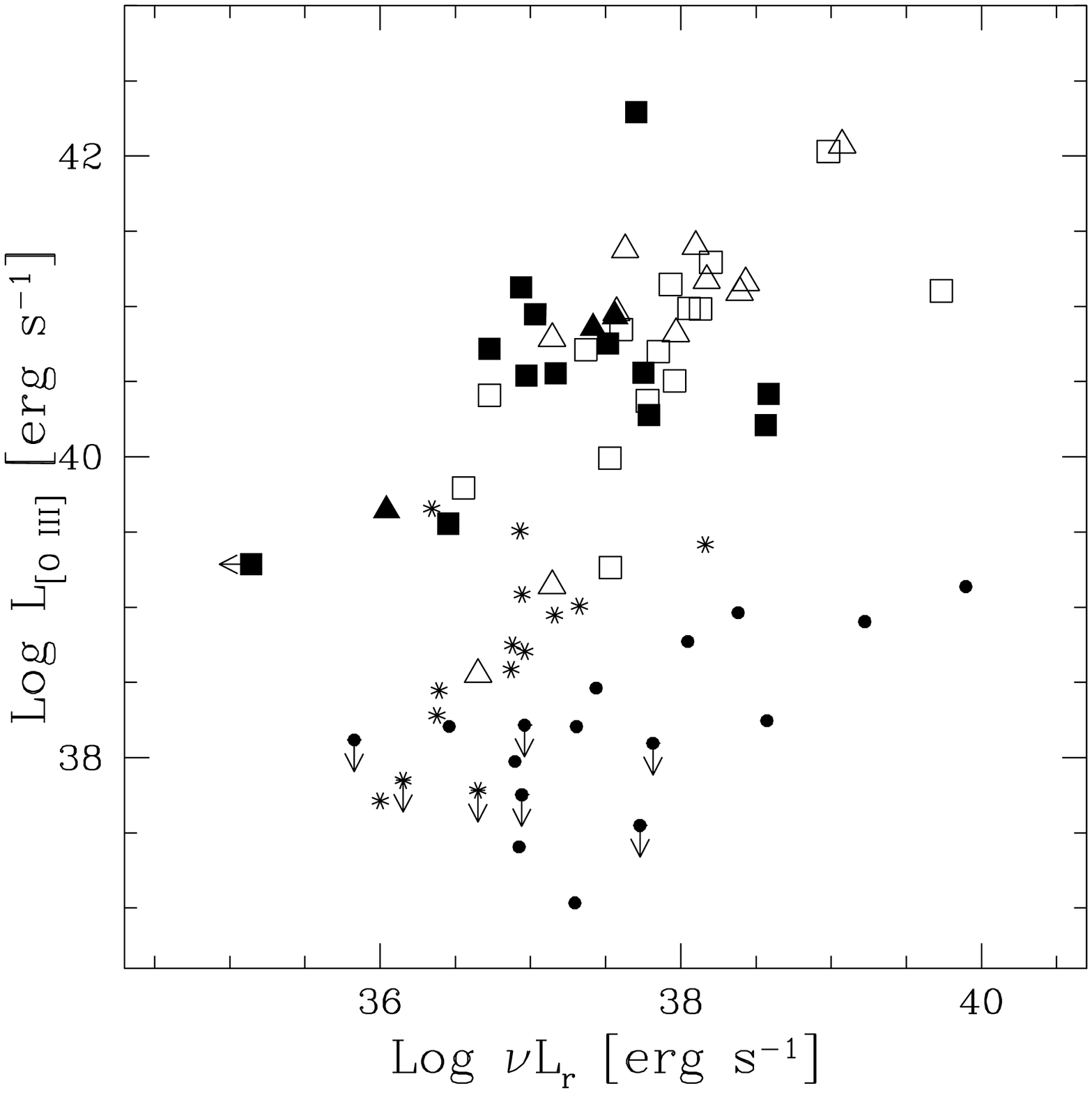,width=0.5\linewidth} }
\caption{Left panel: comparison of radio-core (at 3.6 cm) and X-ray nuclear
luminosity (in the 2-10 keV band) for the sample of Seyfert
galaxies (triangles are Sy 1, squares are Sy 2). Filled symbols
represent Seyfert for which it was possible to obtain and fit
the brightness profile. 
We also report ``power-law'' galaxies (stars) and ``core''
galaxies (filled circles) from \citetalias{paper3}. 
The solid line represents the
correlation derived in \citetalias{paper2} 
between radio and X-ray nuclear luminosity 
of radio-loud AGN, including ``core'' galaxies and 
the 3C/FR~I sample of low luminosity
radio-galaxies from \citet{balmaverde06}.
The dashed line represents the same relation derived by \citet{panessa07}
from their sample of Seyfert galaxies. 
Right panel: comparison of radio and [O III] emission line luminosity,
using the same symbols as in the left panel.}
\label{rx}
\end{figure*}

In the following we compare the nuclear properties of these Seyfert
galaxies with the samples of early-type galaxies we studied in
CB05, BC06, and CB06. In Fig.\ref{rx} (on the left) we
compared X-ray and radio luminosities\footnote{NGC 3516 
AKA UGC 6153 is part both of the Seyfert and ``power-law'' samples.
We mark its representative point in Fig.\ref{rx} only as Seyfert galaxy.}. 
Seyfert galaxies are located
well above the correlation defined by low luminosity radio-galaxies
(LLRG) and ``core'' galaxies (by a factor
$\sim$ 10 - 10$^4$). At the lower luminosities they show a behavior
similar to the radio-quiet ``power-law'' galaxies, but they extend the coverage
by a factor of 100
toward higher radio luminosities and by a factor of 1000 in X-ray 
and line luminosities. 
Furthermore, our Seyfert sample, including only early-type hosts, 
follows a trend
in the  $L_r$ vs $L_x$ plane similar to the Seyfert galaxies from the
Palomar studied by \citet{panessa07}.

In \citetalias{paper3} we showed that radio-loud and radio-quiet AGN,
i.e. LLRG/``core'' 
and ``power-law'' galaxies, are well separated also when comparing
radio and [O III] emission line luminosity, leading to the definition
of a spectroscopic radio loudness parameter, R$_{\rm [O III]}$.
Fig.\ref{rx} (right panel) shows that Seyfert galaxies have a large
excess also of line-emission (at a given radio-core luminosity)
with respect to radio-loud objects.

\begin{figure*}
\centerline{ 
\psfig{figure=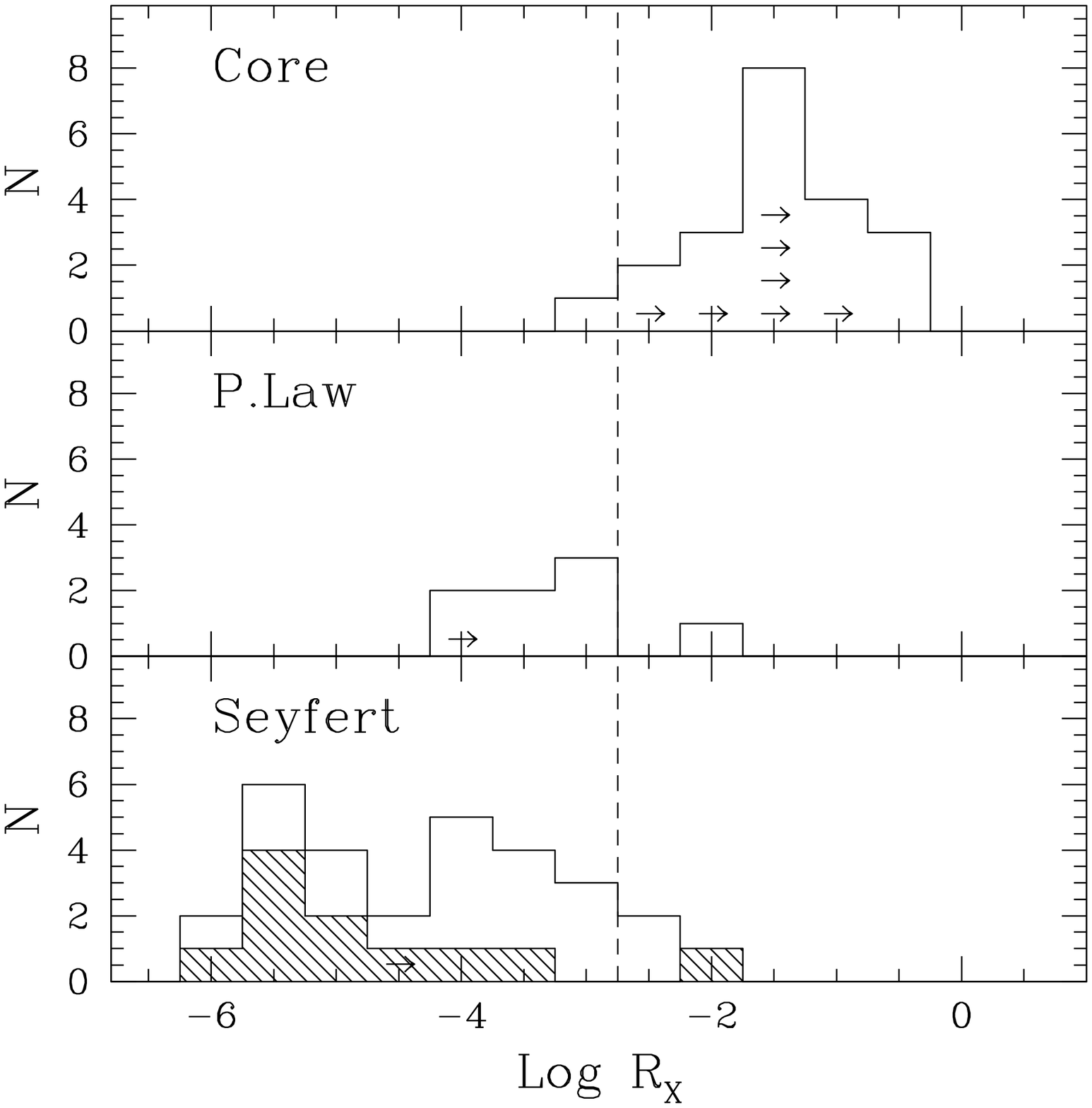,width=0.5\linewidth}
\psfig{figure=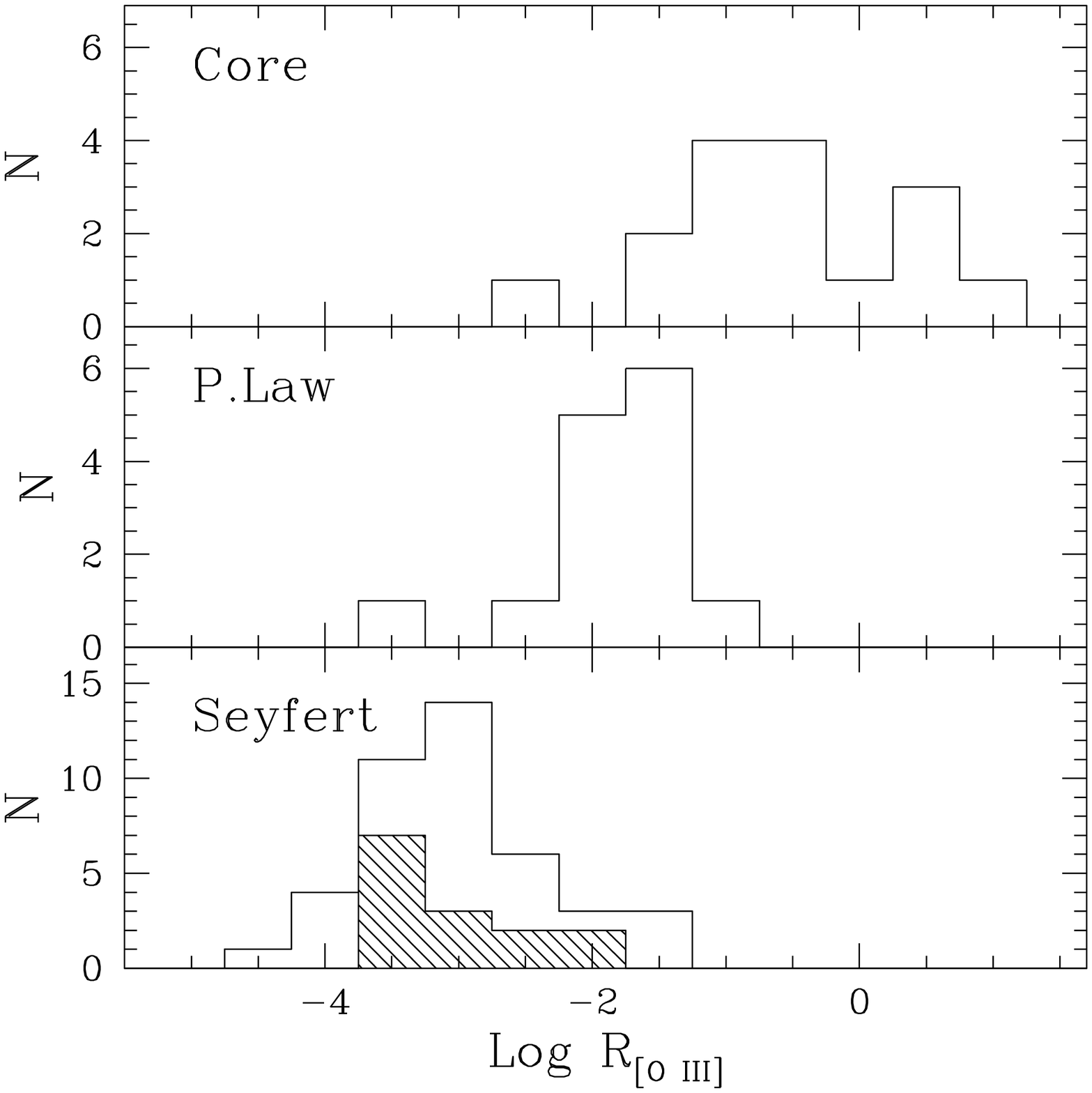,width=0.5\linewidth} }
\caption{
Left panel: distribution of radio loudness parameter
estimated from the ratio of radio and X-ray nuclear luminosity
R$_X=(\nu L_r/L_x)$ compared for the
two sub-samples of early-type galaxies 
we analyzed in \citetalias{paper1} (separated
into ``core'' and ``power-law'' galaxies) and for the present sample
of Seyfert galaxies. The shaded region in the bottom panel marks 
the contribution of type I Seyfert.
The vertical dashed line mark the value of $R_X$ that provides
the best separation between radio-loud and radio-quiet
low luminosity AGN from \citet{panessa07}.
Right panel: same as the left panel, but comparing 
the spectroscopic radio loudness parameter
R$_{[OIII]}=(\nu L_r/L_{[OIII]})$ of the three samples.}
\label{rloud}
\end{figure*}

The distributions of the radio loudness parameters $R_X$ (based on the ratio
between radio luminosity at 5 GHz and the 2-10 keV X-ray luminosity, following
the definition introduced by \citet{terashima03}) \footnote{having converted
 the radio luminosity at 3.6 cm adopting a radio spectral index of $\alpha$
 =1.} and R$_{\rm [O III]}$ estimated from the ratio of [O III] luminosity
and radio
power for the different samples are shown in Fig. \ref{rloud}. Seyfert
galaxies have both radio loudness parameters significantly lower than those
derived for ``power-law'' galaxies, and, a fortiori, of the LLRG and ``core''
galaxies. With only 3 exceptions, Seyfert galaxies
show value of X-ray radio loudness below the
threshold of Log $R_X = -2.8$ derived by \citet{panessa07} that provides
the best separation between radio-loud and radio-quiet
low luminosity AGN.
Furthermore, the dichotomy between radio-quiet and radio-loud AGN becomes much
stronger with the inclusion of Seyfert galaxies. Indeed, in
\citetalias{paper3} we argued that one of our selection criteria for
``power-law'' galaxies, i.e. the detection of a radio source, was likely to
bias the sample toward the inclusion of the radio-quiet AGN with higher
radio loudness parameters and that they were likely to represent only the
tail, toward high values of $R$, of the overall population of radio-quiet AGN.
This is confirmed by the present analysis.

Our results are based on the sub-sample of objects for which are
available HST images suitable for the analysis of the surface
brightness profile, representing about one third of the sample. It is
then 
important to assess whether they provide us with an unbiased representation
of Seyfert galaxies hosted by
early-type galaxies. 
We therefore considered the distributions of $M_{BH}$, $M_K$, 
of the luminosities in radio, line and X-ray, as well as of
the radio loudness parameters $R_X$ and R$_{[OIII]}$. 
According to the Kolmogorov-Smirnov test, 
the probability $P$ that the two samples are
drawn from the same parent distribution is always larger than 0.80,
and the medians differ by at most a factor of 1.3.
The only exception is the radio luminosity (and consequently
the spectroscopic radio loudness parameter
$R_{[O III]}$) for which we found P = 0.38. 
This indicates that the objects for which a surface brightness 
analysis was possible have, on average, a lower radio luminosity. 
However, the medians of the two
distributions (Log L$_r$ = 37.2 and 37.6 respectively) 
differ only by a factor of 2.5, not a substantial offset considering
the range of 4 orders of magnitude in L$_r$ covered by the sample. 
Conversely, we note that the fraction of objects for which it was 
possible to perform
an isophotal analysis differs drastically from objects imaged
in the optical (12\%) and in the infrared (67\%), mostly likely
due to the reduced effects of dust obscuration. 
Apparently, the strongest influence on the possibility
of a successful analysis of the brightness profile is related to
the band in which the HST observations were taken, and not
to the AGN or host properties.
We conclude that the sub-sample of 16 objects with well
behaved brightness profiles 
is well representative of the population of Seyfert hosted
in early-type galaxies, with only a slight preference in favor of
sources with lower radio emission.

\section{Summary and discussion}
\label{summary}

We presented a study of
a sample of 42 nearby (cz$\leq$ 7000 km s$^{-1}$) 
early-type galaxies hosting a Seyfert nucleus.
From the nuclear point of view, they show a large deficit of
radio emission (at given X-ray or narrow line luminosity) with respect
to radio-loud AGN, conforming with their identification with
radio-quiet AGN. With only 3 exceptions, their X-ray based
radio loudness parameter is smaller than the 
threshold value of Log $R_X = -2.8$ introduced by \citet{panessa07} 
to separate radio-loud and radio-quiet low luminosity AGN.

We analyzed their brightness profiles by using archival
HST images. Having discarded complex and highly nucleated
galaxies we were left with a sub-sample of 16 well behaved objects.
By fitting the brightness profiles with a Nukers law,
we found that the nuclear cusps are reproduced with a slope in the range
$\gamma = 0.51 \,-\, 1.07$, typical of ``power-law'' galaxies.

The lack of ``core'' galaxies (i.e. galaxies with 
$\gamma < 0.3 $) is not simply the
consequence of the luminosity of the Seyfert hosts.
In fact, their range of K band absolute magnitudes is 
-22.5 $<$ M$_K$ $<$ -25.2, with only one exception. 
Adopting a color of V-K=3.3 \citep{mannucci01} this translates
into M$_{\rm V}$ = -19.2 -- -21.9 indicating that most objects
lie in the luminosity range 
where ``core'' and ``power-law'' coexist 
\citep[i.e. -20 $<$ M$_{\rm V}$ $<$ -22, ][]{lauer06}.
Nonetheless, only ``power-law'' profiles were found from our analysis.
Although the reference sample of \citeauthor{lauer06} is not complete,
thus preventing us to assess the significance of this result on a statistical
basis, the lack of ``core'' galaxies as hosts of radio-quiet AGN 
confirms the presence of a link between the brightness profile
and the AGN radio loudness.

With respect to the initial series of 3 papers we included here a full
modeling of the brightness profiles also with a S\'ersic model. 
In no object we found evidence for a central light deficit with respect
to a pure S\'ersic model, the defining feature of ``core'' galaxies in
this modeling framework. We also assessed that such a light deficit,
extending over a fiducial core size of 200 pc,
would have been easily seen despite the larger distances of these galaxies 
(relative to other samples studied in the literature)
and the presence of prominent nuclear point sources.

The analysis of the early-type galaxies in the Virgo Cluster Survey
presented in \citet{ferrarese06}
provides us with a useful benchmark for the interpretation
of the results obtained from a S\'ersic fit of the brightness profiles. 
In particular, the correspondence we found 
between the classification as ``power-laws'' 
(in the Nukers scheme) and pure-S\'ersic galaxies naturally emerges
from the inspection of the properties of the \citeauthor{ferrarese06} sample.
Pure-S\'ersic galaxies form a well defined sequence in the 
$\gamma_{0.1}$ vs $n$ plane\footnote{$\gamma_{0.1}$ is the logarithmic slope
of the brightness profile evaluated at 0\farcs1, and it corresponds to
$\gamma_{0.1}=b_n/n \times (r_e/0.1)^{-1/n}$} 
(see their Fig. 166, panel bc) 
with $\gamma_{0.1}$ steadily increasing with $n$,
while ``core'' galaxies stand aside, forming a well separated group;
furthermore, brighter galaxies have larger values of $n$ 
and, on average, also of $\gamma_{0.1}$ (see their Fig. 166, panels ac and ab). 
Thus, in the luminosity range
of our sample, substantially higher than the average for the VCS,
the S\'ersic model predicts rather steep central slopes. 
Indeed, the values of $\gamma_{0.1}$ estimated for the Seyfert hosts
(given in Table \ref{tabser}) 
are in the range 0.36 - 1.02, with a median of 0.48.
Not surprisingly their steep cusps are well reproduced by ``power-laws'' 
profiles in the Nukers scheme.

Similarly, we have already shown in \citetalias{paper2} that galaxies
classified as ``core'' in the Nuker scheme are 
reproduced by a core-S\'ersic profile, i.e. with a well defined
light deficit from a pure S\'ersic model. 
Effectively, when considering
relatively bright galaxies (brighter than $\sim M^{*} +2$), such as
those analyzed here and in our previous papers,
there is apparently a complete correspondence between pure S\'ersic and
``power-law'' galaxies on one side, and 
between core-S\'ersic and ``core'' Nukers galaxies on the other.

It must also be noted that the controversy over the 
analytical form to be used to reproduce the brightness profiles 
is not in whether or not there are two classes of 
early-type galaxies, but in the form of the central structure on the non-core
class. In fact,
despite their different approaches, \citeauthor{lauer06} and
\citeauthor{ferrarese06} both identify ``core'' galaxies
as a separate class of objects. 

Quite reassuringly, the link between the brightness profile
and the AGN radio loudness is 
independent on the fitting scheme since we
recover a unique correspondence between the host's brightness profile and the
AGN properties. Our general results can then be re-phrased as 
{\sl radio-loud nuclei
are hosted by ``core'' galaxies while radio-quiet AGN are found 
in ``non-core'' galaxies.}

In particular, the study presented here enabled us to confirm
that radio-quiet AGN are hosted
by ``non-core'' galaxies. The inclusion of Seyfert galaxies
extends the coverage up to an X-ray luminosity 
of $\sim$ 10$^{43}$ erg s$^{-1}$, a factor of 1000 larger
than the median luminosity of the sample of radio-quiet AGN
we discussed in \citetalias{paper3}. 
Taking the results presented here, in CB06, and
in \citet{deruiter05} (where we showed that low-luminosity radio-galaxies
are hosted by ``core'' galaxies), we have covered the different manifestations
of nuclear activity in the local Universe. We consistently recovered
the association of radio-loud AGN with ``core'' galaxies
and of radio-quiet AGN with ``non-core'' galaxies.
We also confirm that
radio-loud and radio-quiet nuclei cannot be distinguished
on the basis of other parameters, such as the host's luminosity or
the black-hole mass, as they differ only on a statistical basis. 
In particular Seyfert hosts reach an absolute magnitude
of M$_K \sim$ -25 and harbor black holes with masses as high as 
$4 \times 10^8 M_{\sun}$, well into the range of radio-loud AGN.
Only the brightness profiles provide a full separation between the
two classes.

\end{document}